\documentclass[times,authoryear]{elsarticle}

\usepackage[margin=1.5cm]{geometry}
\usepackage{framed,multirow}

\usepackage{amsmath,amssymb}
\usepackage{latexsym}
\usepackage{subfigure}
\usepackage{booktabs,multirow}

\usepackage[switch]{lineno}

\usepackage{url}
\usepackage{xcolor}
\definecolor{newcolor}{rgb}{.8,.349,.1}

\usepackage[citebordercolor=white]{hyperref}

\newcommand{\bbm}{\begin{bmatrix}}
\newcommand{\ebm}{\end{bmatrix}}
\newcommand{\mbs}[1]{{\boldsymbol{#1}}}
\DeclareMathAlphabet{\mbf}{OT1}{ptm}{b}{n}
\newcommand{\trans}{{\ensuremath{\mathsf{T}}}} 

\journal{Advances in Space Research}

\begin{document}


\begin{frontmatter}

\title{Deployable Prototype Testing and Control Allocation of the CABLESSail Concept for Solar Sail Shape Control and Momentum Management}

\author[1]{Soojeong Lee}
\author[1]{Michael States}
\author[1]{Keegan R. Bunker}
\author[1]{Ryan J. Caverly\corref{cor1}}

\ead{rcaverly@umn.edu}
\cortext[cor1]{Corresponding author: 
  Tel.: +1-612-625-8000}

\affiliation[1]{organization={Department of Aerospace Engineering and Mechanics, University of Minnesota, Twin Cities},
                addressline={110 Union St. SE},
                city={Minneapolis, MN},
                postcode={55455},
                country={USA}}


\begin{abstract}
This paper presents prototype testing and a control allocation algorithm for the Cable-Actuated Bio-inspired Lightweight Elastic Solar Sail (CABLESSail) concept aimed at performing momentum management of a solar sail. CABLESSail uses actuated cables routed along the structural booms of the solar sail to control the shape of the solar sail and changes the solar radiation pressure disturbance torques acting on it. Small-scale prototype tests of CABLESSail are presented in this paper, which demonstrate the effectiveness of cable actuation on deployable booms.  A novel control allocation method is also presented in this paper that provides a computationally-efficient manner to determine the deformations required in each of the structural booms to impart the desired momentum management torque on the solar sail. Numerical simulation results with the proposed algorithm demonstrate robustness to uncertainty in the shape of the sail membrane, resulting in reliable generation of momentum management torques that exceed or meet the capabilities of state-of-the-art solar sail actuators. Both the prototype tests and control allocation methods presented in this paper represent key steps in raising the technology readiness level of the CABLESSail concept.
\end{abstract}


\end{frontmatter}


\section{Introduction}

Solar sails enable space exploration in a manner that is beyond the reach of spacecraft with traditional propulsion. Through the use of solar radiation pressure (SRP) for propulsion, solar sails are capable of performing orbital transfer and station-keeping maneuvers without any propellant. Successful development and maturation of solar sail technology will open up exciting opportunities for heliophysics, planetary science, and space exploration.

The technology needed to make solar sails a reality has advanced substantially over the past decades~\citep{Berthet2024-mi}. A number of solar sail flights have been performed with varying degrees of success, including JAXA's IKAROS~\citep{tsuda2013achievement}, the Planetary Society's LightSail~2~\citep{spencer2021lightsail}, Gama Alpha~\citep{ancona2025recent}, as well as NASA's NanoSail-D~\citep{johnson2011nanosail}, NEA Scout~\citep{lockett2020near,pezent2021contingency}, and ACS3~\citep{wilkie2021overview,amodio2025dynamical} solar sails. These flight experiments have largely served as technology demonstration missions to advance the maturity of solar sail technology. Next-generation solar sail designs, such as NASA's Solar Cruiser~\citep{pezent2021preliminary}, Solar Polar Imager~\citep{thomas2020solar}, Space Weather Investigation Frontier (SWIFT)~\citep{Johnson2025-lt}, and HIPERSail~\citep{wilkie2021overview} are notably larger than previously-flown solar sails. This will allow them to move beyond technology demonstrations and start performing impactful science.

Larger solar sails will come with a greater degree of structural flexibility~\citep{pimienta2019heliogyro,brownell2023time,Boni2023-mb}, which leads to non-ideal sail shapes~\citep{Huang2021-ba,Hibbert2021-xg,Wang2025-ql}. Deformation in the sail structure and its membrane can result in a significant misalignment between the solar sail's center of mass and center of pressure, which results in large disturbance torques acting on the spacecraft~\citep{gauvain2023solar}. A typical solar sail will be equipped with a reaction-wheel-based attitude control system~\citep{inness2023momentum} that will saturate in the presence of large disturbance torques, necessitating momentum management~\citep{Tyler2024,inness2023momentum,shen2025solar}. Momentum management is particularly challenging for solar sails, as propellant-free solutions are desired to enable long-duration missions. Less-traditional actuators have been explored for this purpose, including an active mass translator (AMT) that shifts the solar sail's center of mass through a planar translation mechanism between portions of the solar sail's bus~\citep{inness2023momentum} and reflectivity control devices (RCDs) that generate out-of-plane torques by adjusting the reflectivity properties of patches embedded into the sail membrane that are inclined at a tent angle~\citep{heaton2023reflectivity}.
As an example, NASA's Solar Cruiser is designed to use an AMT to generate in-plane (yaw/pitch) momentum management torques and RCDs to generate out-of-plane (roll) momentum management torques~\citep{inness2023momentum}. Other actuators and concepts have been developed for similar purposes, which can be found in~\citet{gong2019review,fu2016solar}. Many of these actuators, including the AMT, face scalability challenges for solar sails larger than Solar Cruiser, as their operating principle involves the indirect cancellation of disturbance torques due to undesirable deformations in the shape of the solar sail~\cite{gong2019review}. The magnitude of these disturbance torques increases with larger solar sails, which requires larger and heavier actuators to indirectly cancel out their effect. There are also technological challenges associated with RCDs, such as embedding them within the solar sail membrane and providing them with power far from the solar sail bus. This points to the pressing need to develop new momentum management actuation technology that will enable the design and flight of the next generation of large solar sails~\citep{spencer2019solar}.

\begin{figure}[t!]
\begin{center}
\includegraphics[width=0.55\textwidth]{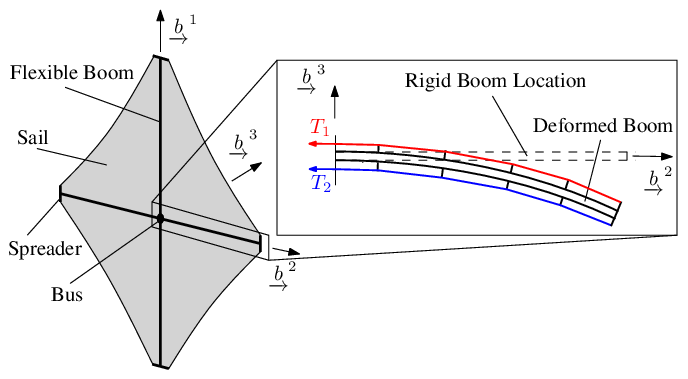}
\end{center}
\caption{The CABLESSail concept, which involves adjusting the tensions in cables routed along the length of the solar sail's booms to control the boom's bending deformation. Each boom has two actuating cables that allow for control of the boom's out-of-plane deformation. The body frame is defined by the $b_1$ (yaw), $b_2$ (pitch), and $b_3$ (roll) axes.}\label{fig:CABLESSailConcept}
\end{figure}

The Cable-Actuated Bio-inspired Lightweight Elastic Solar Sail (CABLESSail) concept was first introduced by~\citet{caverly2023solar} as a means to produce large, scalable momentum management torques through actuated control of the solar sail's shape. Specifically, the CABLESSail concept leverages the fact that disturbance torques acting on the solar sail are predominantly due to unwanted boom deformations~\citep{gauvain2023solar}. By using cables routed along the length of the booms to actively control their deformation, CABLESSail can directly cancel out unwanted boom deformations, and thus cancel out these disturbance torques, or generate momentum management torques by purposefully creating boom deformations. This concept effectively transforms the flexible nature of the solar sail structure from being an undesirable property to a novel means in which to generate torques that can be used for momentum management or attitude control. Piezoelectric shape control of a solar sail's booms has been investigated for a similar purpose~\citep{zhang2021solar,zhang2021three}, although the ability of these actuators to perform significant shape changes in the booms is limited by their small actuation capabilities. In contrast, CABLESSail uses cables routed along the length of the booms that are actuated by motor-driven winches in the bus of the solar sail, which is shown in Fig.~\ref{fig:CABLESSailConcept}. Specifically, cables are routed along each boom to control its out-of-plane deformation in both directions, which results in a scalable and lightweight means to control the boom shape and, thus, the generation of torques. Incidentally, actuated cables have previously been proposed as a means to assist with vibration control of a solar sail's membrane~\citep{Chen2023-nw}.

Although the CABLESSail concept is relatively new, substantial preliminary work towards the development of this technology is found in the literature. Following the initial formulation of the concept by~\citet{caverly2023solar}, a modular multi-body dynamic simulation of CABLESSail was developed by~\citet{bunker2024modular}. A flat sail membrane model was used for the analysis performed by~\citet{bunker2024modular}, which was augmented to account for non-flat sail membranes by~\citet{bunker2025static}. The work of~\citet{bunker2025static} also compared the momentum management torque generation capabilities of CABLESSail to an AMT, demonstrating that reliably large torques can be generated, yet uncovered the challenge in determining appropriate boom deformations to generate a desired momentum management torque. Testing of the CABLESSail concept on a small-scale prototype was performed by~\citet{bodin2025design}, which included both open-loop and closed-loop feedback tests with the CABLESSail actuating cables employing the controllers developed by~\citet{lee2024robust,lee2025passivity}. Although the experiments performed by~\citet{bodin2025design} were promising, they only involved triangular, rollable, and collapsible (TRAC) booms made of two tape measures that were pre-deployed. Further testing on a fully deployable prototype that accommodates other boom types (e.g., lenticular booms) remains a pressing need to validate the CABLESSail concept.

This paper serves as a summary of the current state of the CABLESSail technology and includes novel results that extend upon previously published CABLESSail work~\citep{caverly2023solar,bunker2024modular,lee2024robust,bodin2025design,bunker2025static,lee2025cablessail,lee2025passivity}. In particular, the novel contributions presented in this paper include 1) validation of the CABLESSail technology with a deployable $2$~m composite lenticular boom prototype and 2) the formulation and validation of a control allocation algorithm to determine the boom deformations needed by the CABLESSail technology to generate a desired torque. These contributions fill in key needs towards elevating CABLESSail to technology readiness level (TRL) 3. Note that although a preliminary version of the proposed control allocation algorithm appeared in the work of~\citet{lee2025cablessail}, this prior work did not account for the effect of clock-angle dependence, did not incorporate any constraints to ensure physically-realizable boom tip deformations, and did not assess the range of feasible torques generated by the algorithm. Additionally,~\citet{lee2025cablessail} presented preliminary results with a deployable $1$~m composite lenticular boom that suffered from issues with boom sag due to gravity and did not feature any quantification of the CABLESSail's actuations capabilities. Based on these significant limitations, the contributions of this paper represent a significant extension on the preliminary work of~\citet{lee2025cablessail}.

The remainder of this paper is organized as follows. Section~\ref{sec:Concept} presents an overview of the CABLESSail concept and a detailed summary of prior work towards the development of this technology. A description of the small-scale deployable CABLESSail prototype is provided in Section~\ref{sec:DeployablePrototype}, along with test results with the prototype. A control allocation method that determines the desired boom deformations needed to generated a specified momentum management torque is presented and tested in Section~\ref{sec:ControlAllocation}. Section~\ref{sec:Future} discusses the future outlook of the CABLESSail technology, followed by concluding remarks in Section~\ref{sec:Conclusions}.

\section{CABLESSail Concept: Overview and Summary of Prior Work}
\label{sec:Concept}

This section presents an overview of the CABLESSail concept, followed by a summary of previous simulation and prototyping results, as well as a discussion on the status of CABLESSail prior to the work presented in this paper.

\subsection{CABLESSail Concept Overview}

The CABLESSail concept is centered around the notion that unwanted boom deformations are the main contributor to shifts in the solar sail's center of pressure, and thus, disturbance torques acting on a solar sail~\citep{gauvain2023solar}. The goal of CABLESSail is to either counteract unwanted boom deformations to negate the disturbance torque or purposefully create boom deformations if a non-zero momentum management torque is desired.

To control the deformation of the solar sail's booms, CABLESSail takes inspiration from the area of soft continuum robotics, where cable actuation can be used to control the deformation of a slender flexible structure. The ``bio-inspired'' portion of CABLESSail's name derives from the fact that much of the soft continuum robotics literature is inspired by biological systems, such as the manner in which the trunk of an elephant or the arms of a starfish can be articulated. Given that the primary deformation of the booms of a solar sail is in the out-of-plane direction normal to the sail membrane, CABLESSail involves routing two cables along the length of each boom, as shown in Fig.~\ref{fig:CABLESSailConcept}, where one cable lies above the boom's neutral axis for bending and the other cable lies below this axis.  The cables are controlled by winches mounted inside the solar sail's bus and are connected to the end of the boom.  The out-of-plane deformation of each boom can be controlled by adjusting the tension in its two actuating cables.

A significant advantage to the CABLESSail concept is the ease in which it scales to large solar sails. Very little tension is required in the actuating cable to enact large boom deformations, which means that relatively small cables can be used with minimal added mass to the system. The cables can easily be stored in a wound configuration around a winch and released during deployment of the sail membrane. Another advantage of CABLESSail is that rather than attempting to mitigate the effect of unwanted solar sail shapes like other momentum management actuators, it directly tackles the problem by controlling the boom and sail shape.

\begin{figure}[t!]
    \centering
	\subfigure[]
	{
        \includegraphics[width=.48\linewidth]{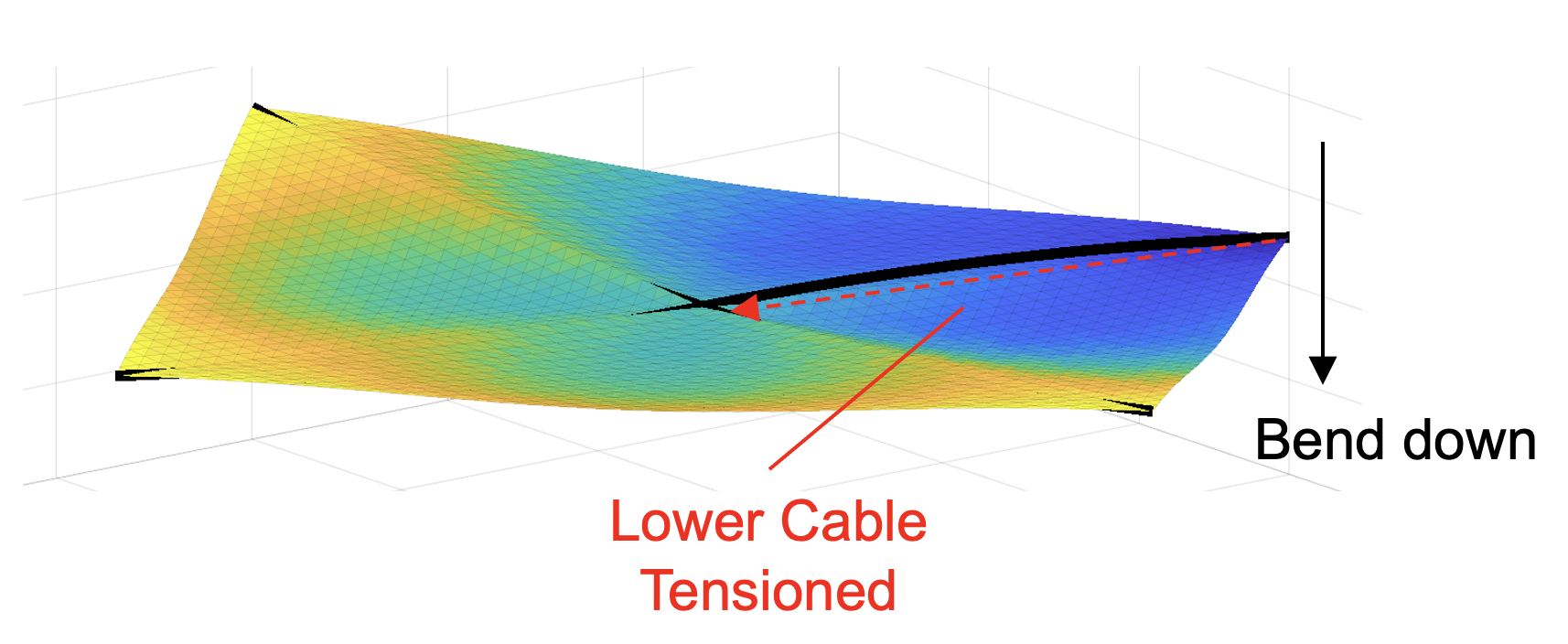}
        \label{subfig:PitchYawMode}
    }
    \subfigure[]
    {
        \includegraphics[width=.48\linewidth]{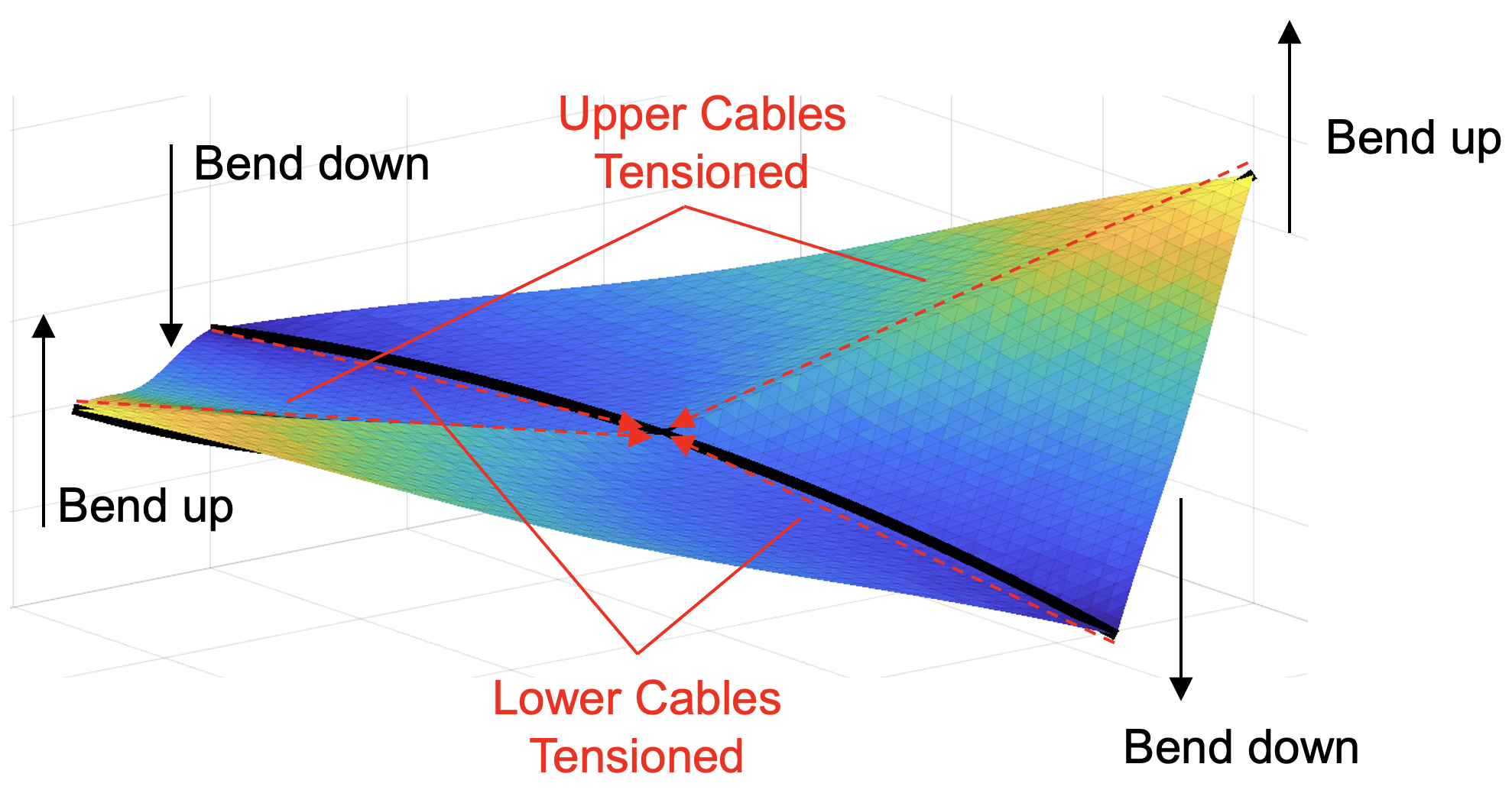}
        \label{subfig:RollMode}
    }    
    \caption{CABLESSail actuation modes: (a) yaw-pitch mode involving a single boom deformation and (b) roll mode involving coordinated deformation of all booms.}
    \label{fig:CABLESSail_Modes}
\end{figure}

As a simple illustrative example of how CABLESSail generates momentum management torques, consider the deformation of an individual boom, as shown in Fig.~\ref{subfig:PitchYawMode}. This boom deformation will change the local Sun incidence angle (SIA) of the sail membrane quadrants attached to the deformed boom, resulting in a shift of the solar sail's center of pressure. This local SIA change is related to the out-of-plane displacement of the membrane, which is visualized by the shading in Fig.~\ref{fig:CABLESSail_Modes}. As shown in Section~\ref{sec:Sim}, this is an effective means to generate torques in the yaw/pitch axes (i.e., the axes aligned with the nominal plane of the sail).  Deforming all booms in alternating directions, as shown in Fig.~\ref{subfig:RollMode}, results in roll torques (i.e., torques in the direction normal to the nominal plane of the sail) for non-zero SIAs.  Although there are many more ways in which the booms can be deformed to generate useful torques, these examples provide some intuition as to how CABLESSail can generate momentum management torques. A more systematic algorithm to determine appropriate doom deformations to create desired torques is proposed in Section~\ref{sec:ControlAllocation}.

\subsection{CABLESSail Simulation} \label{sec:Sim}
A numerical simulation environment has been developed for CABLESSail that can be used to test different design options and configurations, as well as provide benchmarks to other actuation mechanisms, such as the AMT. The source code for this CABLESSail simulation is available in the Aerospace, Robotics, Dynamics, and Control (ARDC) Lab GitHub repository\footnote{\url{https://github.com/ARDCLab/CABLESSail-Modular-NullSpace-Dynamic-Simulation}} and a more detailed overview of the simulation and the modeling approaches used can be found in the work of~\citet{bunker2024modular,bunker2025static}.

\subsubsection{Simulation Features}

\begin{figure}[t!]
\centering
    \includegraphics[width=0.48\columnwidth]{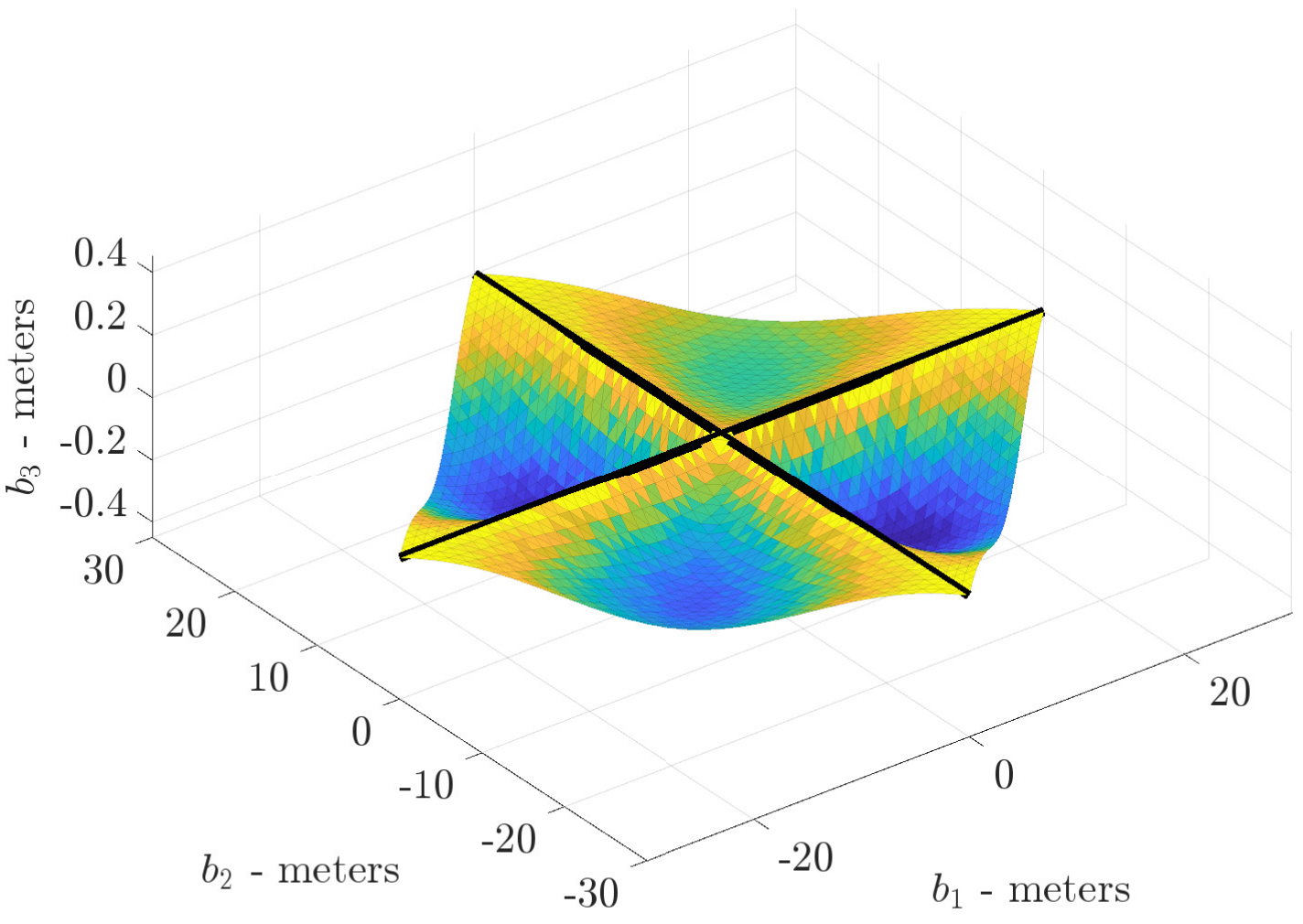}
\caption{Depiction of a simulated sail membrane shape.}
\label{fig:SimShape}
\end{figure}

The CABLESSail simulation models the solar sail bus as a rigid body and its structural booms as flexible bodies connected to the rigid bus. The simulation is designed to be modular to allow for streamlined testing with different solar sail geometries, various CABLESSail designs, and models with varying degrees of complexity and fidelity, as well as creating the ability to make comparisons to other actuation options. 

An option within the simulation is to choose or randomly generate non-flat sail membrane shapes by specifying out-of-plane deformations at the nodes of a triangular mesh. The SRP force and torque are computed at each element of the mesh using the NEA Scout optical properties found in the work of~\citet{heaton2015update}, then the results are summed across all of the elements to obtain the net SRP force and torque. An example of a non-flat sail membrane shape is shown in Fig.~\ref{fig:SimShape}. As described by~\citet{bunker2025static}, the simulation can accommodate any sail membrane shape deformation,
where it is assumed that the corners of each membrane quadrant are co-located with the spacecraft bus and the tips of the neighboring booms. The results presented in this paper use a membrane shape that places the maximum deformation at the quadrant centroid, which matches the approach used by~\citet{gauvain2023solar}.

The numerical parameters used in the simulation results of this paper are given in Table~\ref{table:SimParam}. The parameters are chosen to roughly match those of Solar Cruiser~\citep{JeremyTRAC, nguyen2023solar}. The booms are approximated as Euler-Bernoulli beams with axial, transverse, and out-of-plane deformation using an assumed modes method.

\begin{table}[t!]
\caption{Numerical values for the simulations performed in this paper.}\label{table:SimParam}
\def\arraystretch{1.2}
\begin{center}
\begin{tabular}{l@{\quad}l@{\quad}l}
    \toprule
    Symbol & Parameter & Value\\
     \midrule
    $L$ & Boom length & $29.5$~m\\
    $\rho$ & Linear density & $0.1017$~kg/m\\
    $EI$ & Flexural Rigidity & $1,700$~N$\cdot$m$^2$\\
    $\zeta$ & Damping Ratio & $< 1$\% \\
    \multirow{2}{*}{$h$} & Distance between the & \multirow{2}{*}{$0.1$~m}\\
    & cable and the boom & \\
    \multirow{2}{*}{---} & Number of sail membrane  & \multirow{2}{*}{$3,600$} \\
     & mesh elements &  \\
    --- & Mesh element side length & $1$~m \\
    \bottomrule
\end{tabular}
\def\arraystretch{1.0}
\end{center}	
\end{table}

\subsubsection{Simulation Results}
\label{sec:StaticSim}

To assess CABLESSail's ability to reliably generate large momentum management torques, static simulations were performed by~\citet{bunker2025static} with intuitive boom deformation maneuvers that are within reasonable CABLESSail actuation limits. Three specific maneuvers were tested: a yaw-torque-inducing maneuver where one boom tip is deformed $50$~cm and a pitch-torque-inducing maneuver where one boom tip is deformed $-50$~cm, (shown in Fig.~\ref{subfig:PitchYawMode}), and a roll-torque-inducing maneuver where two opposing boom tips are deformed $50$~cm and the other two opposing boom tips are deformed $-50$~cm (shown in Fig.~\ref{subfig:RollMode}). In the yaw and the pitch maneuvers, each maneuver deforms one of the two booms perpendicular to the torque axis, respectively.

\begin{figure}[t!]
\centering
 \subfigure[]
 	{
\includegraphics[width=0.48\columnwidth]{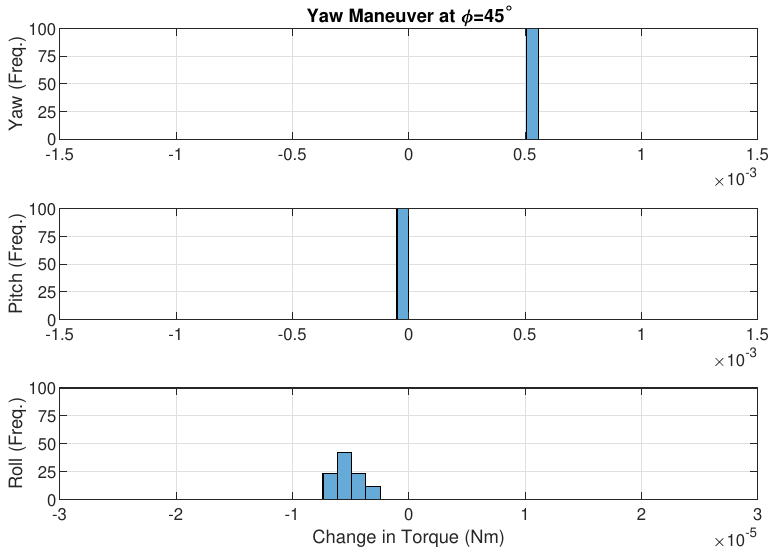} \label{subfig:StaticSim_yaw}}
 \subfigure[]
 	{
\includegraphics[width=0.48\columnwidth]{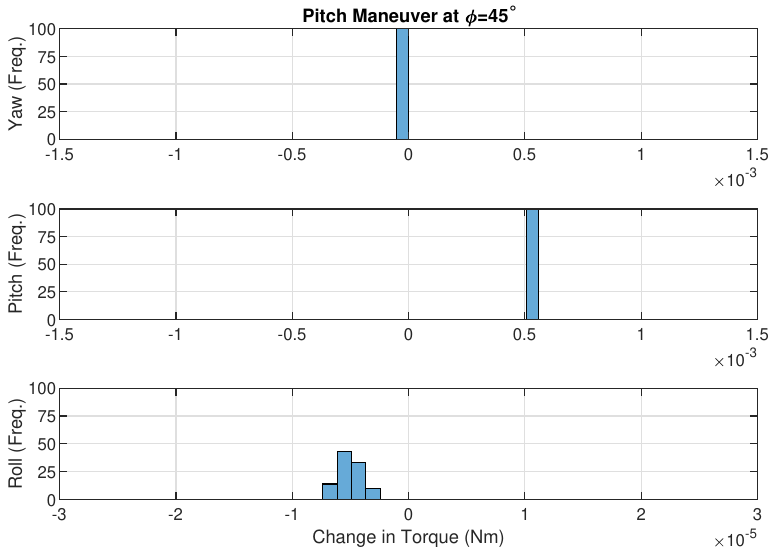} \label{subfig:StaticSim_pitch}}
 \subfigure[]
 	{
\includegraphics[width=0.48\columnwidth]{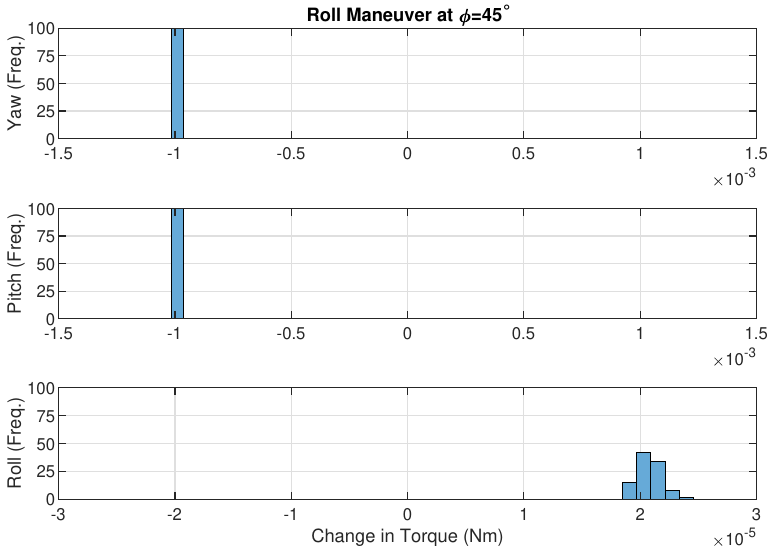}
\label{subfig:StaticSim_roll} }
\caption{Monte Carlo static simulations of (a) yaw, (b) pitch, and (b) roll maneuvers. Histogram of change in torque generated across all simulated sail membrane shapes.}
\label{fig:StaticSim}
\end{figure}

Monte Carlo simulations were performed with these three maneuvers across a range of possible sail membrane shapes, where the maximum membrane deformation was sampled from a uniform distribution of $-15$~cm to $15$~cm in each sail quadrant. Each randomly-generated membrane shape results in a different nominal disturbance torque acting on the vehicle prior to any actuation of the booms. As CABLESSail is primarily intended as a momentum management actuator, its key performance metric is its ability to cancel out unwanted disturbance torques. To assess this, the change in torque acting on the solar sail due to CABLESSail's actuated boom deformation maneuvers was computed. The resulting change in torque histogram plots for each maneuver at a SIA of $17$~degrees and a clock angle of $45$~degrees are shown in Fig.~\ref{fig:StaticSim}.

It is observed in Figs.~\ref{subfig:StaticSim_yaw} and~\ref{subfig:StaticSim_pitch} that the yaw- and pitch-torque maneuvers generate reliably-large yaw and pitch torques across all membrane shapes. 
This is a seemingly robust maneuver that produces momentum management torques that are similar in magnitude to the worst-case disturbance torques predicted for Solar Cruiser~\citep{gauvain2023solar}.
For the roll-torque maneuver in Fig.~\ref{subfig:StaticSim_roll}, a reliably-large roll torque is generated for all membrane shapes. Unfortunately, significant yaw and pitch torques are also generated when performing this maneuver. Although this is an undesirable effect, the result is still notable, as roll torques are substantially more difficult to generate with existing actuator technology, such as RCDs and thrusters. This is highlighted in~\citet{inness2023momentum}, where it is stated that ``understanding each option for roll control is key as one single option for roll momentum management is not sufficient to completely manage the roll axis.'' Moreover, this motivates the need to further optimize the boom deformations to minimize the residual yaw and pitch torques from the roll-torque maneuver and the residual roll torque from the yaw- and pitch-torque maneuvers. A novel control allocation algorithm that is developed for this purpose is presented in Section~\ref{sec:ControlAllocation}.

\subsection{Deployed Prototype Testing}
\label{sec:Prototype}

Small-scale prototype testing serves as a means to assess and develop CABLESSail's technology in complement to the use of numerical simulations. This section outlines prior work involving two fully-deployed prototypes built using metallic tape measures to mimic a TRAC boom.

\begin{figure}[t!]
    \centering
    \subfigure[]
    {
        \includegraphics[width=.48\linewidth]{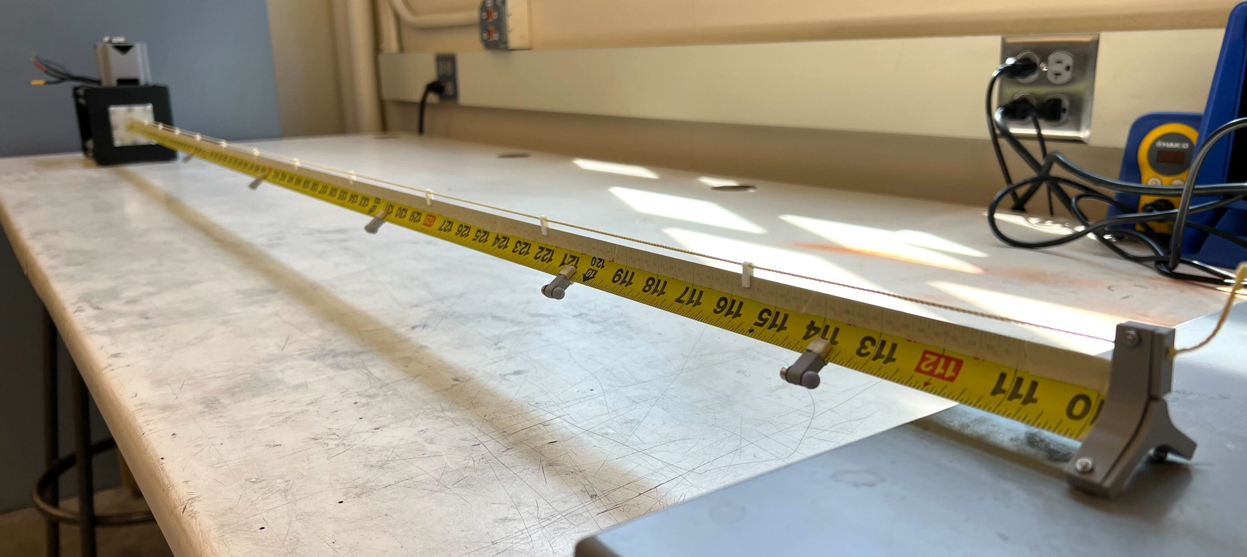}
        \label{subfig:tracboomtapemeasure}
    }
    \subfigure[]
	{    \includegraphics[width=.48\linewidth]{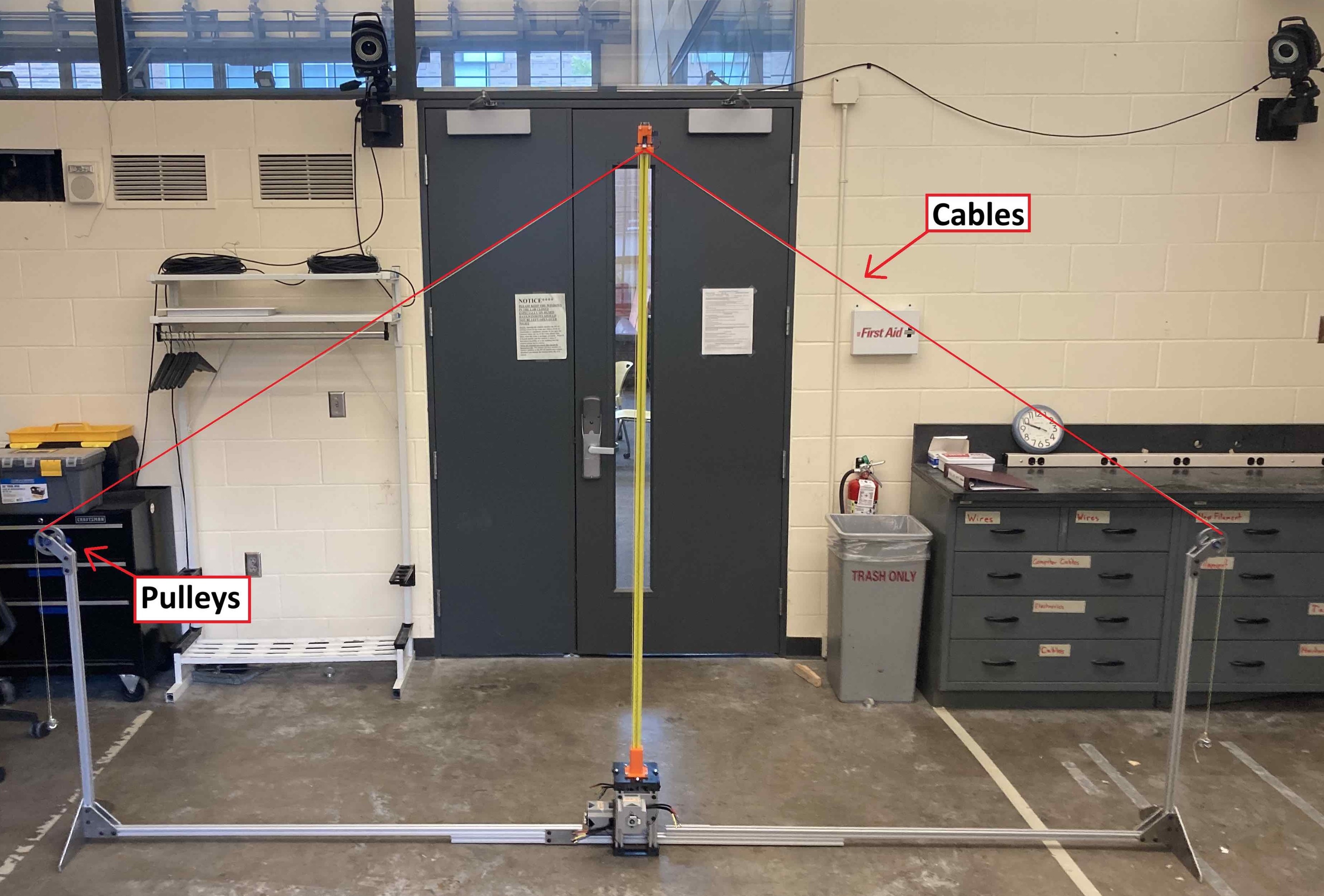}
        \label{subfig:SailTension}
    }
   \caption{Images of (a) a close-up of the TRAC boom prototype with a single actuating cable and (b) the vertical prototype testbed with a sail tension simulation device and the Vicon motion capture system in the background.}
    \label{fig:SailTensionSystem}
\end{figure}

\begin{figure}[t!]
    \centering
	\subfigure[]
	{    \includegraphics[height=5.1cm]{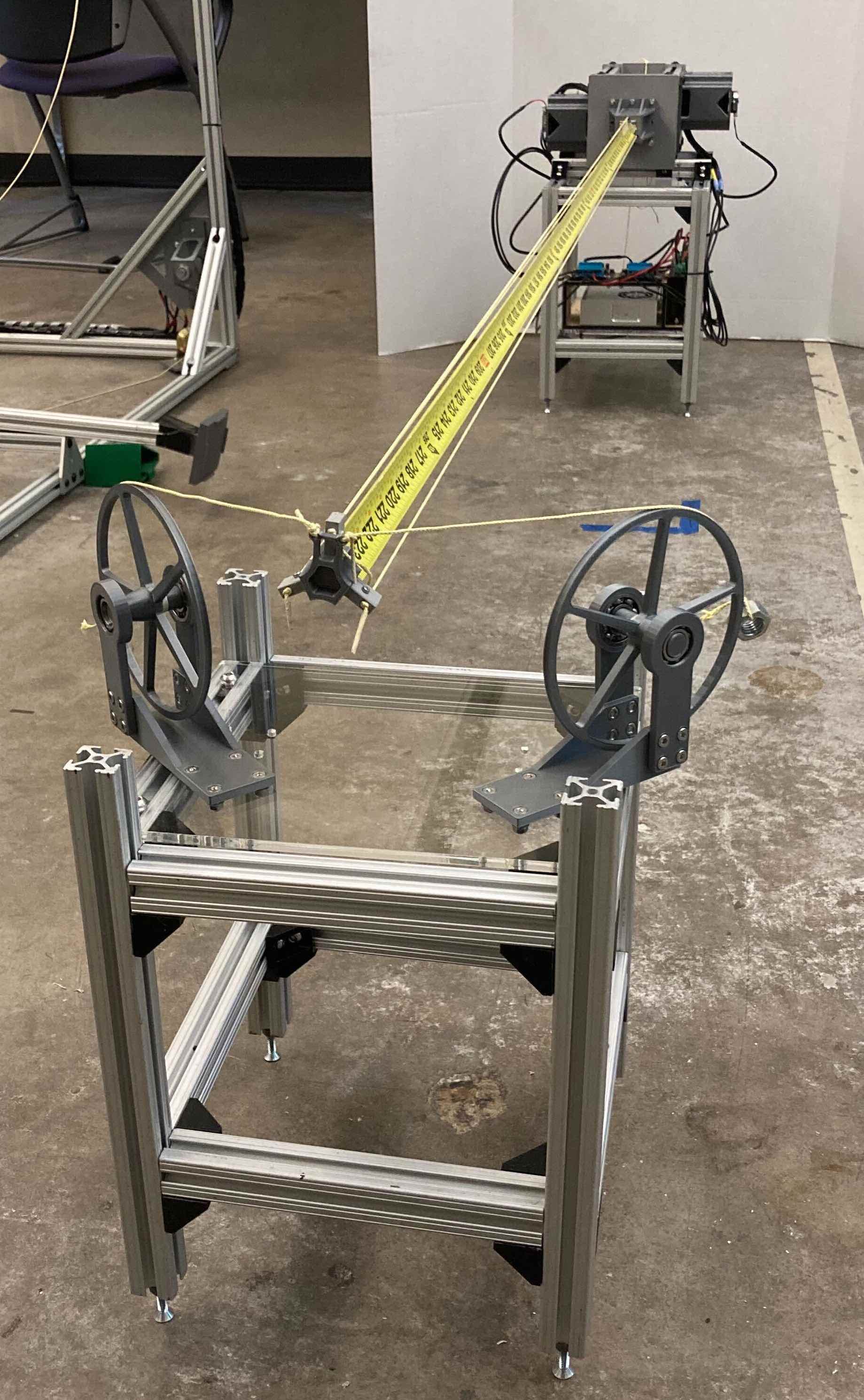}
        \label{subfig:Horiz1}
    }
    \subfigure[]
    {
        \includegraphics[height=5.1cm]{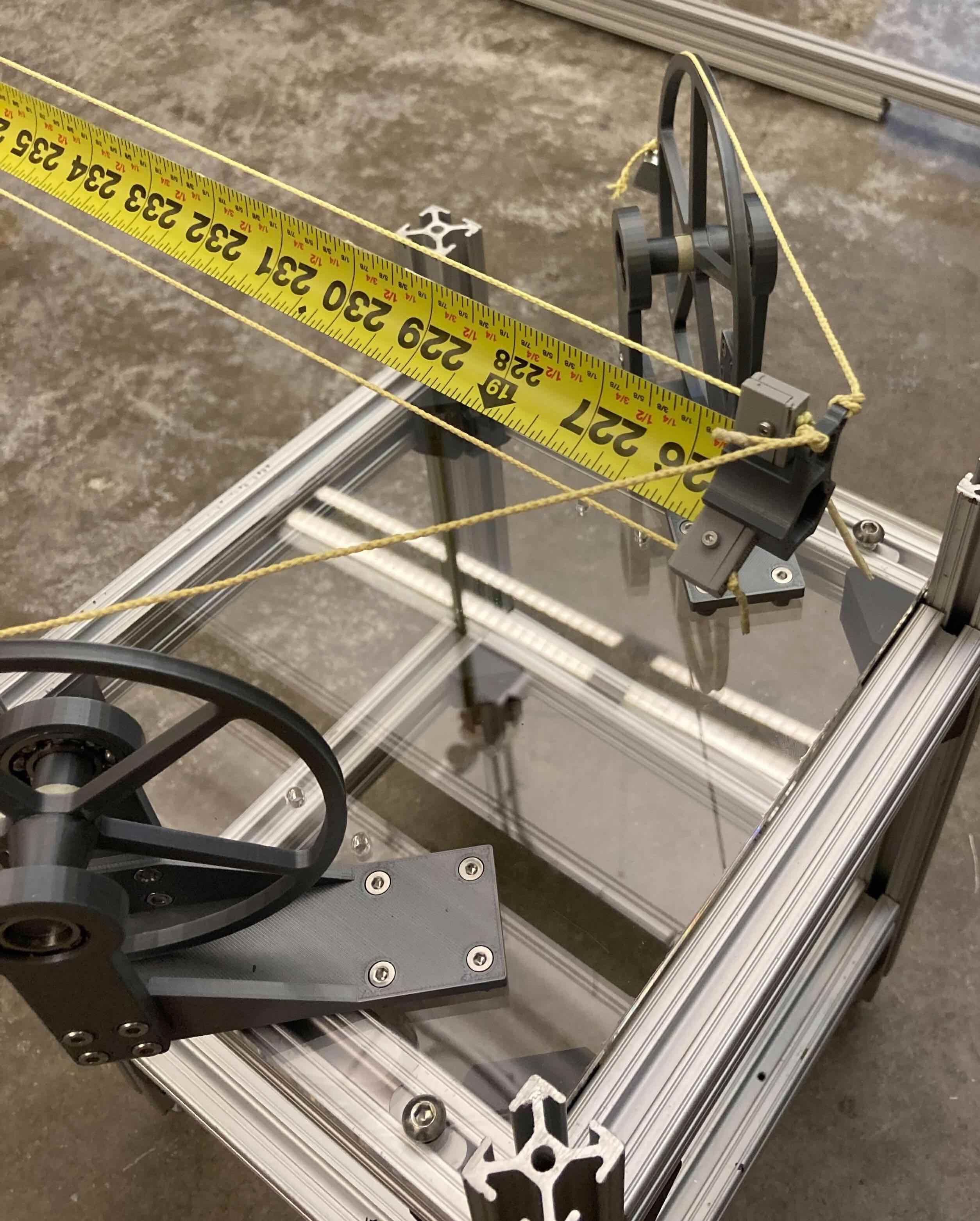}
        \label{subfig:Horiz2}
    }    
        \subfigure[]
    {
        \includegraphics[height=5.1cm]{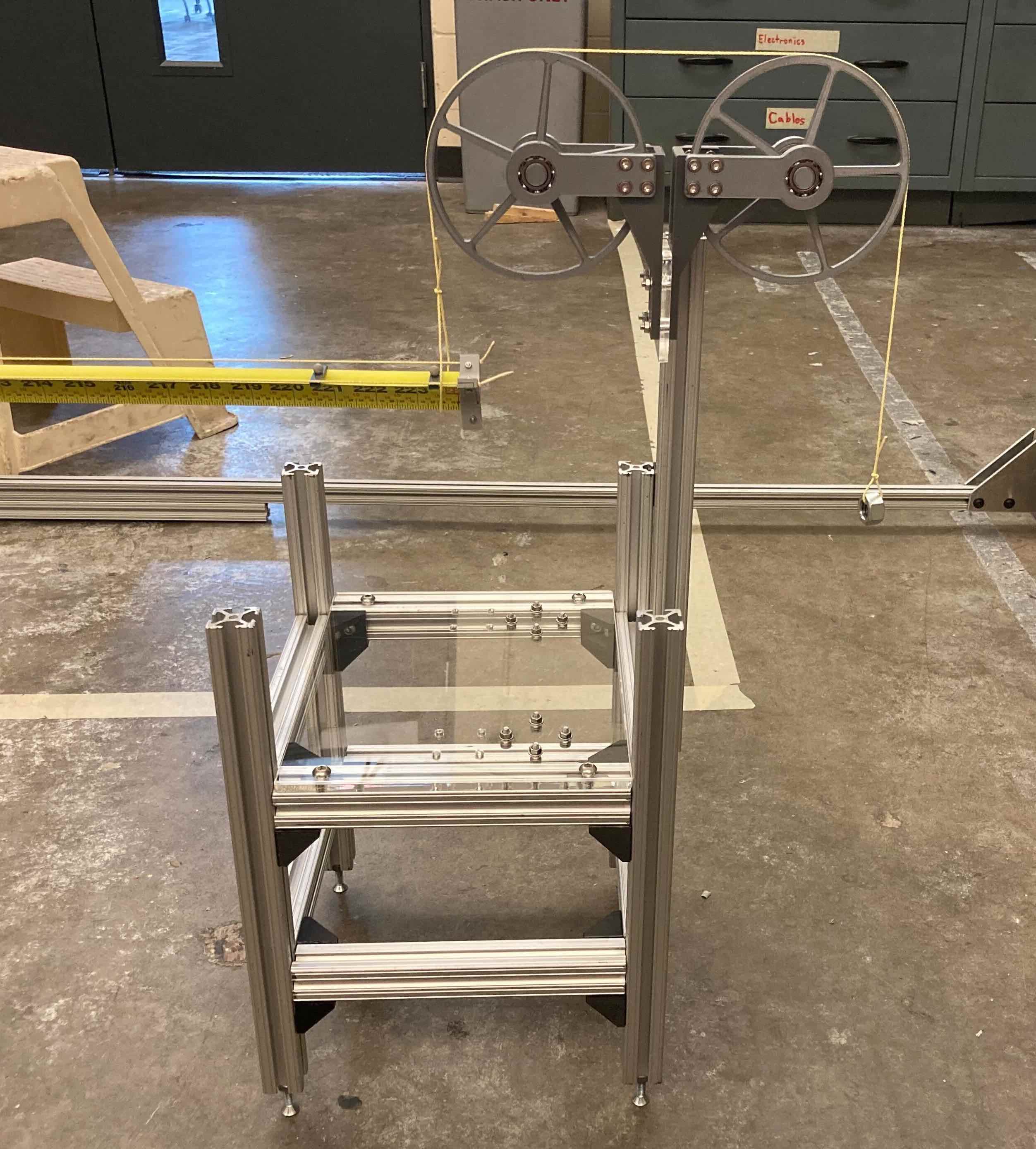}
        \label{subfig:Horiz3}
    }    
    \caption{Images of (a) the deployed tape-measure CABLESSail prototype testbed designed to use gravity to simulate thermal expansion effects on the boom, (b) a close-up of the cable attachment points at the end of the boom and the additional cables routed around pulleys to simulate sail tension, and (c) an optional pulley system attached to the boom tip to offload some of the gravitational forces and decrease the nominal tip deformation.}
    \label{fig:HorizTestbed}
\end{figure}

Preliminary work towards a small-scale CABLESSail prototype was presented in the work of~\citet{bodin2025design}, where metallic TRAC booms were fabricated by gluing two tape measures along one edge, as shown in Fig.~\ref{subfig:tracboomtapemeasure}.
Actuating cables were run along the length of the TRAC booms with one end connected to a 3D-printed cap at the tip of the boom and the other end wrapped around a winch connected to an actuating motor at the base of the boom. The TRAC boom in Fig.~\ref{subfig:tracboomtapemeasure} features a single cable along the web of the boom, while Fig.~\ref{subfig:Horiz2} has a more complete representation of CABLESSail TRAC boom, where one cable runs along the web and two cables rung along the flanges.

Two deployed prototype testbed are presented in the work of~\citet{bodin2025design}: a vertical testbed shown in Fig.~\ref{subfig:SailTension} that minimizes the effect of gravity on the boom deformation and a horizontal testbed shown in Fig.~\ref{fig:HorizTestbed} that purposefully uses gravity to simulate a nominal boom deformation due to thermal effects. The vertical testbed features additional cables attached to the tip of the boom and routed around pulleys with masses hanging from them to simulate the effect of sail membrane tensioning on the TRAC boom. The horizontal testbed has a similar sail membrane tensioning system shown in Fig.~\ref{subfig:Horiz2}, as well as a gravity offloading device shown in Fig.~\ref{subfig:Horiz3} that can adjust the amount of nominal deformation induced in the boom due to gravity. Further details regarding the hardware and electronics used to fabricate and operate the prototypes are found in the work of~\citet{bodin2025design}.

Experimental results with both prototypes are provided by~\citet{bodin2025design}, including open-loop actuation tests that demonstrate the ability to deform the boom in both directions on the horizontal testbed and illustrate the effect that eyelets guiding the actuating cables along the boom have on its response using the vertical testbed. These tests provide insight into the design of the CABLESSail concept that complement the analysis performed through numerical simulations. Scaling laws based on Euler-Bernoulli beam theory are also presented in the work of~\citet{bodin2025design} to better relate these small-scale prototype results to full-scale numerical values.

\subsection{Estimation \& Control}
\label{sec:EstimationControl}

\begin{figure}[t!]
    \centering
        \includegraphics[width=.85\linewidth]{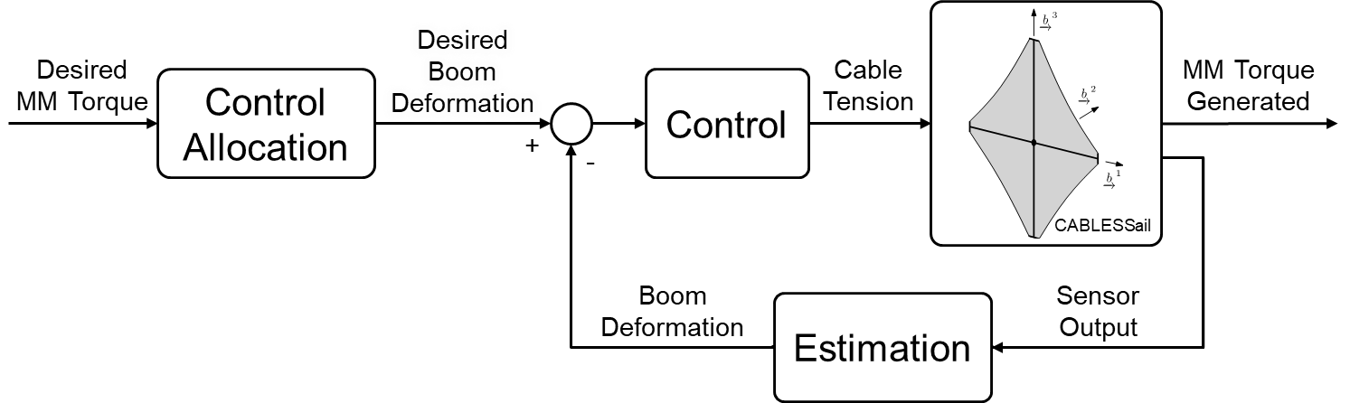}
    \caption{Block diagram outlining the high-level estimation and control architecture. The abbreviation MM in this figure stands for momentum management.}
    \label{fig:BlockDiagram}
\end{figure}

The CABLESSail concept relies on precise control of the solar sail's booms in order to generate desired torques through shape control. This necessitates the ability to estimate deformations in the booms and the design of a robust feedback controller that will ensure the booms track their desired deformation values.
CABLESSail's estimation and control architecture is shown in Fig.~\ref{fig:BlockDiagram}. Within this architecture, the desired torque is mapped to desired boom tip deformations through a control allocation algorithm, the estimated boom tip deformations are then subtracted from the desired deformations to generate an error that is regulated by a feedback controller that determines the actuating tension in each cable.

Prior work has focused on the design and testing of the control and estimation techniques. Specifically, a feedback controller that ensures each boom tracks the desired boom tip deformation was originally developed by~\citet{lee2024robust} and tested experimentally on the vertical small-scale prototype by~\citet{lee2025passivity,bodin2025design}. The main challenge in designing this controller is ensuring robust closed-loop stability in the presence of substantial uncertainty in the structural properties and dynamics of the boom, the sail membrane, and the actuating cables. It was shown by~\citet{lee2024robust,lee2025passivity} that a linearized model of an Euler-Bernoulli solar sail beam actuated with a cable using the CABLESSail concept is passive from the cable tension input to the boom tip's transverse deformation rate. This result was shown to hold for large variations in the structural properties of the boom, which motivated the use of passivity-based control. The numerical and experimental results in the work of~\citet{lee2024robust,lee2025passivity} demonstrate that accurate boom tip deformation tracking can be achieved with this controller, even in the presence of notable measurement noise and with the use of low-cost prototype hardware.

Prior work on the estimation algorithm assumed that the sensors available include encoders on the winches actuating the cables and IMUs at the boom tips. Two different Kalman filters were investigated by~\citet{bodin2025design} to fuse together this information to obtain an accurate boom tip deformation estimate. Specifically, individual Kalman filters were implemented to estimate each boom's tip deformation. The Kalman filters presented in~\citet{bodin2025design} used kinematic process models that are driven by the rate gyroscope of the boom-tip-mounted IMU and a measurement model that assumes the winch encoders can be related to the angular deflection at the boom tip. Experimental results presented in the work of~\citet{bodin2025design} with the deployed prototype described in Section~\ref{sec:Prototype} demonstrated that the boom tip deformation can be estimated with low-cost hardware, although it highlighted some of the challenges associated with calibrating the winch encoder measurements to a corresponding boom tip deformation measurement.

Most notably, all prior work on CABLESSail's control and estimation algorithms assumed that the desired boom tip deformations were known. In practice, a control allocation algorithm is required to determine the boom tip deformations required to achieve a desired momentum management torque. In response to this need, a novel control allocation algorithm is developed and tested in Section~\ref{sec:ControlAllocation}.

\subsection{Status of the CABLESSail Technology}

The numerical simulations and small-scale prototype testing presented in prior publications demonstrate that controlled deformations of a solar sail's booms with the CABLESSail concept are possible and can result in the ability to reliably generate large momentum management torques. Although this is an important step in the maturation of the CABLESSail technology, there remain two critical barriers before TRL 3 can be achieved: testing on a deployable prototype and the development of a control allocation algorithm. The remainder of this paper focuses on work towards these two areas, which amounts to the novel contributions of this paper.

\section{Deployable Prototype Development}
\label{sec:DeployablePrototype}

Testing on a small-scale deployable prototype is needed to assess the CABLESSail technology on a structure analogous to a solar sail boom. This section outlines the development and testing of a deployable prototype that incorporates composite lenticular booms.

\subsection{Deployable Prototype Fabrication}

\begin{figure}[t!]
    \centering
        \subfigure[]
	{    \includegraphics[height=5.3cm]{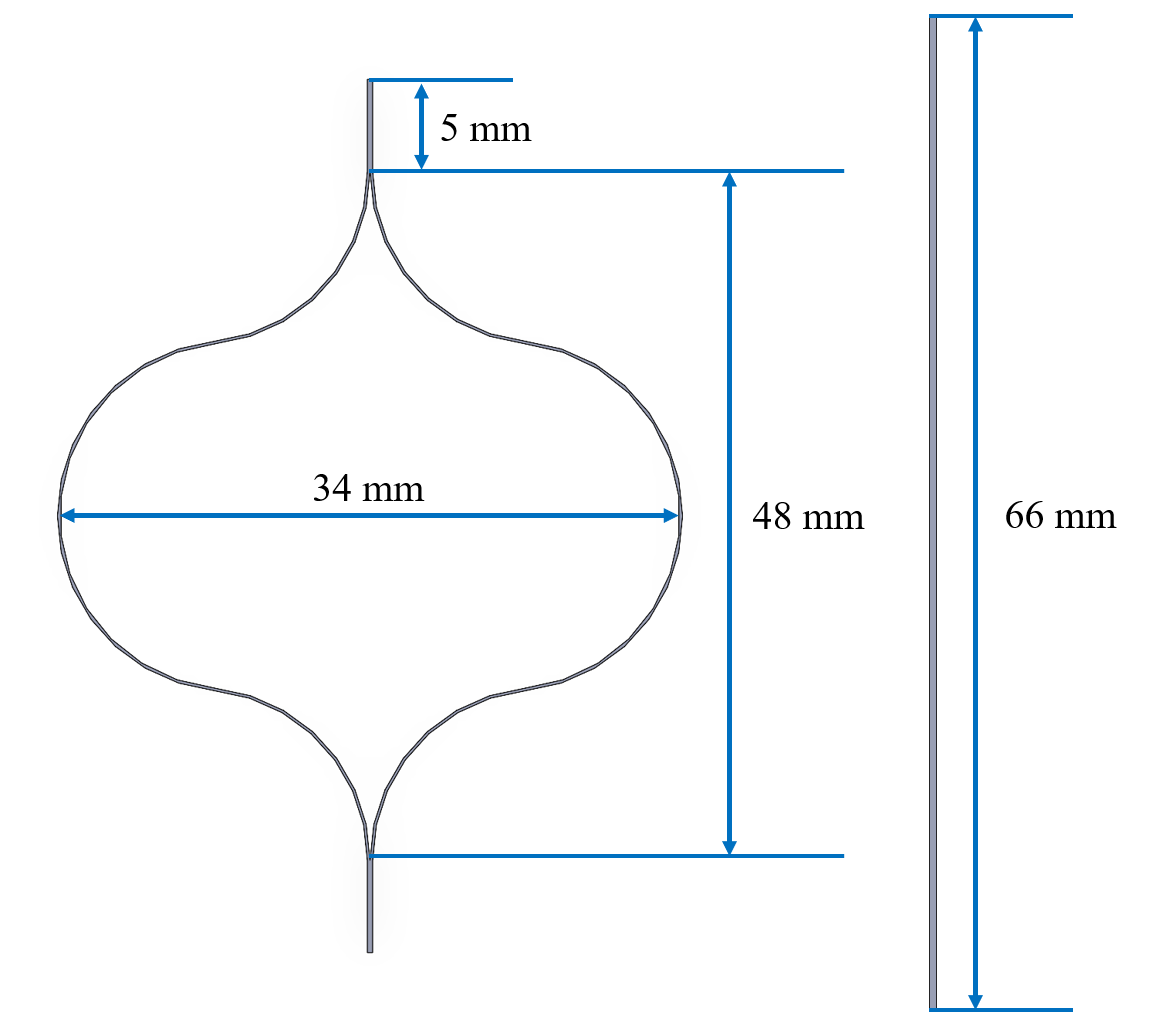}
    \label{subfig:ACS3_dimension}
    }
    \subfigure[]
    {
        \includegraphics[height=5.3cm]{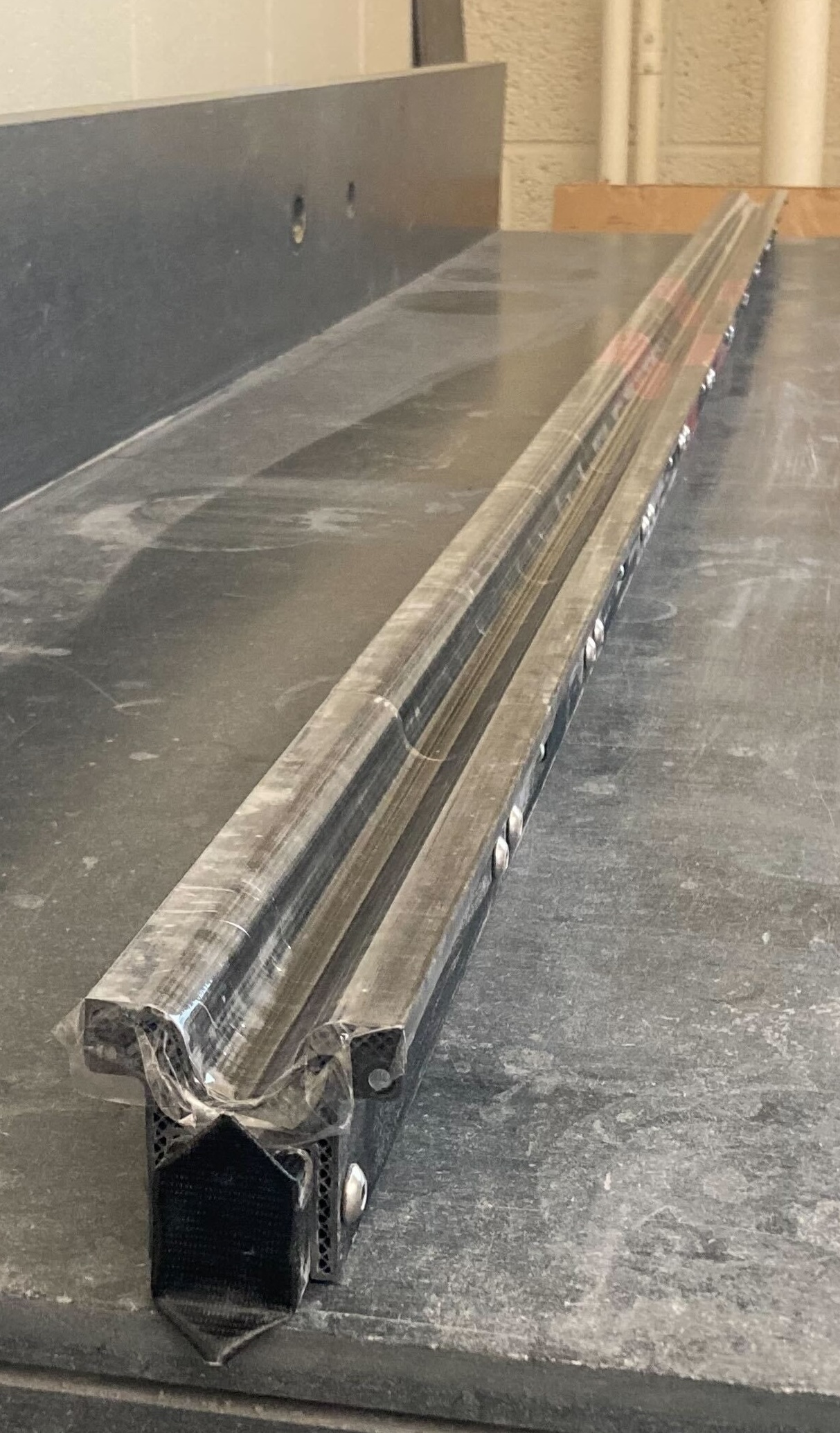}
        \label{subfig:Mold}
    }
    \subfigure[]
	{    \includegraphics[height=5.3cm]{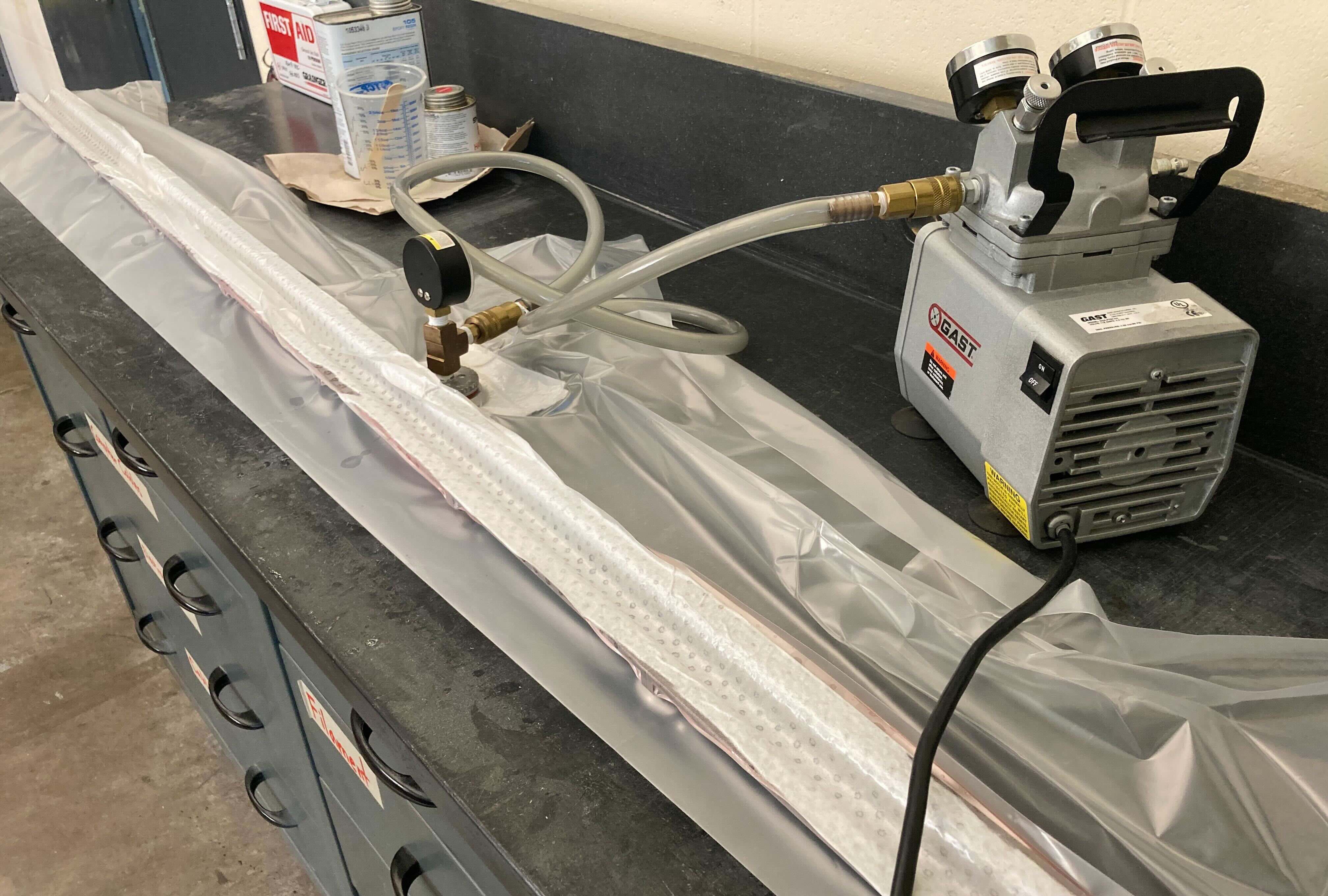}
    \label{subfig:Curing}
    }
    \subfigure[]
	{    \includegraphics[height=4.5cm]{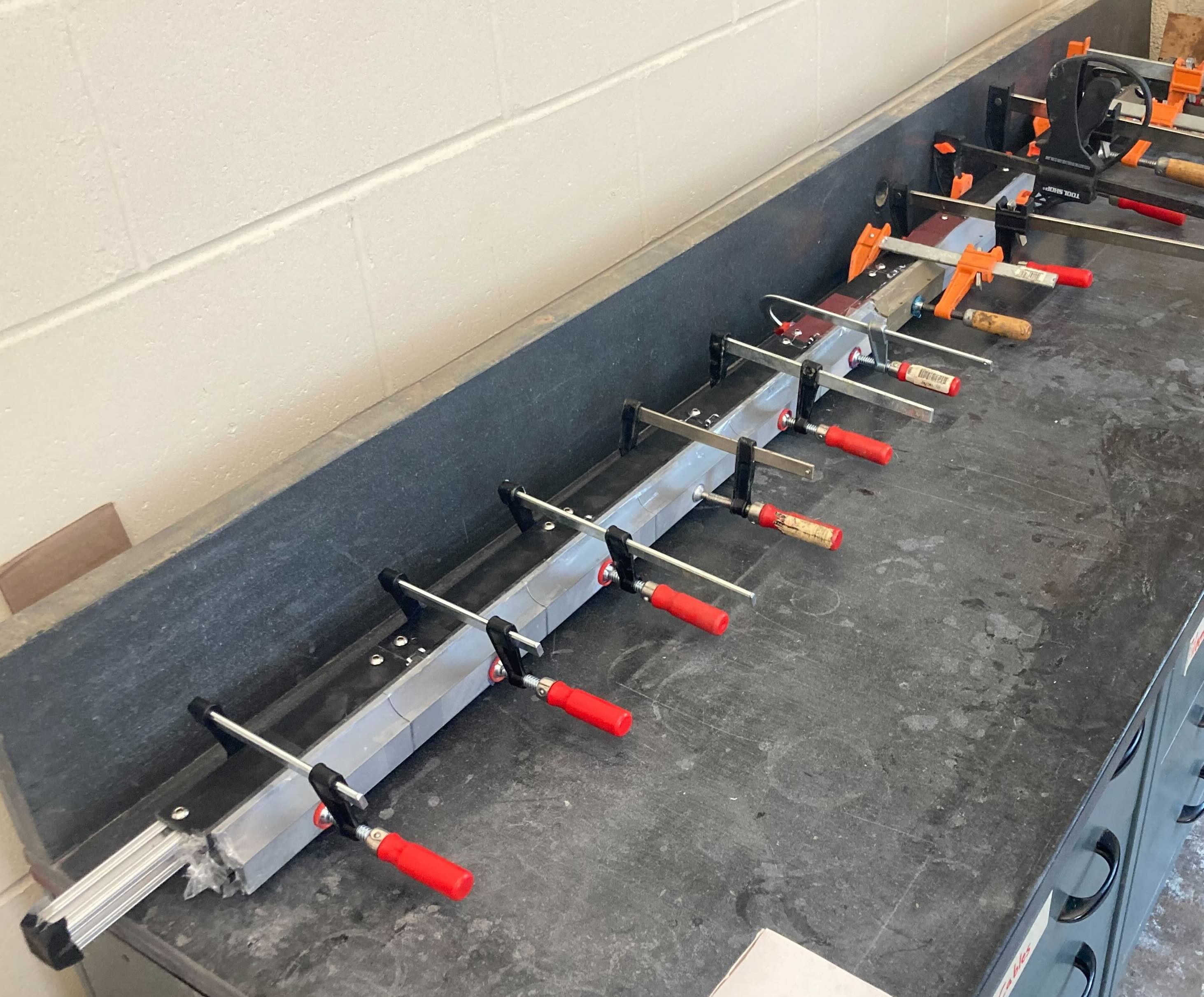}
    \label{subfig:Joining}
    }
    \subfigure[]
	{    \includegraphics[height=4.5cm]{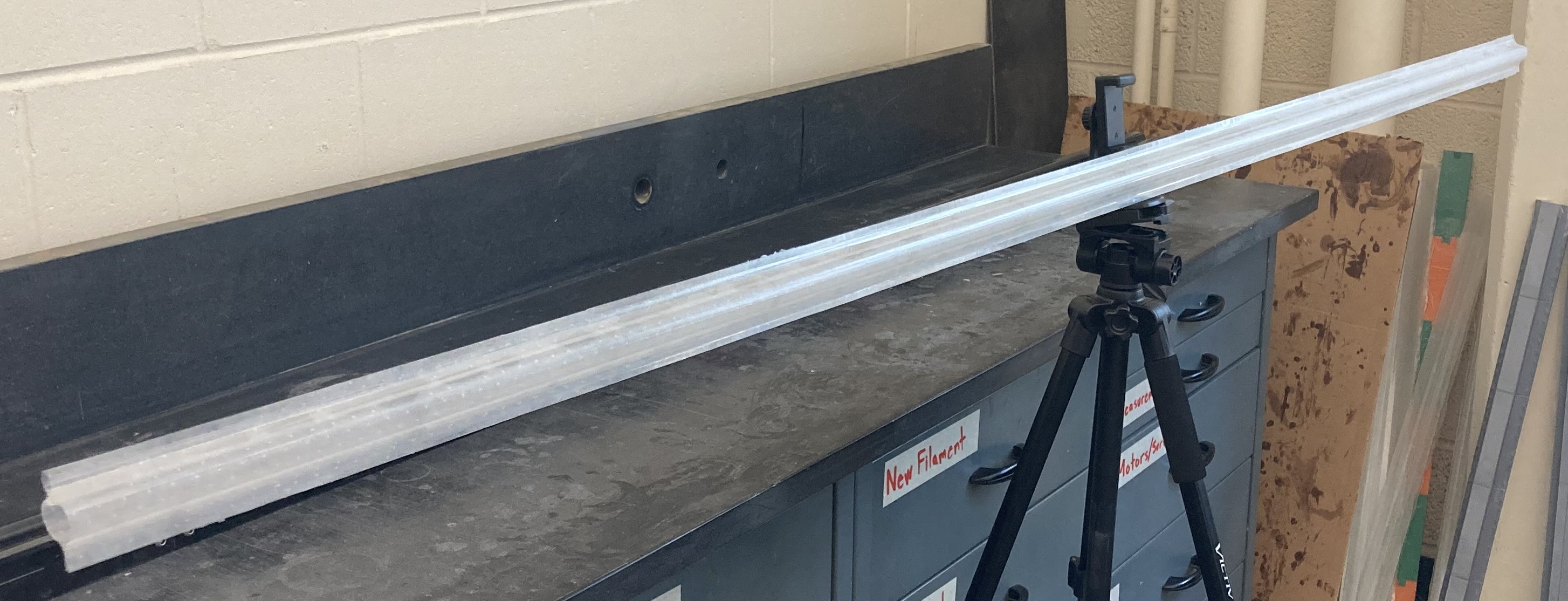}
    \label{subfig:FinishedBoom}
    }
    \caption{Images of the composite boom and manufacturing process, including (a) its cross section and dimensions in deployed and flattened configurations, (b) the layup mold, (c) the vacuum-bagged curing process, (d) the joining of two boom halves through clamping, and (e) the finished boom.}
\end{figure}

To better assess CABLESSail's integration with a deployable boom, fiberglass composite lenticular booms were fabricated to match the cross-sectional dimensions of the ACS3 booms, as shown in Fig.~\ref{subfig:ACS3_dimension}. Note that glass fiber reinforced polymer is used for manufacturing convenience in this work, rather than the carbon fiber reinforced polymer used for the ACS3 booms. The fabrication procedure involves laying up each half of the boom into a 3D-printed negative mold as separate parts then joining them together using epoxy. The layup schedule consists of a single $0$~degree orientation layer of Fibre Glast plain weave $2$~ounces per square yard ($67.8$~grams per square meter) fiberglass with a matrix of West System 105 epoxy resin and 206 slow hardener, giving a post-cure thickness of $0.005$~inches ($0.127$~mm). Vacuum bagging is used during the curing process to improve the consistency and reduce the thickness of the layup by removing air bubbles, pressing the layup into the mold, and extracting excess resin. Images of the layup mold and the vacuum bag curing process are provided in Figs.~\ref{subfig:Mold} and~\ref{subfig:Curing}. The two halves are joined by applying the epoxy to the flat portions of the booms and clamping them together, as shown in Fig.~\ref{subfig:Joining}. An insert is placed between the two halves during the clamping process to ensure that the desired boom cross section is maintained during the clamping process. Once this joining process is complete, the boom flanges are trimmed to the correct height, and holes are drilled at the root for mounting to the spool. An image of a completed boom is shown in Fig.~\ref{subfig:FinishedBoom}. Booms of various lengths have been manufactured using this process. A length of 2 meters was found to be the maximum length at which the joining process can be performed without risking the insert getting stuck inside the boom.

\begin{figure}[t!]
    \centering
    \subfigure[]
    {
        \includegraphics[height=7cm]{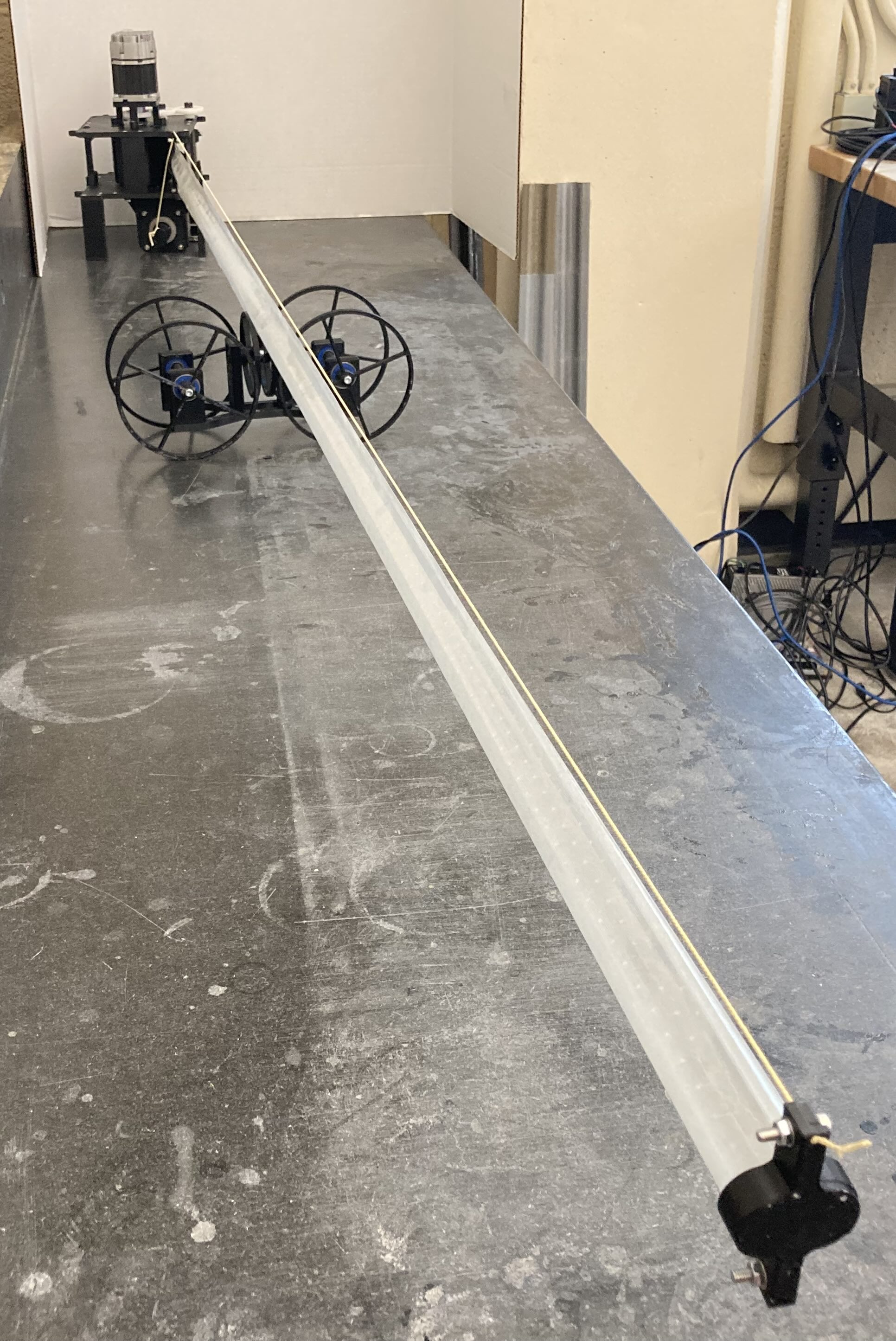}
        \label{subfig:Prototype1}
    }  
    \subfigure[]
    {
        \includegraphics[height=7cm]{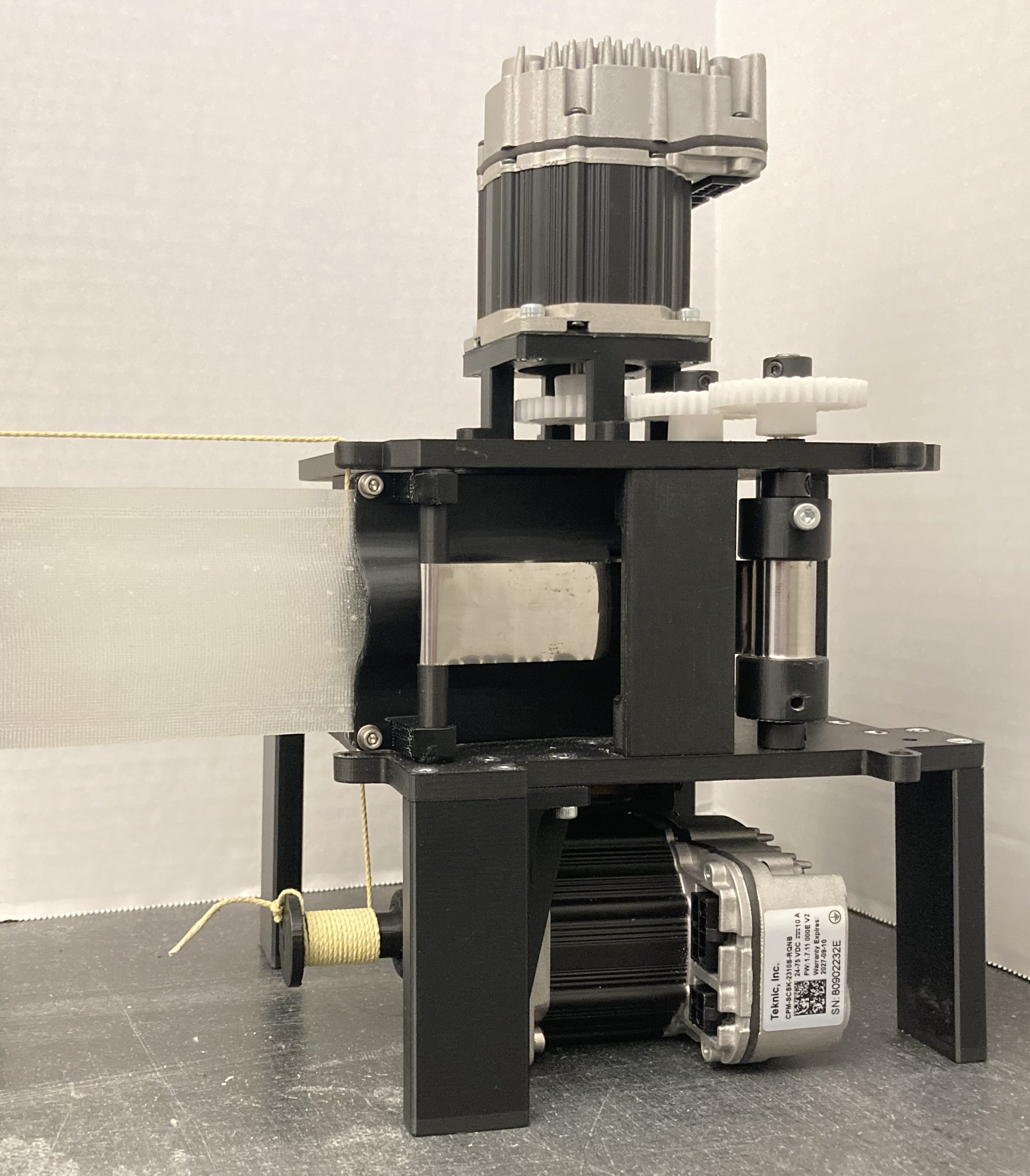}
        \label{subfig:Prototype2}
    }  
    \subfigure[]
    {
        \includegraphics[height=7cm]{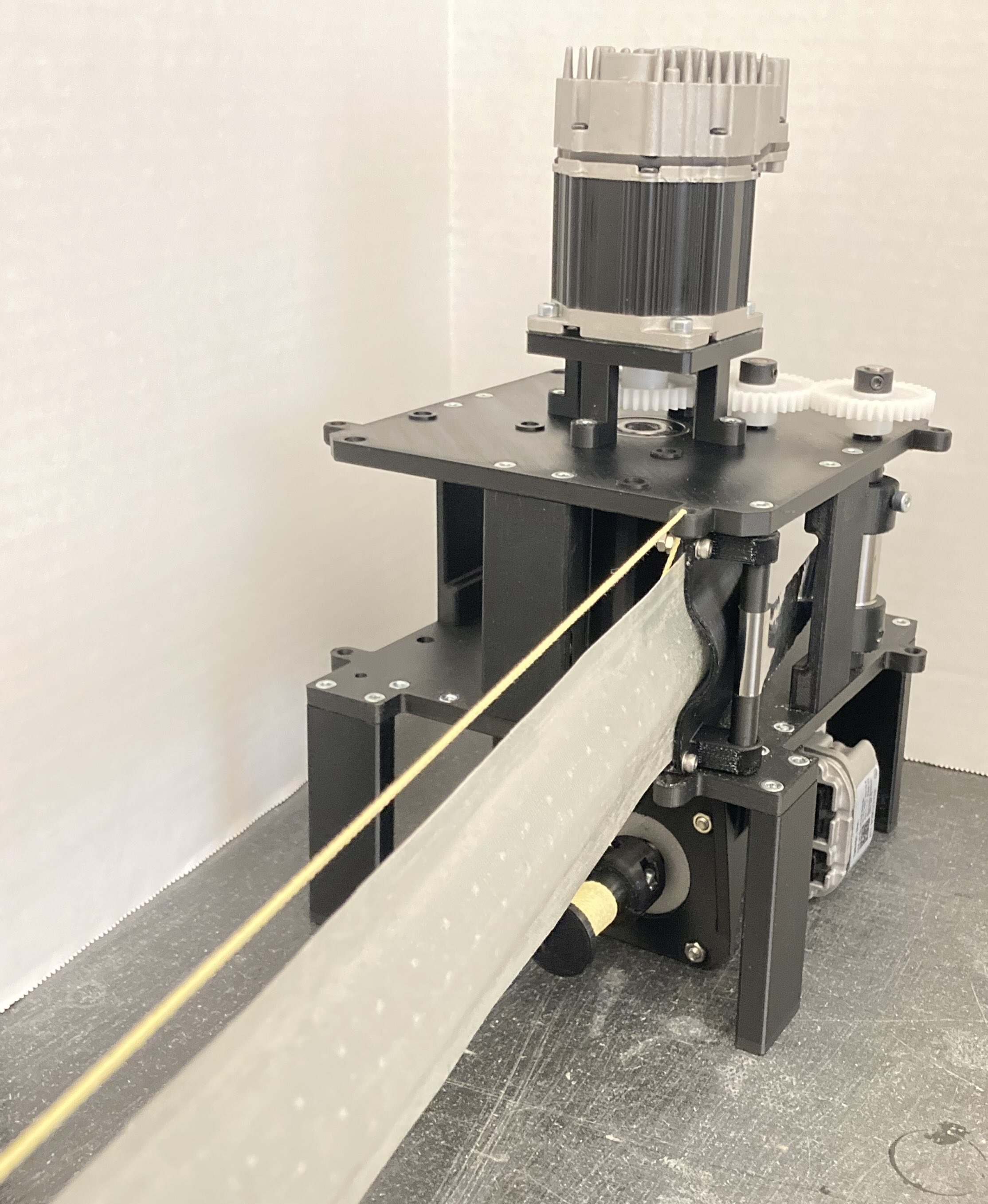}
        \label{subfig:Prototype3}
    }  
    \caption{Images of the deployable CABLESSail prototype that features a 2-meter fiberglass composite lenticular boom.}
    \label{fig:Deployable2}
\end{figure}

\begin{figure}[t!]
    \centering
    \subfigure[]
    {
        \includegraphics[height=4.6cm]{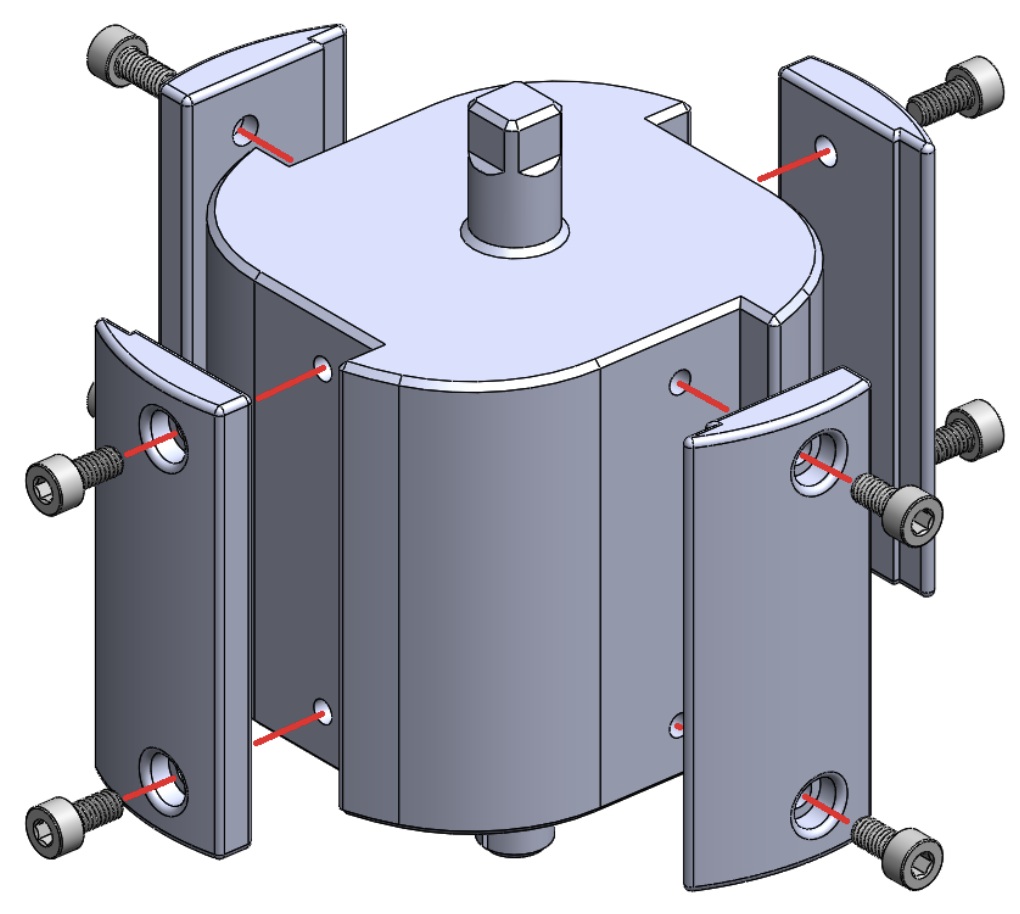}
        \label{subfig:SpoolRibbon1}
    }  
    \subfigure[]
    {
        \includegraphics[height=4.6cm]{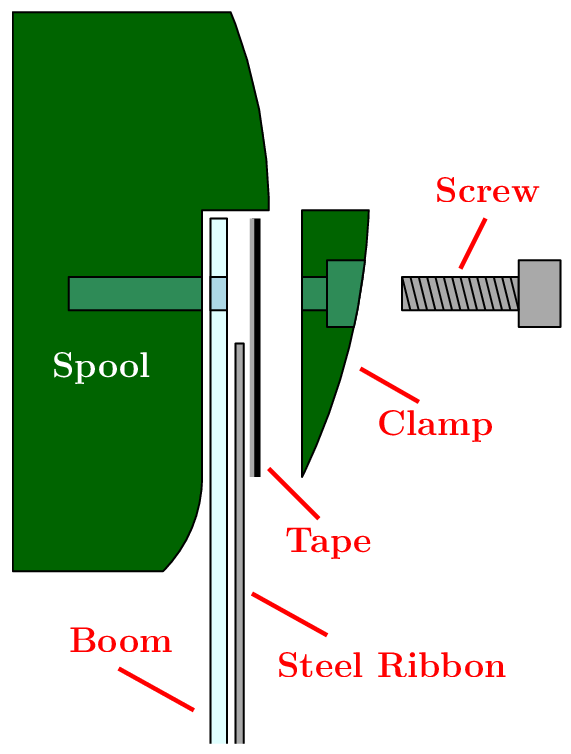}
        \label{subfig:SpoolRibbon2}
    }  
    \subfigure[]
    {
        \includegraphics[height=4.6cm]{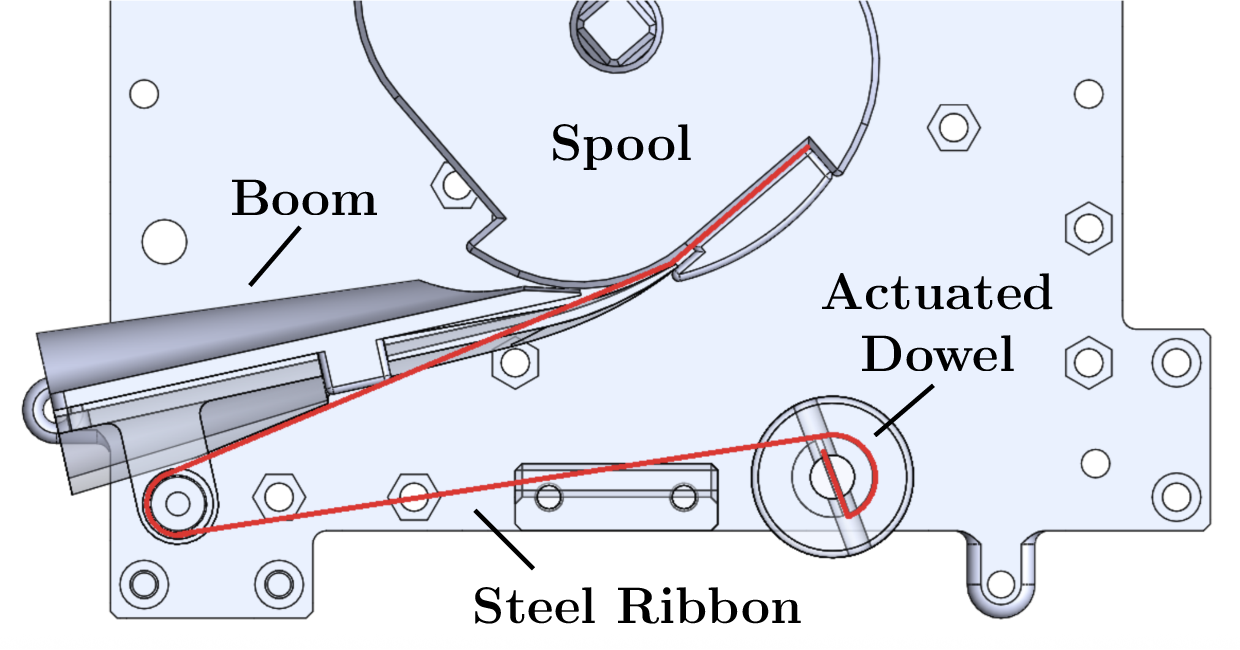}
        \label{subfig:SpoolRibbon3}
    }  
    \caption{Depictions of the attachments between the spool of the deployable mechanism and the boom, as well as the steel ribbon that is used for deployment. Specifically, (a) a CAD model of the spool and its clamps; (b) the attachment between the spool, the boom, and the steel ribbon; and (c) an overhead view of the steel ribbon's routing from the spool to the actuated dowel.}
    \label{fig:SpoolRibbon}
\end{figure}

A deployment mechanism for the composite lenticular boom is designed to allow for testing. The fabricated 3D-printed mechanism integrated with a 2-meter composite boom is shown in Fig.~\ref{fig:Deployable2}. The design of this mechanism is similar to that used by ACS3, where a steel ribbon is wrapped around the stored boom and used to pull the boom out for deployment. The other end of the steel ribbon is wound around a dowel connected to an actuating motor through a series of gears. The connection between the steel ribbon, the boom, and the spool that drives the boom deployment, as well as the steel ribbon's routing to the actuated dowel is visualized in Fig.~\ref{fig:SpoolRibbon}. Although the mechanism fabricated for testing in this work includes only a single boom, the design allows for four booms to be integrated into the prototype, as shown in Fig.~\ref{subfig:SpoolRibbon1}.

The motor used to drive the boom deployment is a Teknic CPM-SCSK-2310S-RLNB servo motor that has an integrated motor controller. This motor, shown at the top of the deployment mechanism in Fig.~\ref{fig:Deployable2}, is much larger than what is needed to drive the boom deployment and was chosen due to its availability from a prior project. In practice, a much more compact motor can be used to drive the deployment mechanism, although this is left to be implemented in future work.

A spiral torsion spring is attached to the spool upon which the boom is stored to provide a small torque that opposes the deployment of the boom. The torsion spring is situated underneath the spool in the images of Fig.~\ref{fig:Deployable2}. This torsion spring helps ensure that the boom does not expand or bloom in an undesirable fashion during deployment.

A second motor is included in the prototype to actuate the CABLESSail cable and provide a means to deform the boom in the out-of-plane direction. One end of the actuating cable is wrapped around a winch connected to the motor, while the other end is routed through a small eyelet above the boom and then is connected to the tip of the boom. A Teknic CPM-SCSK-2310S-RLNB servo motor is used for this purpose, which provides far more actuation capability than is required for this prototype. As with the boom deployment motor, a much more compact motor could be implemented in practice.

A gravity-offloading cart is used during deployment tests, as shown in Fig.~\ref{subfig:Prototype1}. This cart has two sets of wheels to allow for free movement of the boom during deployment. One set of wheels is located at the base of the cart and allows the cart/boom to move in the direction perpendicular to the boom deployment direction. The boom lies on the second set of wheels at the top of the cart, which allow for the boom to slide along the cart with little resistance.

\subsection{Deployable Prototype Test Results}
\label{subsec:DeployablePrototypeTestResults}
Two experimental tests are performed with the deployable prototype to assess CABLESSail's performance. The first test investigates its deployment and subsequent actuation capability, while the second test investigates its deformation performance in a configuration where it does not need to overcome gravity.

\begin{figure*}[t!]
    \centering
	\subfigure[]
    {    \includegraphics[width=.98\linewidth]{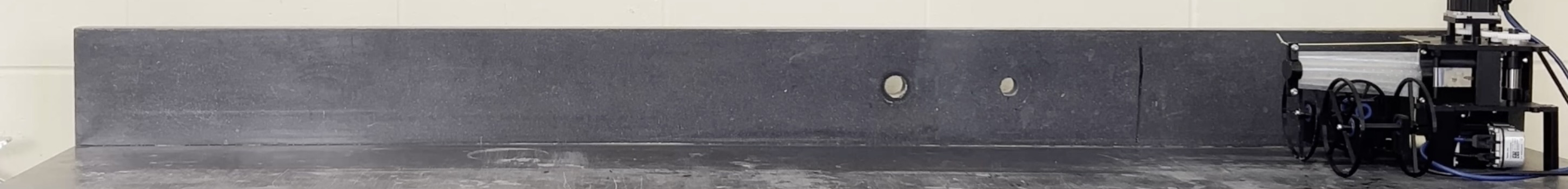}
        \label{subfig:Deploy1}
    }
	\subfigure[]
    {    \includegraphics[width=.98\linewidth]{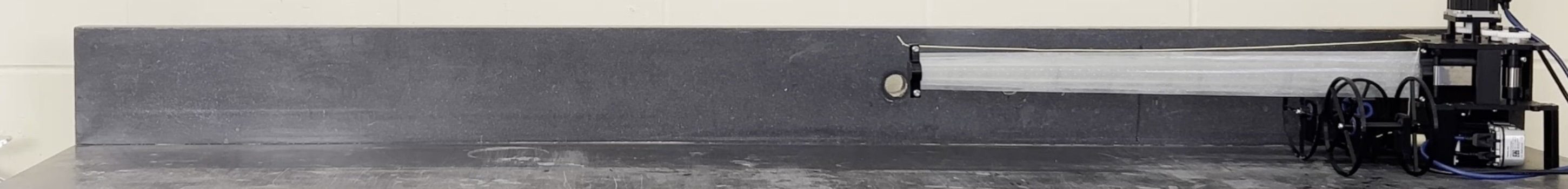}
        \label{subfig:Deploy2}
    }
    \subfigure[]
    {    \includegraphics[width=.98\linewidth]{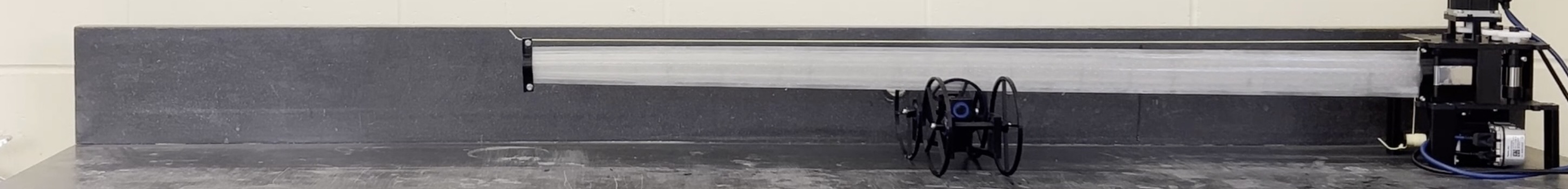}
        \label{subfig:Deploy2b}
    }
    \subfigure[]
    {    \includegraphics[width=.98\linewidth]{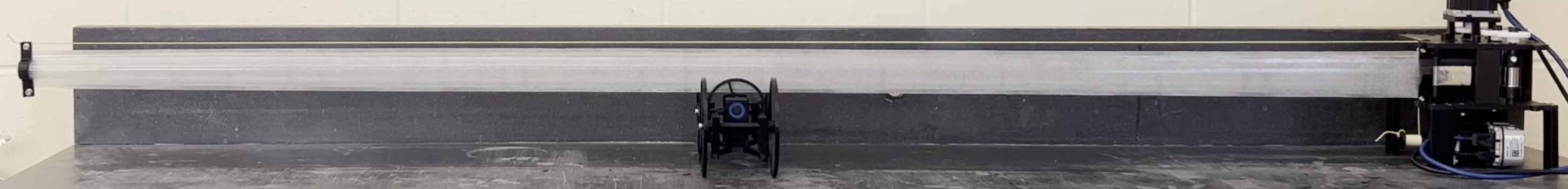}
        \label{subfig:Deploy3}
    }
    \subfigure[]
    {    \includegraphics[width=.98\linewidth]{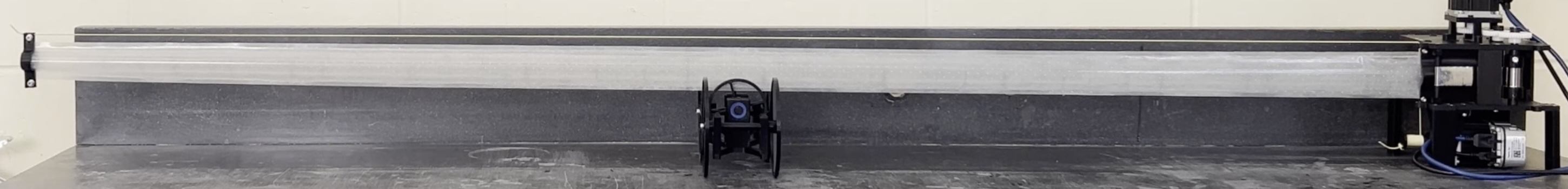}
        \label{subfig:Deploy4}
    }
    \subfigure[]
    {    \includegraphics[width=.98\linewidth]{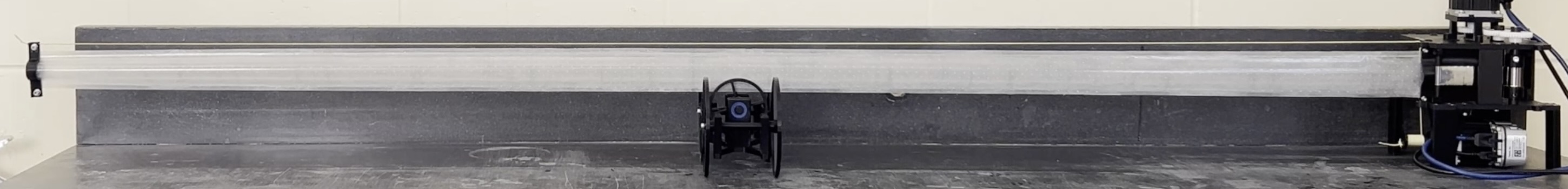}
        \label{subfig:Deploy5}
    }
    \caption{Still frames from a test with the deployable CABLESSail prototype. The composite lenticular boom is deployed in frames (a)-(d), then the CABLESSail actuator is used to deform the boom in the upward direction in (e), followed by the CABLESSail actuator being deactivated and the boom returning to its nominal deployed configuration in (f). Note that the gravity-offloading cart is manually moved down the length of the boom throughout the course of the deployment to avoid sagging of the boom.}
    \label{fig:Deployable}
\end{figure*}

The first experimental test is performed where the boom is deployed, then the CABLESSail cable is used to deform the boom in the upwards direction. Still frames from a video of this test are included in Fig.~\ref{fig:Deployable}, where the entire sequence of deployment and actuation occurs over a 6-minute period. It is shown in Fig.~\ref{fig:Deployable} that the boom deploys properly from Fig.~\ref{subfig:Deploy1} to Fig.~\ref{subfig:Deploy3} without any interference from the actuating cable. The actuation of the CABLESSail cable is used in Fig.~\ref{subfig:Deploy4}, where a slight, yet noticeable, upwards deformation of the boom is induced. The tension in the actuation cable is then released and the boom returns to its nominal deployed configuration in Fig.~\ref{subfig:Deploy5}. This is a promising result, as the CABLESSail actuating cable is able to counteract and overcome the gravitational force acting on the boom when it deforms the boom above the horizontal plane.

\begin{figure}[t!]
\centering
 \subfigure[]
 	{
    \includegraphics[width=0.48\columnwidth]{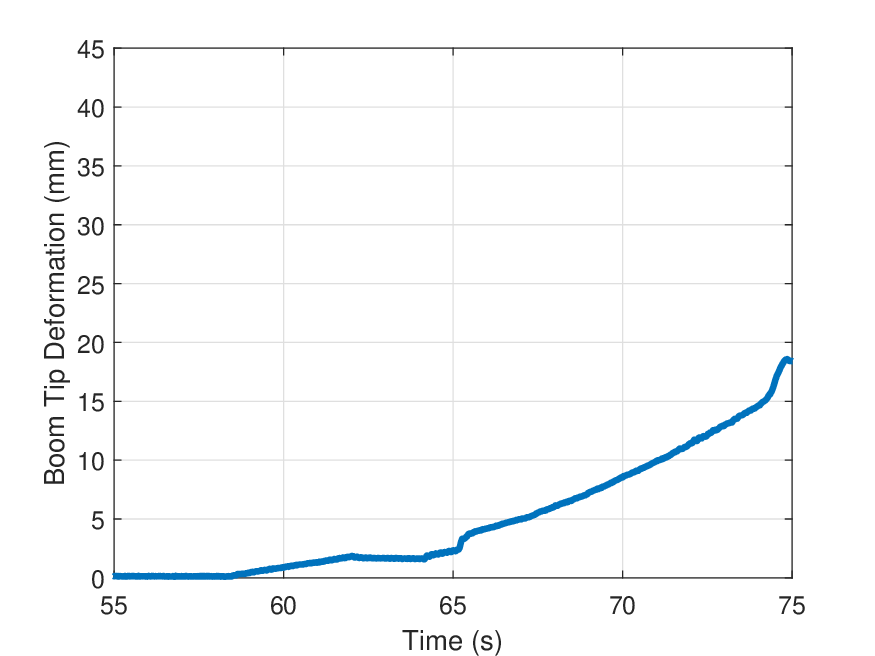}
    \label{fig:VerticalDisp}
    }
\subfigure[]
{
    \includegraphics[width=0.48\columnwidth]{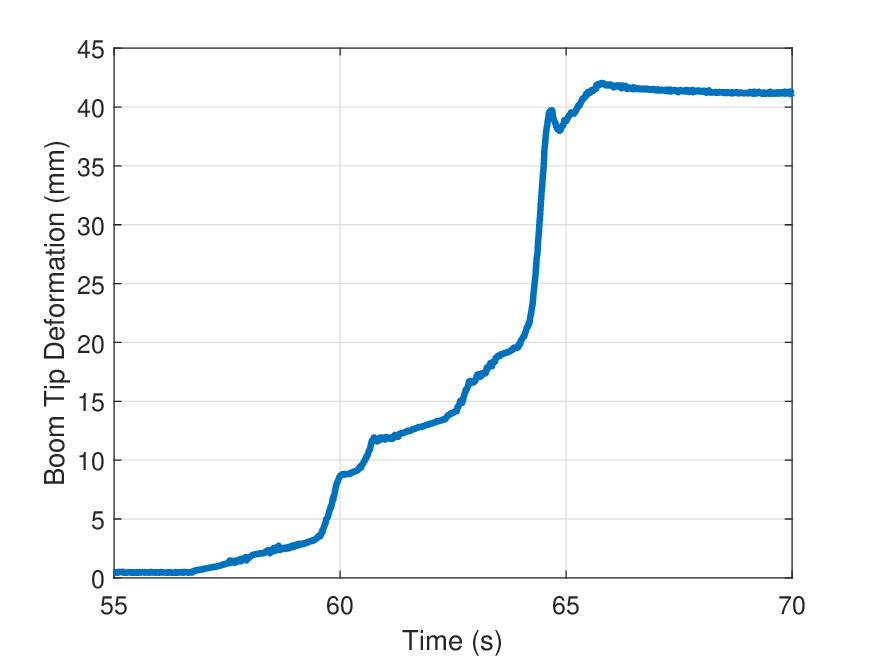}
    \label{fig:SidewaysDisp}
    }
\caption{Boom tip deformation versus time for experimental tests with the deployable prototype where the deformation is performed in (a) the vertical direction and (b) the horizontal direction.}
\label{fig:ViconDisp}
\end{figure}

A Vicon motion capture system is used to further quantify the boom tip deformation obtained when the CABLESSail actuating cable tensioned. Using the same conditions as the deployment test, the actuating cable is incrementally tensioned until the boom resists any further deformation. The results of this test are shown in Fig.~\ref{fig:VerticalDisp}, where a maximum boom tip deformation of 18~mm is achieved within 20 seconds. The time scale, which is directly related to the motor speed, is arbitrarily chosen to demonstrate the cable actuation capability clearly. In other words, this test serves as a proof-of-concept, corresponding to a TRL 3 demonstration. For the CABLESSail concept to advance towards higher TRLs, a full-scale test must be conducted that further investigates its limitations in actuation speed and magnitude. Nonetheless, a full actuation of CABLESSail in 20 seconds is notable, as this scales to roughly 5 minutes of actuation time for a 29.5~m Solar Cruiser-scale boom. In comparison, Solar Cruiser's AMT travels at a maximum speed of 0.5~mm/s, allowing it to travel its full 30~cm distance in about 20~minutes.

\begin{figure}[t!]
\centering
    \includegraphics[width=0.58\columnwidth]{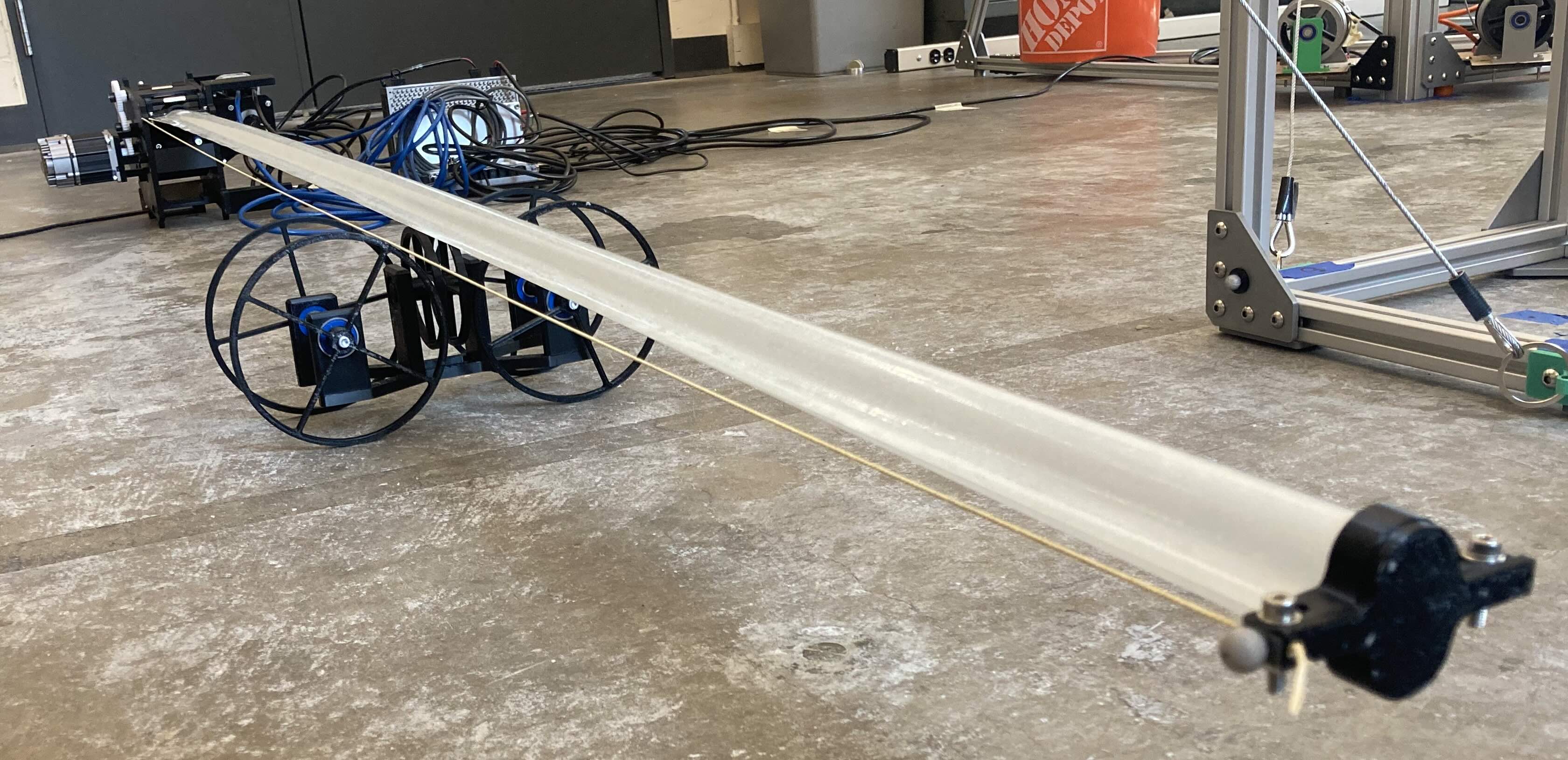}
\caption{The deployable prototype on its side to test actuation of the boom in the horizontal plane with a Vicon marker placed at the cap of the boom.}
\label{fig:Sideways}
\end{figure}

A substantial challenge when performing the preceding tensioning maneuvers is overcoming the gravitational pull acting on the cantilevered boom. To better assess the performance of the CABLESSail actuation without this effect, the prototype is turned sideways, as shown in Fig.~\ref{fig:Sideways}, where tensioning the actuating cable results in the boom deforming in the horizontal plane. Performing the same tensioning maneuver as in the previous tests in this new configuration results in the boom tip deformation shown in Fig.~\ref{fig:SidewaysDisp}. In this case, a 40~mm boom tip deformation is achieved, which is more than two times the deformation obtained in the vertical direction. Following the scaling laws derived by~\citet{bodin2025design}, 40~mm of tip deformation with a 2~m boom is equivalent to 59~cm of tip deformation on Solar Cruiser's 29.5~m long booms. Although larger-scale testing is required to verify CABLESSail's performance on a full-scale boom, this is a promising result and provides preliminary confidence that boom tip actuation in the range of 50-75~cm for a Solar-Cruiser-class solar sail is plausible.

In addition to the fabrication challenges of full-scale booms, several challenges must be overcome if full-scale booms are used for ground testing. Gravity-offloading becomes more difficult, because longer boom increases both weight and the torque at the boom tip. Furthermore, the spiral torsion spring needs to provide larger torque to prevent the boom from expanding in an undesirable fashion during deployment. However, implementation in a space environment should not present significant challenges, as the booms and deployment mechanism used in the small-scale prototype study of this work are directly modeled after the booms and deployment mechanism used by the ACS3 mission.

\section{Control Allocation}
\label{sec:ControlAllocation}

Although the yaw, pitch, and roll maneuvers presented in Section~\ref{sec:StaticSim} are promising, a more systematic approach to determining appropriate maneuver shapes is needed. For example, the roll maneuver results in Fig.~\ref{fig:StaticSim} feature substantial residual yaw and pitch torques. Moreover, it is unclear how to generate torques with components in all three axes simultaneously. To account for this, a control allocation algorithm is devised to determine the boom tip deformations needed to generate a given momentum management torque. Solving for the torque generated by a particular set of boom tip deformations can be achieved using the static simulation described in Section~\ref{sec:StaticSim}. This is a simple process, as the four boom tip deformations map to a unique three-dimensional torque in the presence of no sail membrane deformation. Solving the control allocation problem requires solving the inverse problem, which is much more challenging, as the mapping from the three-dimensional torque to the four boom tip deformations is not unique. Moreover, the mapping from the boom tip deformations to the torque generated is nonlinear, which further complicates solving this inverse problem. A data-driven approach to developing a practical control allocation algorithm is proposed. Although the method outlined in this paper focuses on a single SIA, it can be extended to other SIAs by creating SIA-dependent mappings that are stored in a lookup table. This section presents the methodology of the proposed control allocation approach, followed by numerical simulation results demonstrating its performance.

\subsection{Torque Modeling Approach}

To obtain a model of the torque generated by different combinations of boom tip deformations, static simulations at an SIA of $17$ degrees are performed using the same setup as in Section~\ref{sec:StaticSim}, with an undeformed sail membrane and with a sweep through all possible combinations of boom tip deformations within the range $\pm 50$~cm at all clock angles with 5 degree increments. Starting from one fixed clock angle, the torque generated by each maneuver is recorded as $\mbs{\tau}_{\text{data},k} = \bbm \tau_{\text{yaw},k} & \tau_{\text{pitch},k} & \tau_{\text{roll},k} \ebm$ and associated with the boom tip deformations $\mbf{w}_{\text{data},k} = \bbm w_{1,k} & w_{2,k} & w_{3,k} & w_{4,k} \ebm$, where the subscript $k$ denotes the $k^\mathrm{th}$ data point and $w_{i,k}$, $i=1,\ldots,4$ is the deformation of the $i^\mathrm{th}$ boom at the $k^\mathrm{th}$ data point. Next, a set of basis functions is chosen that are used to map the boom tip deformations to the torque generated. Testing demonstrated that the yaw and pitch torques are linear combinations of the boom tip deformations, while the roll torque is a nonlinear function of boom tip deformations and clock angles. The model of torque generated by the boom tip deformations is expressed in the form 
\begin{equation}
\label{eq:TorqueModel}
\mbs{\tau} = \mbf{f}(\mbf{w},\phi),
\end{equation}
where 
$$
\mbs{\tau} = \bbm \tau_{\text{yaw}} \\ \tau_{\text{pitch}} \\ \tau_{\text{roll}} \ebm, \hspace{20pt}
\mbf{f}(\mbf{w},\phi) = \bbm \mbf{f}_{\text{yaw}}(\mbf{w},\phi) \\ \mbf{f}_{\text{pitch}}(\mbf{w},\phi) \\ \mbf{f}_{\text{roll}}(\mbf{w},\phi)\ebm,
$$
while $\mbf{f}_{\text{yaw}}(\mbf{w},\phi)$ and $\mbf{f}_{\text{pitch}}(\mbf{w},\phi)$ are linear functions of $\mbf{w}$, and $\mbf{f}_{\text{roll}}(\mbf{w},\phi)$ is a nonlinear function of $\mbf{w}$ and $\phi$.

The linear functions are expressed as
\begin{align}
\label{eq:YawTorqueModel}
\mbf{f}_{\text{yaw}}(\mbf{w},\phi) &= \sin(\phi)\mbf{A}_{\text{yaw}} \mbf{w},\\
\label{eq:PitchTorqueModel}
\mbf{f}_{\text{pitch}}(\mbf{w},\phi) &= \cos(\phi)\mbf{A}_{\text{pitch}} \mbf{w},
\end{align}
where $\mbf{A}_{\text{yaw}}$ and $\mbf{A}_{\text{pitch}}$ are matrices that contain linear coefficients of each boom's deformations and constant at all clock angles, the sinusoidal functions for yaw and pitch torques are obtained by an analysis of single boom deformation.
The matrices $\mbf{A}_{\text{yaw}}$ and $\mbf{A}_{\text{pitch}}$ are obtained through a least-squares computation as
\begin{equation}
\bbm \mbf{A}_{\text{yaw}} \\ \mbf{A}_{\text{pitch}} \ebm =\bbm \mbs{\tau}_{\text{data},\text{yaw}}^\trans \\ \mbs{\tau}_{\text{data},\text{pitch}}^\trans \ebm \mbf{w}_{\text{data}}\left(\mbf{w}_{\text{data}}^\trans\mbf{w}_{\text{data}}\right)^{-1},
\end{equation}
where 
\begin{align*}
\mbf{w}_{\text{data}} &=\bbm\mbf{w}_{\text{data},1} \\ \vdots \\ \mbf{w}_{\text{data},n} \ebm , \\
\mbs{\tau}_{\text{data}} &=\bbm \mbs{\tau}_{\text{data},1} \\ \vdots \\ \mbs{\tau}_{\text{data},n}\ebm = \bbm \mbs{\tau}_{\text{data},\text{yaw}} & \mbs{\tau}_{\text{data},\text{pitch}} & \mbs{\tau}_{\text{data},\text{roll}} \ebm,
\end{align*}
$\mbs{\tau}_{\text{data},\text{yaw}} = \bbm \tau_{\text{yaw},1} & \cdots & \tau_{\text{yaw},n} \ebm^\trans$, $\mbs{\tau}_{\text{data},\text{pitch}} = \bbm \tau_{\text{pitch},1} & \cdots & \tau_{\text{pitch},n} \ebm^\trans$, $\mbs{\tau}_{\text{data},\text{roll}} = \bbm \tau_{\text{roll},1} & \cdots & \tau_{\text{roll},n} \ebm^\trans$, and $n$ is the number of data points collected.

The nonlinear function $\mbf{f}_{\text{roll}}(\mbf{w},\phi)$ consists of basis functions $\mbf{F}_{\text{roll}}(\mbf{w})$ that applies boom tip deformations and coefficients $\mbf{q}_{\text{roll}}(\phi)$ related to clock angles as
\begin{equation}
\label{eq:RollTorqueModel}
\mbf{f}_{\text{roll}}(\mbf{w},\phi)=\mbf{F}_{\text{roll}}(\mbf{w})\mbf{q}_{\text{roll}}(\phi).
\end{equation}
The polynomial basis functions span all combinations of linear to cubic terms of boom tip deformations as
\begin{equation}
\mbf{F}_{\text{roll}}(\mbf{w})=\mbf{F}_{\text{r},\text{lin}}(\mbf{w}) + \mbf{F}_{\text{r},\text{qua}}(\mbf{w}) + \mbf{F}_{r,cub}(\mbf{w}),
\end{equation}
where
\begin{align*}
\mbf{F}_{\text{r},\text{lin}}(\mbf{w}) &= \bbm w_1&w_2&w_3&w_4\ebm \mbf{q}_{\text{r},\text{lin}}(\phi), \\
\mbf{F}_{\text{r},\text{qua}}(\mbf{w}) &= \bbm w_1^2&\cdots
&w_1w_2&\cdots\ebm \mbf{q}_{\text{r},\text{qua}}(\phi), \\
\mbf{F}_{\text{r},\text{cub}}(\mbf{w}) &= \bbm w_1^3&\cdots
&w_1^2w_2&\cdots&w_1w_2w_3&\cdots\ebm \mbf{q}_{r,cub}(\phi),
\end{align*}
$\mbf{q}_{\text{r},\text{lin}}(\phi)\in\mathbb{R}^4$, $\mbf{q}_{\text{r},\text{qua}}(\phi)\in\mathbb{R}^{10}$, and $\mbf{q}_{\text{r},\text{cub}}(\phi)\in\mathbb{R}^{20}$ are trigonometric coefficients for each basis function that are expressed as
$$
\mbf{q}_{\text{roll}}(\phi)=\bbm \mbf{q}_{\text{r},\text{lin}}^\trans(\phi) & \mbf{q}_{\text{r},\text{qua}}^\trans(\phi) & \mbf{q}_{\text{r},\text{cub}}^\trans(\phi)\ebm^\trans
=\mbf{q}_\phi\sin(2\phi)\bbm 1 & \csc(\phi) & \sec(\phi) & \csc(2\phi) \ebm^\trans,
$$
and $\mbf{q}_\phi\in\mathbb{R}^{34\times4}$ is a coefficient for each trigonometric function that is determined by a linear regression using the data collected at 5 degrees intervals over the entire 360 degrees range.

\subsection{Control Allocation Methodology}

Now that a fit for the nonlinear mapping from boom tip deformations to torque has been found, performing control allocation amounts to solving for the roots of the expression $\mbs{\tau}_\text{des} - \mbf{f}(\mbf{w},\phi)$, where $\mbs{\tau}_\text{des}$ is the desired momentum management torque to be generated. To ensure that this can be performed onboard a solar sail's flight computer, the proposed control allocation algorithm makes use of the Levenberg-Marquardt method to solve the nonlinear weighted least-squares optimization problem
\begin{align}
\min_{\mbf{w} \in \mathbb{R}^4} \quad & (\mbs{\tau}_\text{des} - \mbf{f}(\mbf{w},\phi))^\trans \mbf{W}(\mbs{\tau}_\text{des} - \mbf{f}(\mbf{w},\phi))\\
\textrm{s.t.} \quad & |w_i| \leq w_{\text{max}}, \,\, i = 1,2,3,4,
\end{align}
where $\mbf{w} = \bbm w_{1} & w_{2} & w_{3} & w_{4} \ebm^\trans$ is the design variable representing the boom tip deformations and $\mbf{W} = \mbf{W}^\trans > 0$ is a positive definite weighting matrix that can be used to emphasize which components of $\mbs{\tau}_\text{des}$ are more important to match. This weighting matrix can also be used to normalize the desired torque, which is useful in this application, since roll torques are typically two orders of magnitude smaller than yaw/pitch torques. 
Based on the expected actuation limits of CABLESSail, a maximum feasible boom tip deformation length $w_{\text{max}}$ is applied as a boundary constraint to the optimization problem.
The algorithm proceeds with the following iterative steps:

\begin{itemize}

\item \textbf{Step 1:} Initialize $\mbf{w}^{(0)}$ and set $j = 0$.

\item \textbf{Step 2:} Solve for $\mbf{w}^{(j+1)}$ with the update law
    \begin{equation}
    \label{eq:ControlAllocationUpdate}
    \mbf{w}^{(j+1)} = \mbf{w}^{(j)} + \left(\mbf{J}^\trans \mbf{W}^{-1}\mbf{J} + \eta \mbf{I} \right)^{-1} \mbf{J}^\trans \mbf{W}^{-1}(\mbs{\tau}_\text{des} - \mbf{f}(\mbf{w}^{(j)},\phi)),
    \end{equation}
    where $\mbf{J} = \partial \mbf{f}/\partial \mbf{w}|_{\mbf{w}^{(j)}}$ is the Jacobian of $\mbf{f}(\mbf{w},\phi)$ evaluated at $\mbf{w}^{(j)}$ and $\eta > 0$ is a term used to add numerical damping to the computations. If no boom tip deformations have been constrained in previous iterations, then $\mbf{f}(\mbf{w},\phi)$ is defined based on Eqs.~\eqref{eq:TorqueModel},~\eqref{eq:YawTorqueModel},~\eqref{eq:PitchTorqueModel}, and~\eqref{eq:RollTorqueModel}, and $\mbf{w}^{(j+1)}$ is updated according to Eq.~\eqref{eq:ControlAllocationUpdate}. If any boom tip deformations have been constrained in previous iterations, then the associated entries of $\mbf{w}^{(j)}$ are fixed at their constrained values when computing $\mbf{f}(\mbf{w}^{(j)},\phi)$, the columns of $\mbf{J} = \partial \mbf{f}/\partial \mbf{w}|_{\mbf{w}^{(j)}}$ associated with these entries are removed, and the update law in Eq.~\eqref{eq:ControlAllocationUpdate} is only used to update the remaining unconstrained boom tip deformations.

\item \textbf{Step 3:} Jump to Step 6 if $\max\left(|w_1^{(j+1)}|, |w_2^{(j+1)}|, |w_3^{(j+1)}|, |w_4^{(j+1)}|\right) \leq w_{\text{max}}$, where $w_{\text{max}}>0$ is a user-defined maximum feasible boom tip deformation length. Else, go to Step 4.

\item \textbf{Step 4:} Choose the boom tip deformation with the largest absolute value, and scale it down to the constraint boundary $w_{\text{max}}$ while keeping its sign. For example, if $|w_1^{(j+1)}| = \max\left(|w_1^{(j+1)}|, |w_2^{(j+1)}|, |w_3^{(j+1)}|, |w_4^{(j+1)}|\right)$, then the solution is updated as $\mbf{w}^{(j+1)} = \bbm w_{1,\text{max}} & w_{2}^{(j+1)} & w_{3}^{(j+1)} & w_{4}^{(j+1)} \ebm^\trans$, where
\begin{equation}
w_{1,\text{max}} = \text{sign}(w_1^{(j+1)})~w_{\text{max}}.
\end{equation}

\item \textbf{Step 5:} Set $j=j+1$ and return to Step 2 with the boom tip deformation constrained in Step 4 removed from the optimization problem.

\item \textbf{Step 6:} Exit if $(\mbs{\tau}_\text{des} - \mbf{f}(\mbf{w}^{(j+1)},\phi))^\trans \mbf{W}(\mbs{\tau}_\text{des} - \mbf{f}(\mbf{w}^{(j+1)},\phi)) < \epsilon$ or the maximum number of iterations is exceeded, where $\epsilon > 0$ is a user-defined tolerance of convergence. Else, set $j = j + 1$ and return to Step 2.

\end{itemize}

Note that the solution and convergence properties of this algorithm depend on the initial guess $\mbf{w}^{(0)}$. An initial guess closer to the optimal value will likely reduce the number of iterations required to meet the convergence criteria and potentially improve the quality of the solution. Results in this section are generated with $\mbf{w}^{(0)} = \mbf{0}$ to demonstrate the performance of the algorithm with a relatively poor initial guess. It is also worth noting that this algorithm transforms the nonlinear weighted least-squares into a sequence of simple computations that only involve a matrix inverse and matrix multiplications. This increases the likelihood that this algorithm could be performed onboard a solar sail flight computer.

\subsection{Control Allocation Simulation Results}

The proposed control allocation algorithm is validated through numerical static simulation studies. The control allocation method is first applied to improve upon the intuitive momentum management maneuvers tested in Section~\ref{sec:StaticSim} with the intent of reducing unwanted residual torques. To better understand the range of desired momentum management torques that can be accurately achieved by the proposed algorithm, a second set of results is presented in this section that explores the accuracy of the torques generated across a range of desired values to determine the range of feasible torques.

\subsubsection{Comparison to Intuitive Maneuvers} \label{subsec:CompareIntuitive}

The proposed control allocation algorithm is first tested by attempting to improve upon the intuitive maneuvers performed in Section~\ref{sec:StaticSim}. Individual yaw, pitch, and roll maneuvers are tested. Desired torques at each clock angle in Table~\ref{table:CA_condition} are provided to the algorithm with an initial guess of $\mbf{w}^{(0)} = \mbf{0}$, numerical damping $\eta = 10^{-6}$, and a convergence tolerance of $\epsilon = 10^{-5}$. A weighting matrix $\mbf{W}$ used for each case is provided in Table~\ref{table:CA_condition}. Based on the discussion in Section~\ref{subsec:DeployablePrototypeTestResults}, constraints of $50$~cm and $75$~cm are chosen for the maximum allowable boom tip deformations. With those conditions, the resulting boom tip deformations computed by the proposed control allocation algorithm, with and without constraints, are shown in Table~\ref{table:CA_result}.

\begin{table}[t!]
\caption{Conditions for the numerical tests of the proposed control allocation algorithm for specific individual yaw, pitch, and roll maneuvers.}\label{table:CA_condition}
\def\arraystretch{1.2}
\begin{center}
\begin{tabular}{c@{\quad}c@{\quad}c@{\quad}l}
    \toprule
    $\phi$ (deg) & Axis & $\mbs{\tau}_{\text{des}}$ (N$\cdot$m) & $\mbf{W}$\\
     \midrule
     \midrule
    & Yaw & $\bbm3.7\times10^{-4}&0&0\ebm^\trans$ & $\text{diag}\{1,1,10^2\}$\\
    30 & Pitch & $\bbm0&6.3\times10^{-4}&0\ebm^\trans$ & $\text{diag}\{1,1,10^2\}$\\
    & Roll & $\bbm0&0&1.8\times10^{-5}\ebm^\trans$ & $\text{diag}\{1,1,10^3\}$\\
    \midrule
    & Yaw & $\bbm5.2\times10^{-4}&0&0\ebm^\trans$ & $\text{diag}\{1,1,10^2\}$\\
    45 & Pitch & $\bbm0&5.2\times10^{-4}&0\ebm^\trans$ & $\text{diag}\{1,1,10^2\}$\\
    & Roll & $\bbm0&0&2.1\times10^{-5}\ebm^\trans$ & $\text{diag}\{1,1,10^3\}$\\
    \bottomrule
\end{tabular}
\def\arraystretch{1.0}
\end{center}
\end{table}

\begin{table}[t!]
\caption{Results of the proposed control allocation algorithm for specific individual yaw, pitch, and roll maneuvers.}\label{table:CA_result}
\def\arraystretch{1.2}
\begin{center}
\begin{tabular}{c@{\quad}c|@{\quad}c@{\quad}c@{\quad}c}
    \toprule
    & & & Optimal $\mbf{w}$ (cm) & \\
    $\phi$ (deg) & Axis & w/ $50$~cm Constraint & w/ $75$~cm Constraint & No Constraint\\
     \midrule
     \midrule
    & Yaw & $\bbm 41.6 & 14.5 & -42.9 & 36.1 \ebm^\trans$ & - & - \\
    30 & Pitch & $\bbm -29.0 & 24.1 & -20.8 & -23.2 \ebm^\trans$ & - & - \\
    & Roll & $\bbm -44.2 & -50.0 & 50.0 & -39.4 \ebm^\trans$ & $\bbm -53.2 & -75.0 & 75.0 & 18.1 \ebm^\trans$ & $\bbm -81.4 & -76.7 & 99.3 & 27.7 \ebm^\trans$ \\
    \midrule
    & Yaw & $\bbm 33.3 & 16.9 & -34.4 & 33.5 \ebm^\trans$ & - & - \\
    45 & Pitch & $\bbm -33.5 & 34.4 & -16.9 & -33.3 \ebm^\trans$ & - & -\\
    & Roll & $\bbm -42.4 & -50.0 & 50.0 & -30.7 \ebm^\trans$ & $\bbm -35.0 & -75.0 & 75.0 & 35.0 \ebm^\trans$ & $\bbm -67.2 & -98.7 & 98.7 & 67.2 \ebm^\trans$ \\
    \bottomrule
\end{tabular}
\def\arraystretch{1.0}
\end{center}
\end{table}

\begin{figure}[t!]
\centering
\subfigure[]
	{
    \includegraphics[width=0.48\columnwidth]{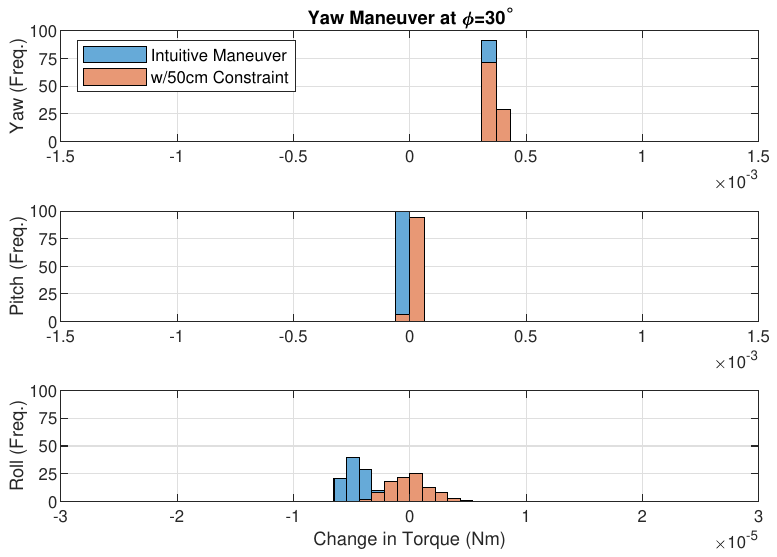}
    \label{subfig:Alloc_yaw_30}
}
\subfigure[]
	{
    \includegraphics[width=0.48\columnwidth]{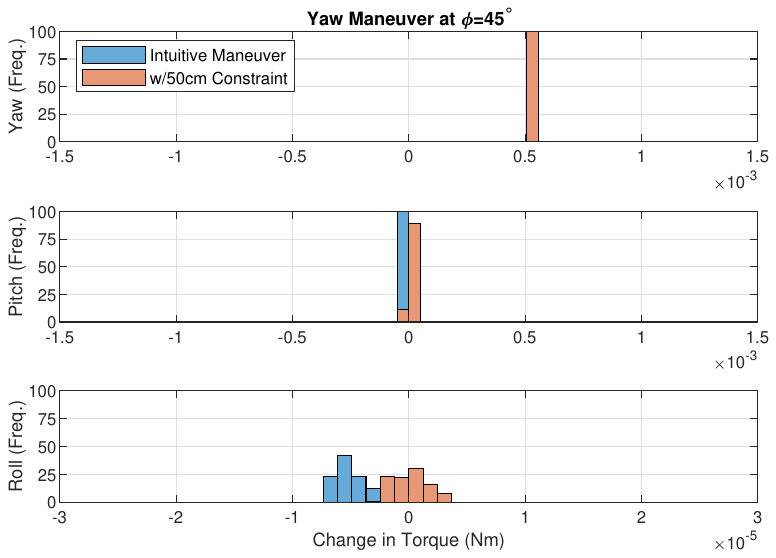}
\label{subfig:Alloc_yaw_45}
}
\subfigure[]
	{
    \includegraphics[width=0.48\columnwidth]{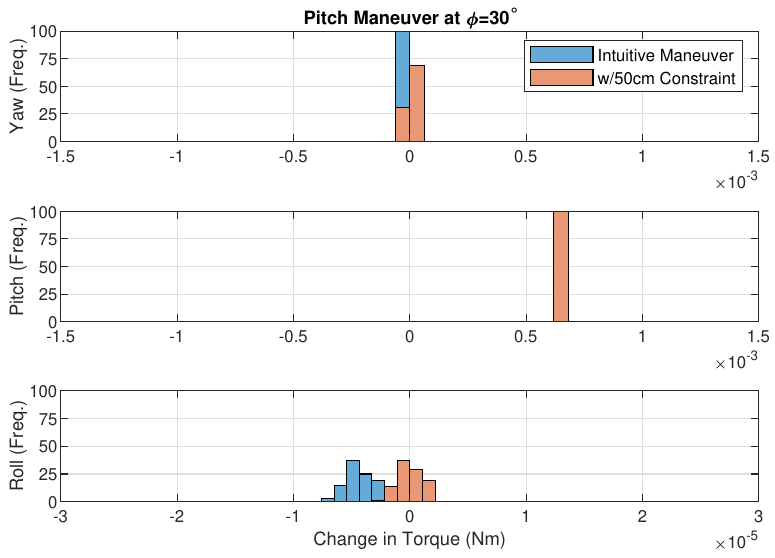}
    \label{subfig:Alloc_pitch_30}
}
\subfigure[]
	{
    \includegraphics[width=0.48\columnwidth]{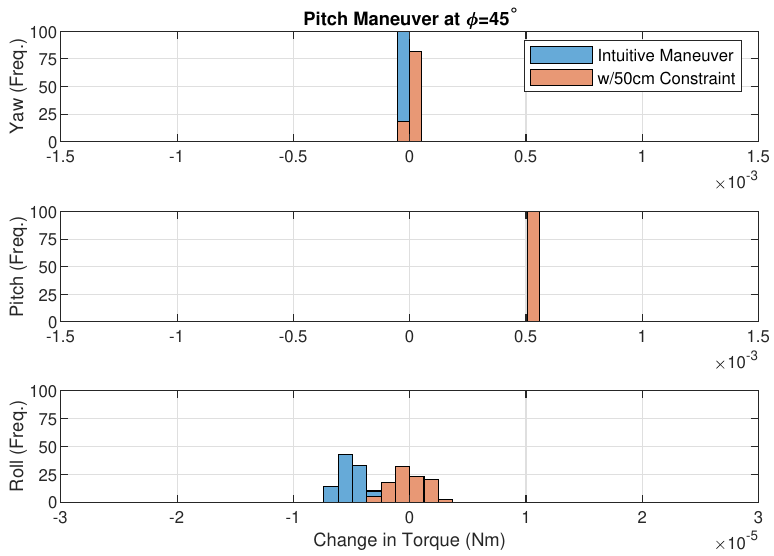}
\label{subfig:Alloc_pitch_45}
}
\caption{Static simulations of (a, b) pure yaw and (c, d) pure pitch torque generation designed using the intuitive maneuver from Section~\ref{sec:StaticSim} and the optimized Control Allocation Maneuver with a constraint. Histograms of change in torque generated across all simulated sail membrane shapes at the clock angles of (a, c) $30$ and (b, d) $45$~degrees.}
\label{fig:Alloc_yaw_pitch}
\end{figure}

\begin{figure}[t!]
\centering
\subfigure[]
	{
    \includegraphics[width=0.48\columnwidth]{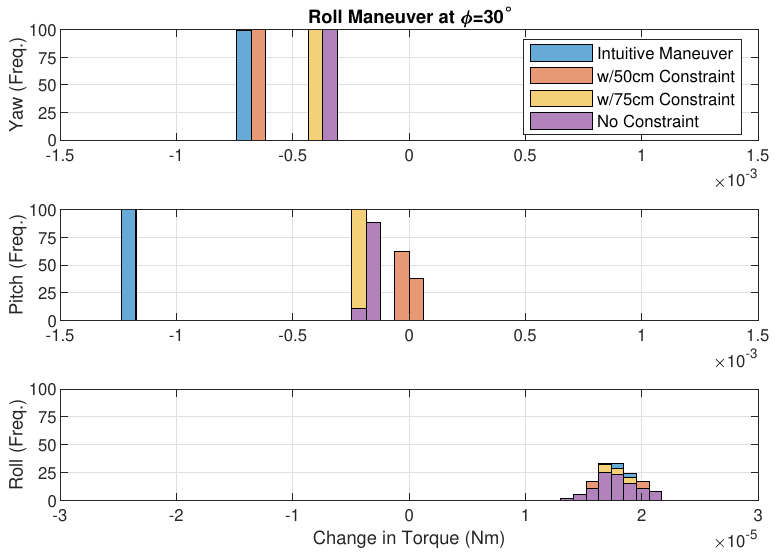}
    \label{subfig:Alloc_roll_30}
}
\subfigure[]
	{
    \includegraphics[width=0.48\columnwidth]{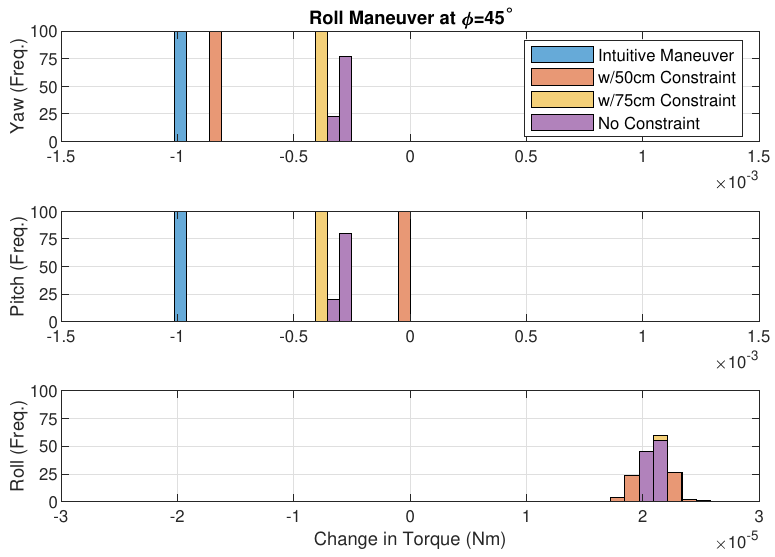}
\label{subfig:Alloc_roll_45}
}
\caption{Static simulations of pure roll torque generation designed using the intuitive maneuver from Section~\ref{sec:StaticSim} and the optimized Control Allocation Maneuver with and without constraints. Histograms of change in torque generated across all simulated sail membrane shapes at the clock angles of (a) $30$ and (b) $45$~degrees.}
\label{fig:Alloc_roll}
\end{figure}

These boom tip deformations are then used as inputs to the same Monte Carlo simulations performed in Section~\ref{sec:StaticSim} that compute the change in torque generated under varying sail membrane deformations, which results in the torque distributions shown in Figs.~\ref{fig:Alloc_yaw_pitch} and~\ref{fig:Alloc_roll}. These figures present histograms of the change in torques induced by the control allocation maneuvers, and are directly compared to the intuitive maneuvers tested in Section~\ref{sec:StaticSim}. Specifically, the yaw and pitch maneuver results are shown in Fig.~\ref{fig:Alloc_yaw_pitch}, while the roll maneuver results are in Fig.~\ref{fig:Alloc_roll}.
The histograms in Fig.~\ref{fig:Alloc_yaw_pitch} only include the $50$~cm constraint, as the yaw and pitch maneuvers do not violate the $50$~cm constraint, removing the need to test the other case.
The control allocation maneuver is shown to reliably generate torques with very similar magnitude to the intuitive maneuver in the desired axes.
Notably, in the case of the roll maneuver in Fig.~\ref{fig:Alloc_roll}, the residual yaw and pitch torques are decreased by roughly a factor of five, which highlights a substantial improvement with the control allocation maneuver compared to the intuitive maneuver. Also, in the case of the yaw and pitch maneuver in Fig.~\ref{fig:Alloc_yaw_pitch}, the residual roll torque is decreased significantly.
The histograms in Fig.~\ref{fig:Alloc_roll} show that the boom tip deformation constraints included in the proposed control allocation algorithm affect the residual yaw and pitch torques while reliably maintaining the roll torque generation. More strict constraints increase the residual yaw and pitch torques, but still show a significant improvement compared to the intuitive maneuver.

\subsubsection{Range of Feasible Torques}\label{subsec:RangeFeasibleTorques}

To find the range of feasible momentum management torques that CABLESSail can generate with the proposed control allocation algorithm, additional tests are performed with a range of commanded desired yaw, pitch, and roll torques.
Two sets of tests are performed to assess the generation of either a combined yaw and pitch torque, or a pure roll torque. In each case, the proposed control allocation algorithm is tested across a grid of desired torques, where the accuracy of the CABLESSail torque generated is assessed as a percent error relative to the desired torque and the unwanted residual torques in the other axis or axes are quantified. The resolutions of the test grids are chosen as $2\times10^{-5}$ and $2\times10^{-6}$~N$\cdot$m for the yaw/pitch and roll torque tests, respectively. These resolutions are chosen based on the spread in torques found in Figs.~\ref{fig:Alloc_yaw_pitch} and~\ref{fig:Alloc_roll} when testing across sail membrane shape variation, as in practice membrane shape uncertainty will limit the resolution of achievable torques. The numerical parameters for the control allocation method, such as its initial guess, numerical damping, convergence tolerance, and weighting matrix, match those used in Section~\ref{subsec:CompareIntuitive} for the yaw/pitch maneuver and pure roll maneuver, respectively. A maximum boom tip deformation constraint of $75$~cm is used, along with a flat sail membrane shape. An SIA of $17$ degrees is considered for all tests and results at clock angles of $5$~degrees, $15$~degrees, $30$~degrees, $45$~degrees, $60$~degrees, and $75$~degrees are included. Results at clock angles outside this range are not included, as symmetry of the sail results in a repeating pattern of results that can be extrapolated from tests within this range.

\begin{figure}[t!]
\centering
\subfigure[]
	{
    \includegraphics[width=0.627\columnwidth]{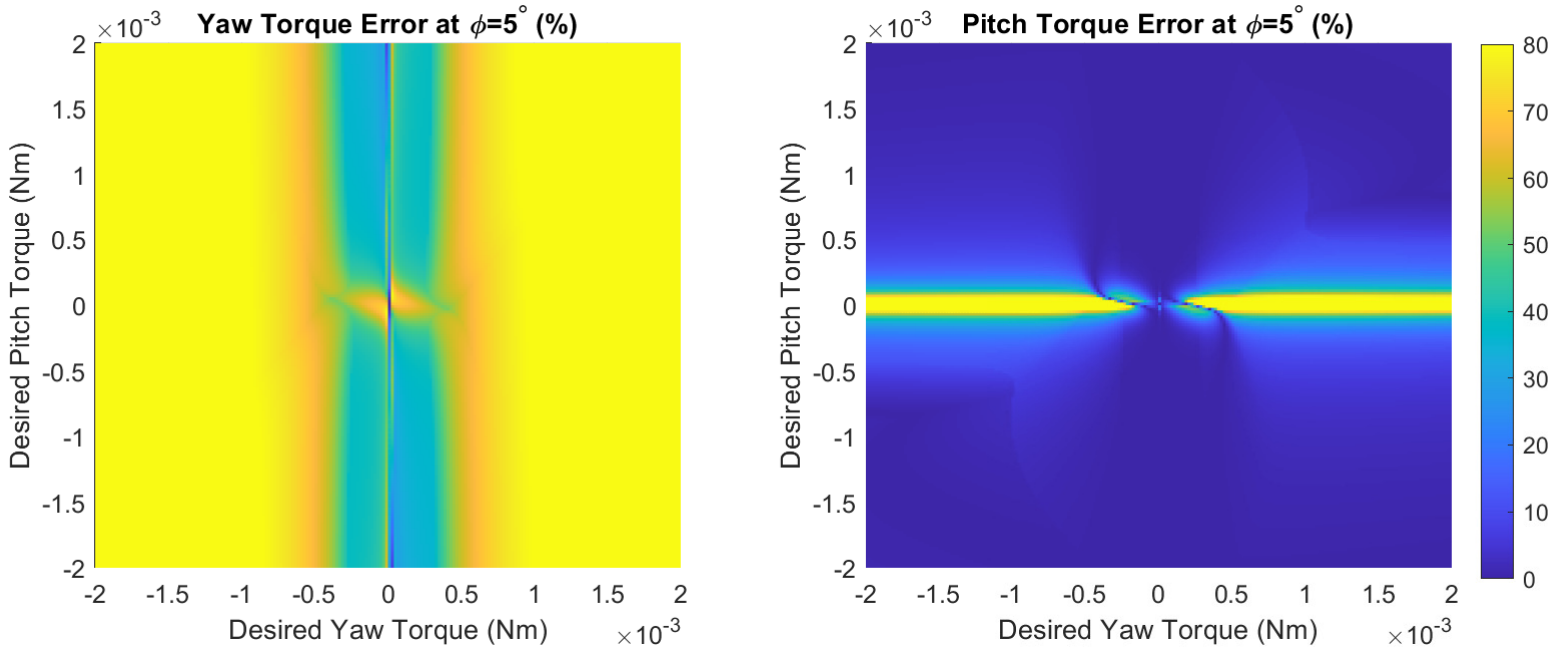}
    \label{subfig:yawpitch_ca5}
}
\subfigure[]
	{
    \includegraphics[width=0.333\columnwidth]{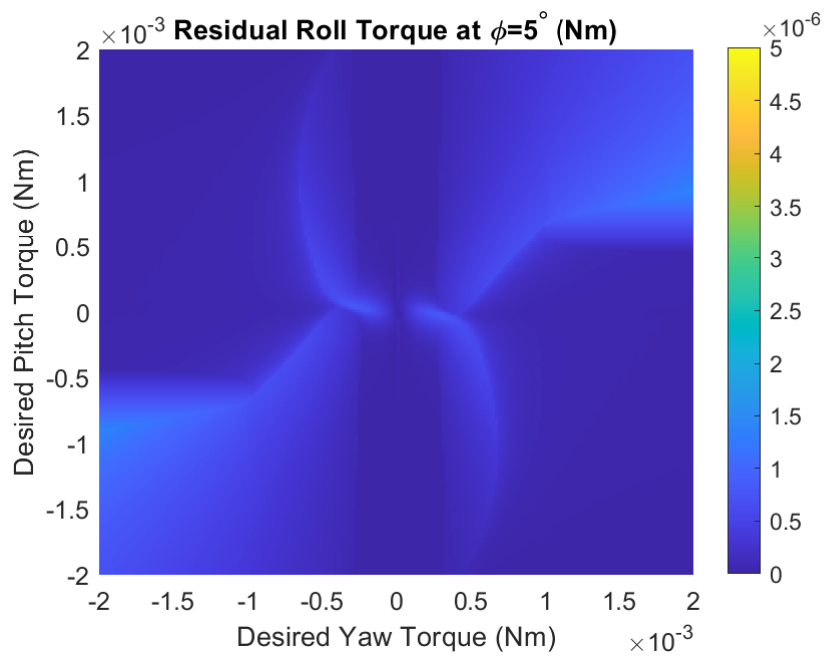}
\label{subfig:yawpitch_res_roll_ca5}
}
\subfigure[]
	{
    \includegraphics[width=0.627\columnwidth]{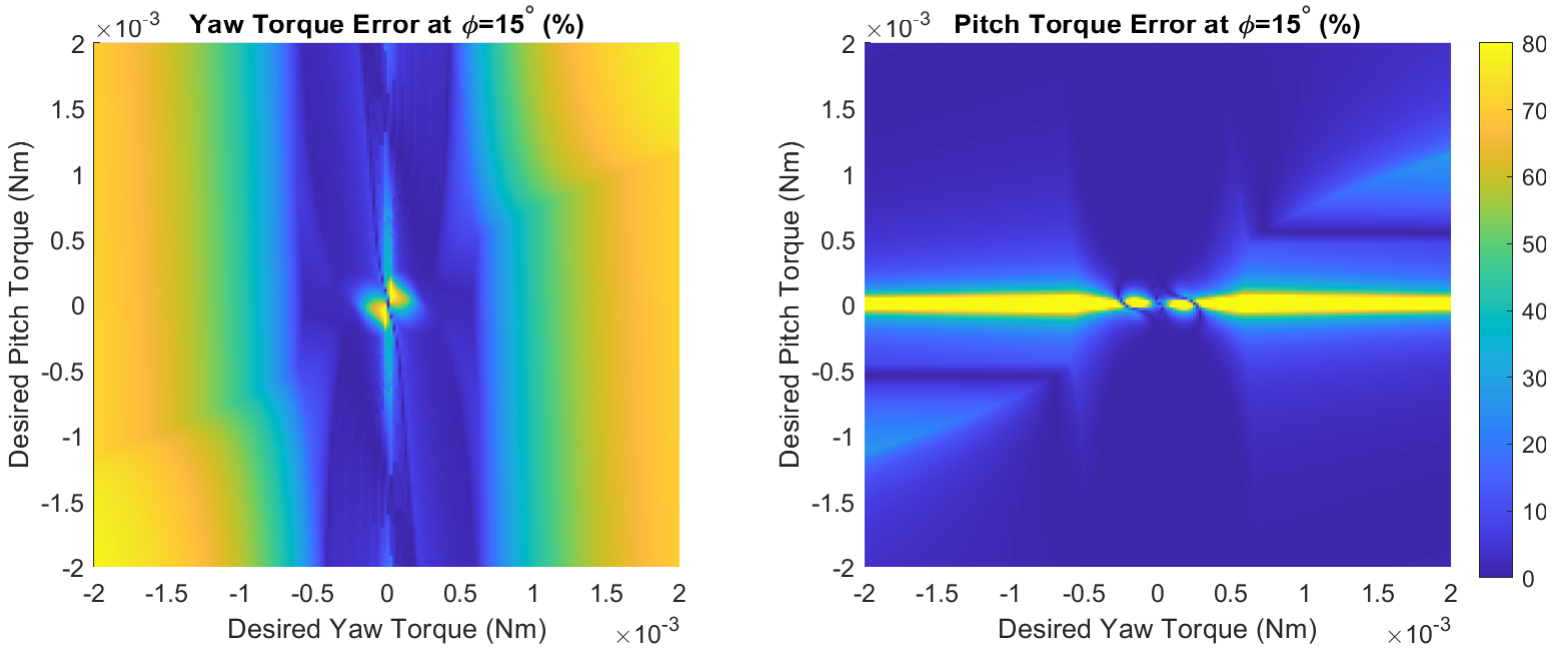}
    \label{subfig:yawpitch_ca15}
}
\subfigure[]
	{
    \includegraphics[width=0.333\columnwidth]{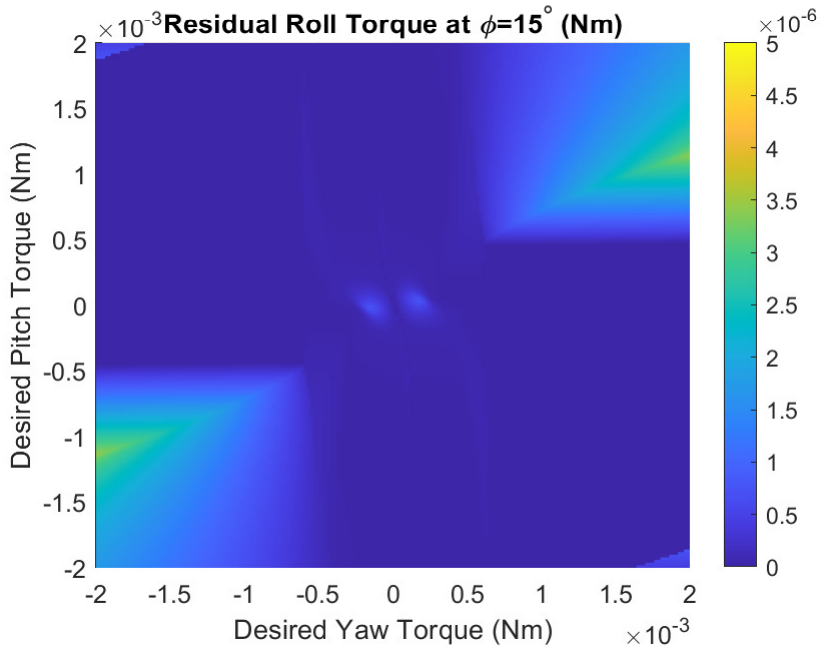}
\label{subfig:yawpitch_res_roll_ca15}
}
\subfigure[]
	{
    \includegraphics[width=0.627\columnwidth]{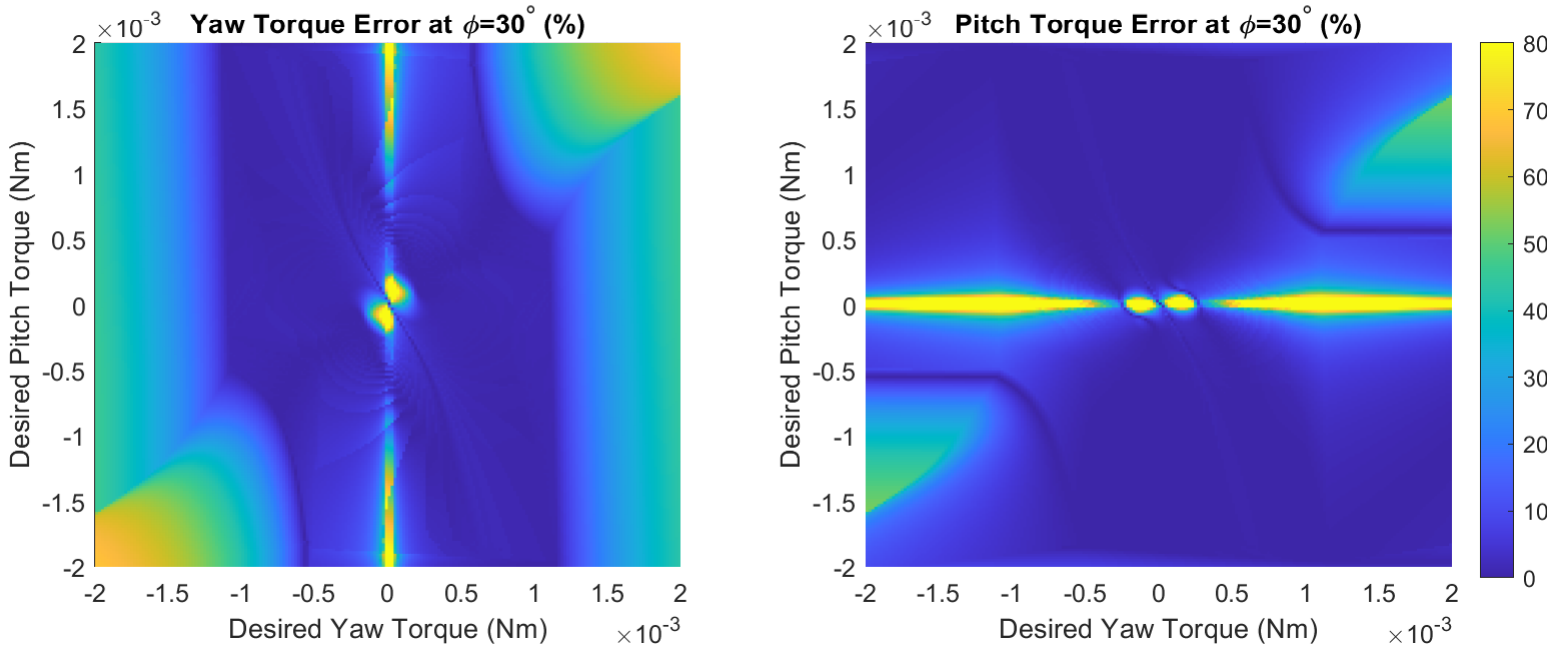}
    \label{subfig:yawpitch_ca30}
}
\subfigure[]
	{
    \includegraphics[width=0.333\columnwidth]{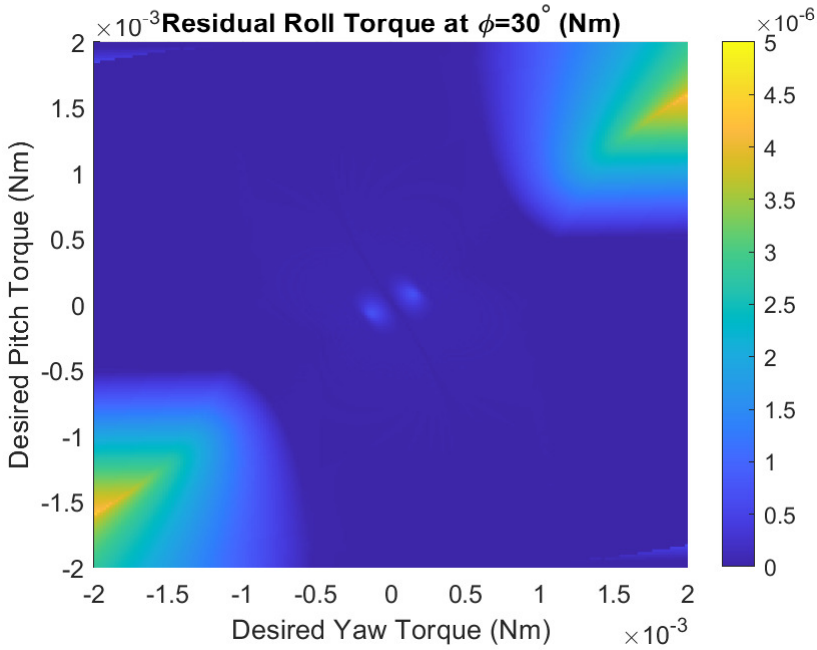}
\label{subfig:yawpitch_res_roll_ca30}
}
\caption{Errors of the torque generated compared to the desired torques for yaw/pitch combined maneuvers with a constraint of $75$~cm. Colormaps of (a, c, e) yaw and pitch torque errors in percent and (b, d, f) residual roll torques in N$\cdot$m are included at clock angles of (a, b) 5, (c, d) 15, and (e, f) 30 degrees.}
\label{fig:range_ca5to30}
\end{figure}

\begin{figure}[t!]
\centering
\subfigure[]
	{
    \includegraphics[width=0.627\columnwidth]{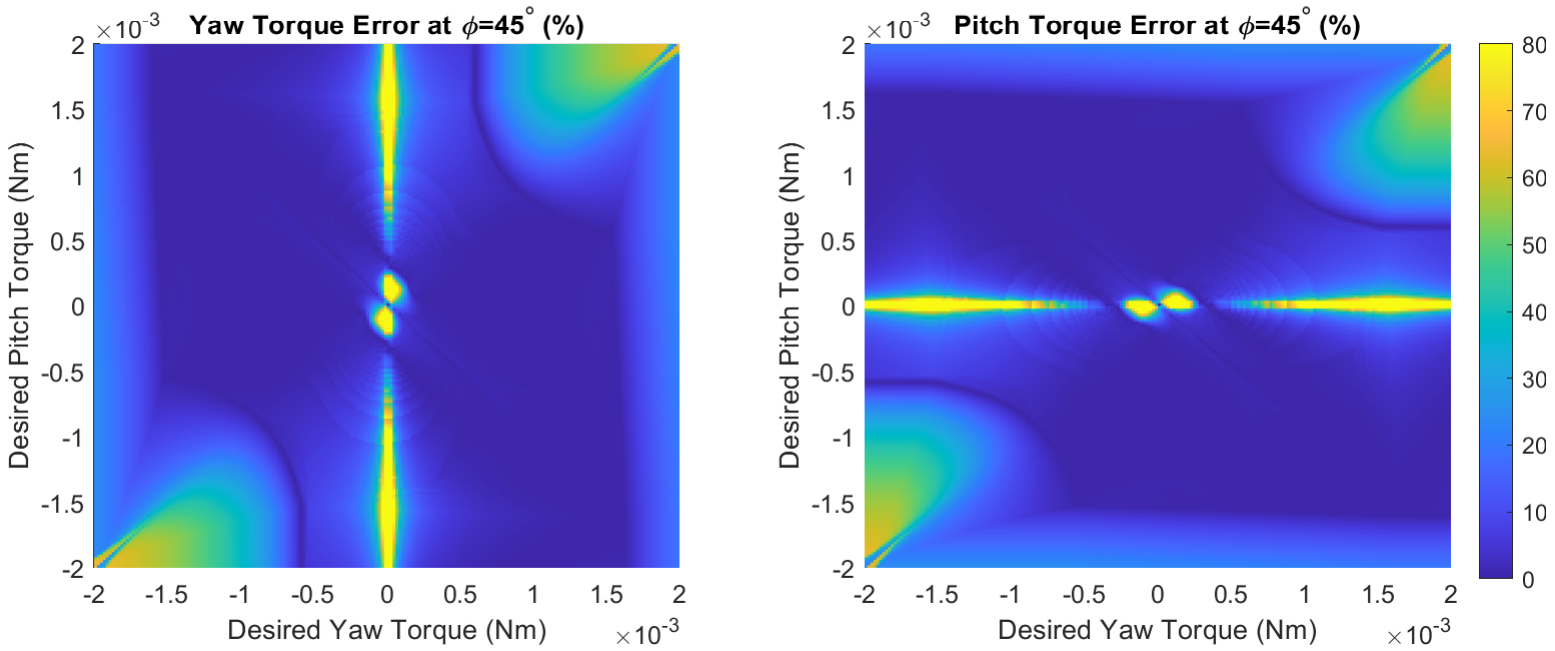}
    \label{subfig:yawpitch_ca45}
}
\subfigure[]
	{
    \includegraphics[width=0.333\columnwidth]{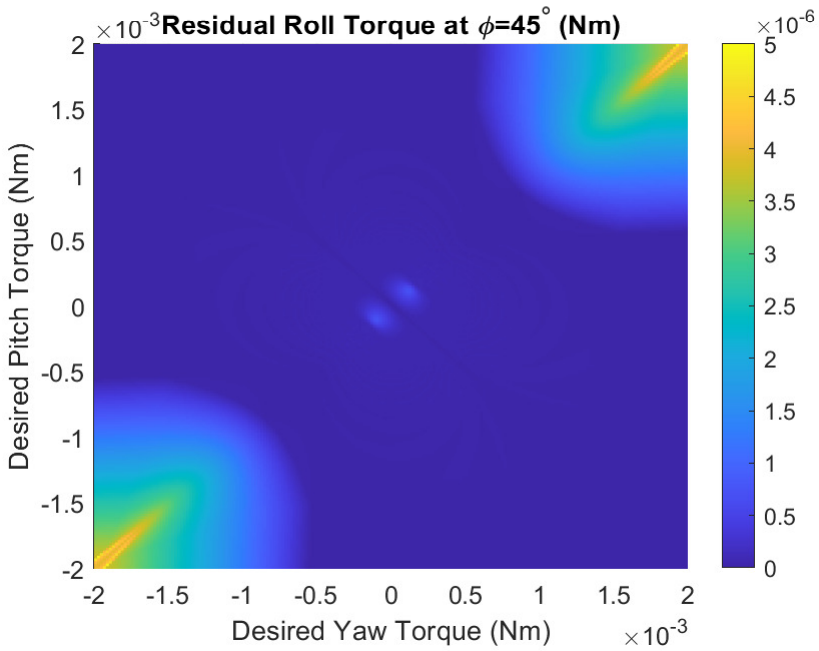}
\label{subfig:yawpitch_res_roll_ca45}
}
\subfigure[]
	{
    \includegraphics[width=0.627\columnwidth]{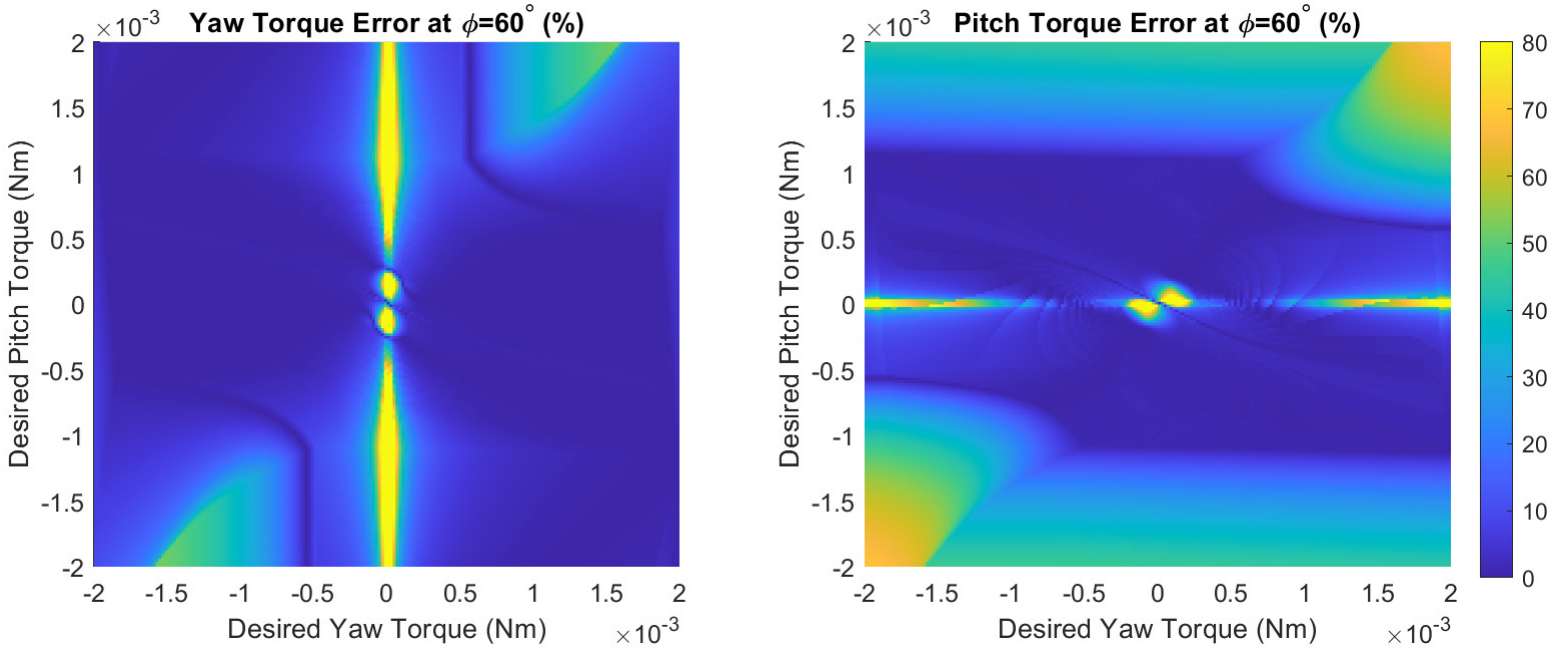}
    \label{subfig:yawpitch_ca60}
}
\subfigure[]
	{
    \includegraphics[width=0.333\columnwidth]{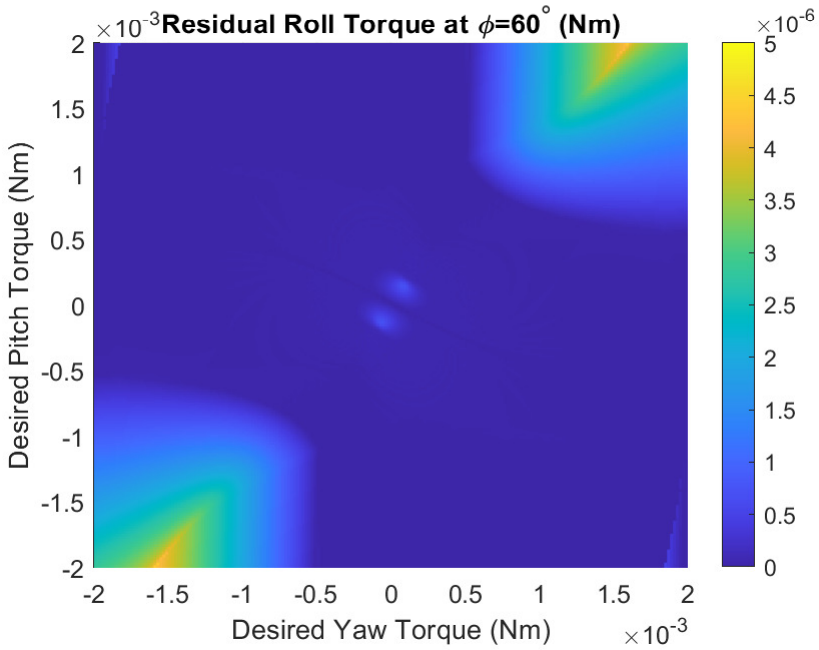}
\label{subfig:yawpitch_res_roll_ca60}
}
\subfigure[]
	{
    \includegraphics[width=0.627\columnwidth]{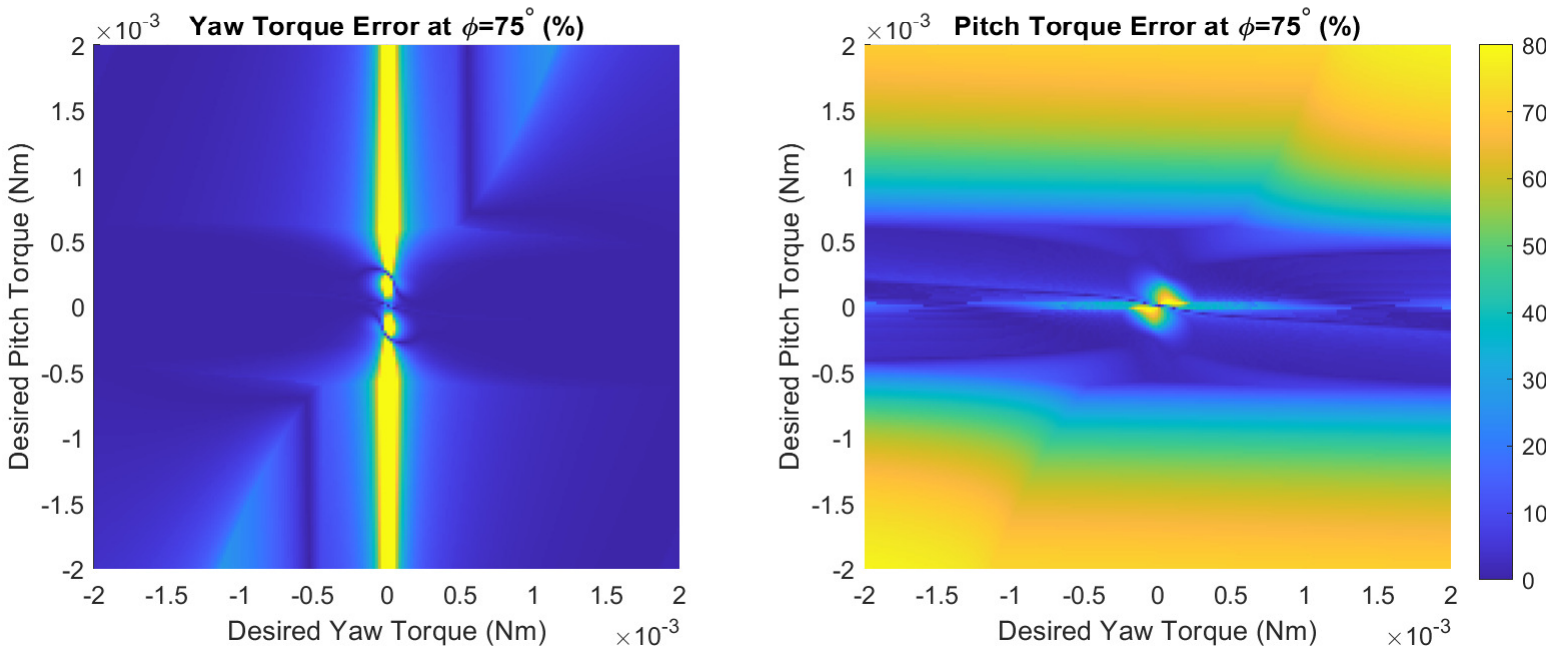}
    \label{subfig:yawpitch_ca75}
}
\subfigure[]
	{
    \includegraphics[width=0.333\columnwidth]{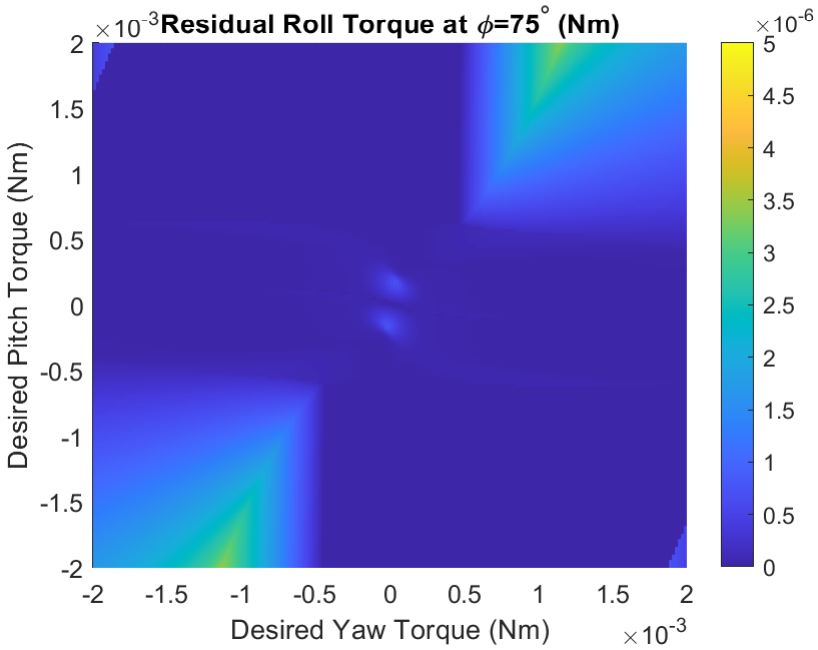}
\label{subfig:yawpitch_res_roll_ca75}
}
\caption{Errors of the torque generated compared to the desired torques for yaw/pitch combined maneuvers with a constraint of $75$~cm. Colormaps of (a, c, e) yaw and pitch torque errors in percent and (b, d, f) residual roll torques in N$\cdot$m are included at clock angles of (a, b) 45, (c, d) 60, and (e, f) 75 degrees.}
\label{fig:range_ca45to75}
\end{figure}

The range of feasible yaw and pitch torques across the clock angles is shown in Figs.~\ref{fig:range_ca5to30} and~\ref{fig:range_ca45to75}. In these figures, the dark regions depict areas of low percentage error, which correspond to the feasible torques that can be generated at the respective clock angles. For yaw torques, it is observed that as the clock angle increases from $5$~degrees to $75$~degrees, the range of feasible torques gets wider. On the other hand, for pitch torques, as the clock angle increases, the range of feasible torques gets narrower. This highlights the clock-angle-dependency of the yaw/pitch torques generated with CABLESSail. This dependency is periodic, which is observed by examining the distributions of the yaw and pitch torque errors in Figs.~\ref{subfig:yawpitch_ca15} and~\ref{subfig:yawpitch_ca75}, where the yaw and pitch errors are swapped at clock angles of $15$~degrees and $75$~degrees. The same relationship is observed between Figs.~\ref{subfig:yawpitch_ca30} and~\ref{subfig:yawpitch_ca60} when comparing clock angles of $30$~degrees and $60$~degrees. The large percentage error along zero desired yaw/pitch torques is slightly misleading, as the torque errors are small in magnitude in this region, but the desired torque is also very small.
In practice, it is inadvisable to set the desired momentum management torque very close to zero due to this error. Figs.~\ref{fig:range_ca5to30} and~\ref{fig:range_ca45to75} also demonstrate that the residual roll torque generated with all of the desired yaw/pitch torques remains very small. Most instances have residual roll torques less than $1 \times 10^{-6}$~N$\cdot$m.

\begin{figure}[t!]
\centering
\subfigure[]
	{
    \includegraphics[width=0.32\columnwidth]{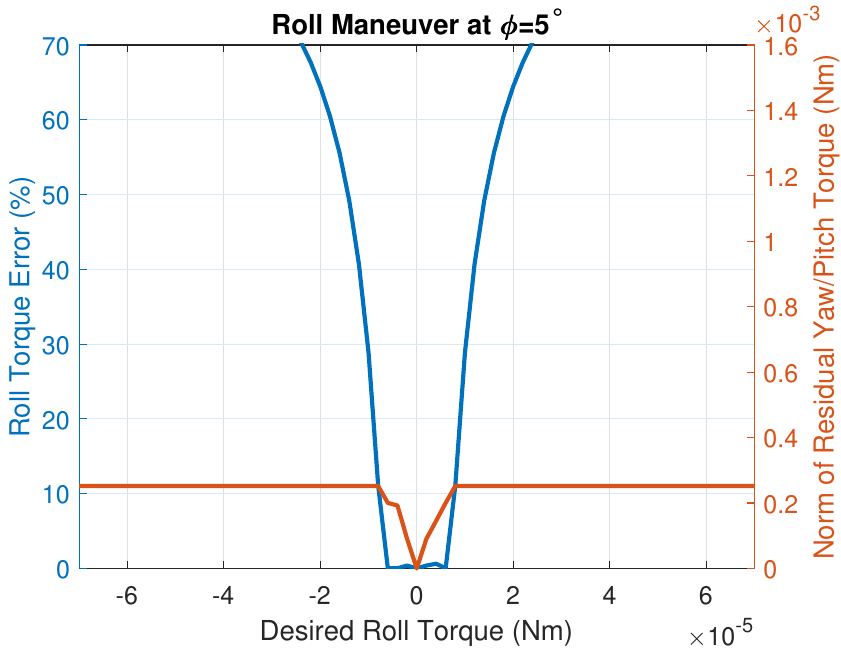}
\label{subfig:roll_ca5}
}
\subfigure[]
	{
    \includegraphics[width=0.32\columnwidth]{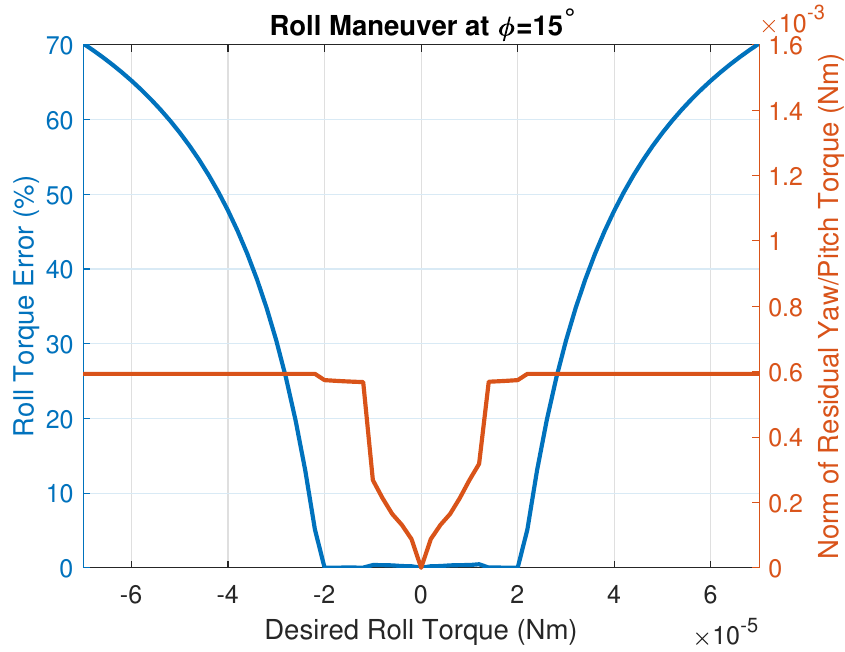}
\label{subfig:roll_ca15}
}
\subfigure[]
	{
    \includegraphics[width=0.32\columnwidth]{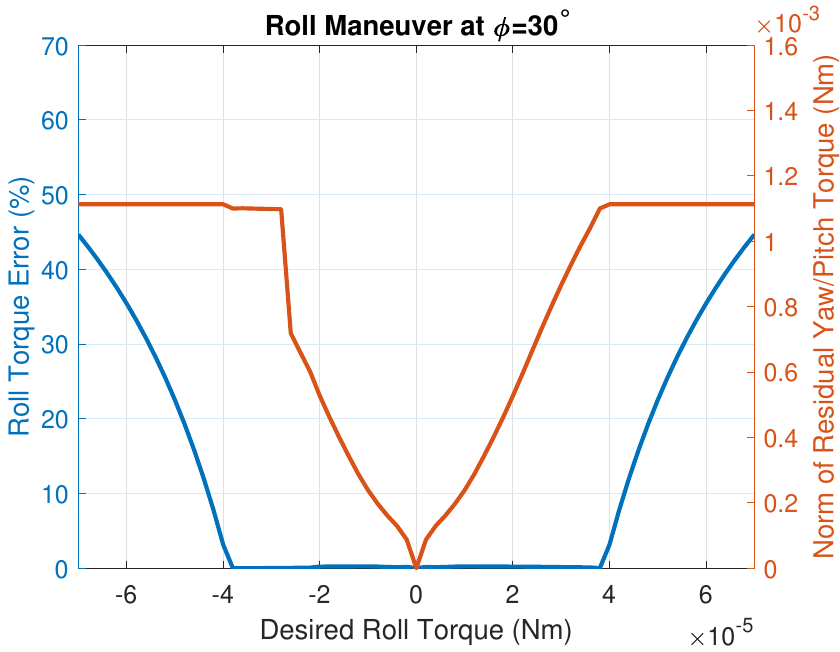}
\label{subfig:roll_ca30}
}
\subfigure[]
	{
    \includegraphics[width=0.32\columnwidth]{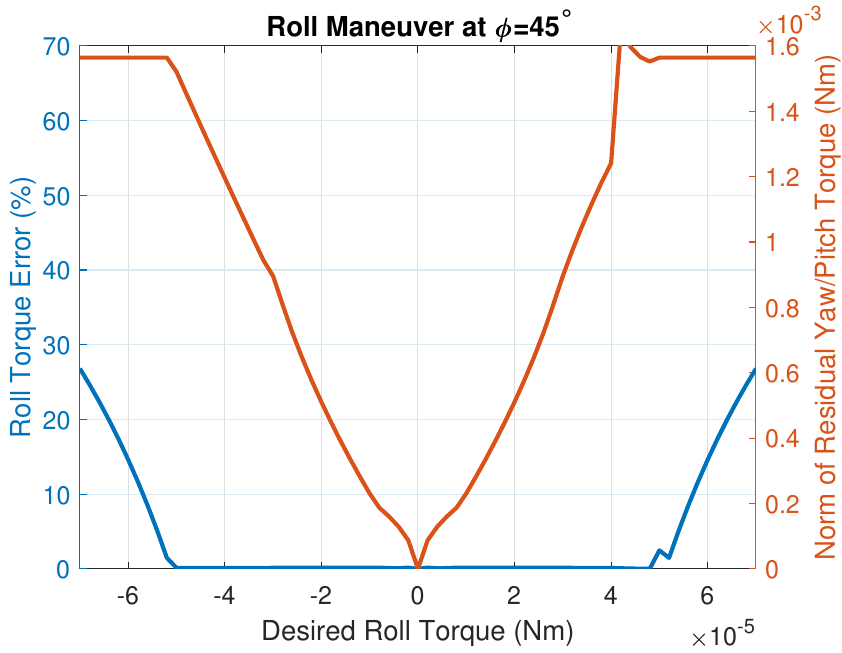}
    \label{subfig:roll_ca45}
}
\subfigure[]
	{
    \includegraphics[width=0.32\columnwidth]{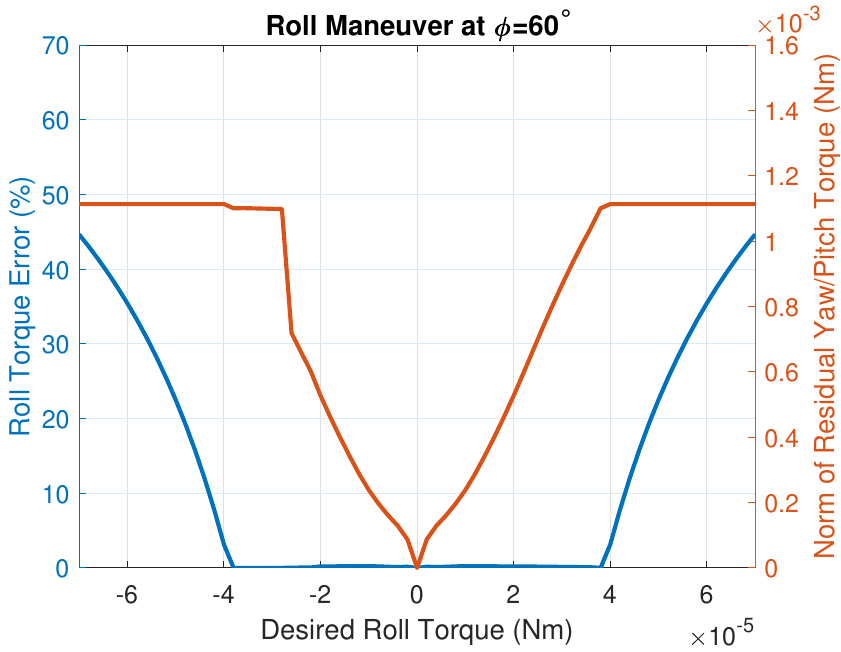}
    \label{subfig:roll_ca60}
}
\subfigure[]
	{
    \includegraphics[width=0.32\columnwidth]{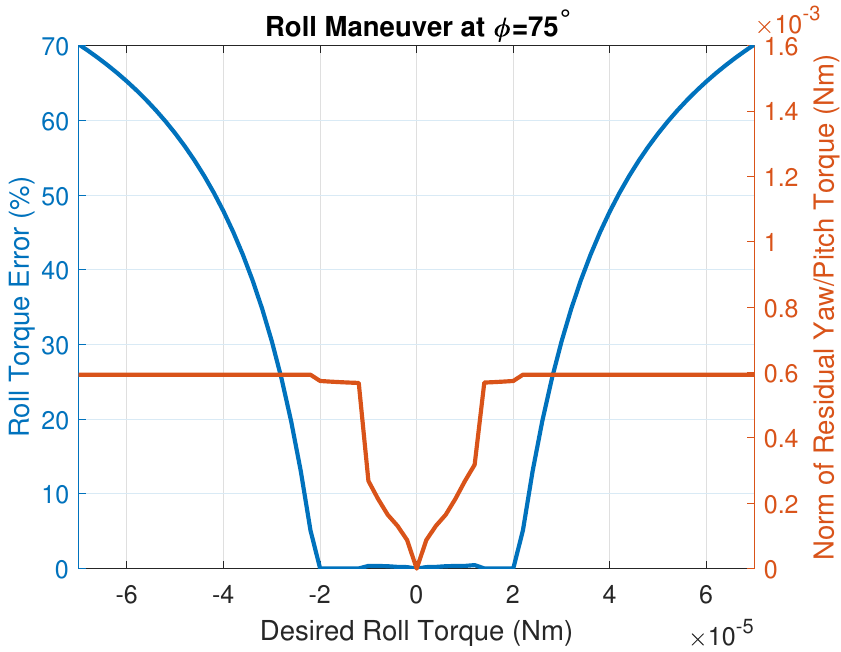}
    \label{subfig:roll_ca75}
}
\caption{Errors of the torque generated compared to the desired torques for roll maneuvers with a constraint of $75$~cm. The left axis shows roll torque error in percent, and the right axis shows norm of residual yaw/pitch torque in N$\cdot$m at clock angles of (a) 5, (b) 15, (c) 30, (d) 45, (e) 60, and (f) 75 degrees.}
\label{fig:range_roll}
\end{figure}

The range of feasible roll torques at each clock angle is shown in Fig.~\ref{fig:range_roll}. Percentage roll torque error is shown on the left axis in blue, and the norm of the residual yaw/pitch torque in N$\cdot$m is shown on the right axis in orange. At a clock angle of $45$~degrees, Fig.~\ref{subfig:roll_ca45} shows that roll torques within the range of $-5\times10^{-5}$ and $5\times10^{-5}$~N$\cdot$m can be generated without any visible error. This clock angle is also where the largest disturbance torques are expected~\citep{gauvain2023solar}. The norm of the residual yaw/pitch torque increases for larger roll torques.
As the clock angle moves away from $45$~degrees, the range of feasible roll torques gets narrower and the norm of residual yaw/pitch torques gets smaller. 

\subsection{Discussion}

The results of Sections~\ref{subsec:CompareIntuitive} and~\ref{subsec:RangeFeasibleTorques} demonstrate CABLESSail's ability to reliably generate large momentum management torques with the proposed control allocation method.
The tests performed in Section~\ref{subsec:CompareIntuitive} highlight the robustness of the control allocation method to uncertainty in the sail membrane shape, as the control allocation algorithm assumes an undeformed membrane. Unwanted residual torques are clearly minimized in these tests when compared to the intuitive maneuvers, even with a significant constraint on the boom tip deformation of $50$~cm. The simulations of Section~\ref{subsec:RangeFeasibleTorques} illustrate the significant clock-angle-dependence of the feasible torque generated with CABLESSail. For example, at clock angles smaller than $5$~degrees in magnitude, it is not feasible to generate a yaw torque with a flat sail membrane due to the solar sail's geometry relative to the Sun. However, this coincides with clock angles where the expected disturbance torque is minimal~\citep{gauvain2023solar}, which potentially makes this effect manageable. Although the residual yaw/pitch torque can be non-negligible when generating large roll torques, its magnitude stays within the range of feasible yaw/pitch torques that can be generated at the same clock angle, as shown in Figs.~\ref{fig:range_ca5to30} and~\ref{fig:range_ca45to75}. This indicates the possibility of performing a yaw/pitch maneuver following a roll maneuver when the magnitude of residual yaw/pitch torque is too large and must be counteracted.

It is also worth noting that the results presented in this section highlight CABLESSail as a promising momentum management actuator when compared to state-of-the-art technology, such as an AMT and RCDs. Solar Cruiser's RCDs were sized to produce roll torques of $3\times10^{-5}$~N$\cdot$m in magnitude~\citep{heaton2023RCD}. Fig.~\ref{fig:range_roll} demonstrates CABLESSail's ability to accurately generate roll torques up to $4\times10^{-5}$~N$\cdot$m for clock angles between $30$~degrees and $60$~degrees. Although not shown in the plots, similar results are generated for clock angles between $120$~degrees to $150$~degrees, $210$~degrees to $240$~degrees, and $300$~degrees to $330$~degrees. The range of accurate roll torque generation decreases for clock angles outside these intervals, although roll disturbance torques are also lower in these intervals~\citep{gauvain2023solar}. Solar Cruiser's AMT is capable of generating relatively large yaw/pitch torques on the order of $1.9\times10^{-3}$~N$\cdot$m~\citep{inness2023momentum,shenISSS2025}. It is shown in Figs.~\ref{fig:range_ca5to30} and~\ref{fig:range_ca45to75} that CABLESSail can generate yaw/pitch torques of a similar magnitude with minimal error at most clock angles. It is worth noting the difference in the residual roll torque generated when using CABLESSail in comparison to an AMT. Figs.~\ref{fig:range_ca5to30} and~\ref{fig:range_ca45to75} illustrate that very little residual roll torque is generated for nearly all combinations of desired yaw/pitch torques and across all clock angles, with most instances resulting in less than $1\times10^{-6}$~N$\cdot$m of residual roll torque. In contrast, tangential SRP forces due to imperfect reflectivity coupled with a center of mass shift due to AMT motion can result in up to $4.5\times10^{-5}$~N$\cdot$m of residual roll torque. This is found by multiplying the expected tangential SRP force of $3\times10^{-4}$~N found from NEA Scout's optical properties~\citep{heaton2015update} and Solar Cruiser's dimensions with the $0.15$~m maximum center of mass shift enabled by Solar Cruiser's AMT~\citep{JohnsonLes2020SCTM}. This worst-case residual roll torque generated by the AMT is significant, as it surpasses the roll torque capability of its RCDs and of CABLESSail. CABLESSail's ability to generate large yaw/pitch torques without inducing large residual roll torques could represent a significant advancement over the use of an AMT.

\section{Future Outlook of CABLESSail}
\label{sec:Future}

The numerical simulations presented in this paper demonstrate that controlled deformations of a solar sail's booms with the CABLESSail concept result in the ability to generate significant momentum management torques in the presence of sail membrane uncertainty. Small-scale prototype testing has confirmed that CABLESSail's cable actuation is capable of reliably deforming deployable lenticular booms. These results, along with prior CABLESSail studies~\citep{caverly2023solar,bunker2024modular,lee2024robust,bodin2025design,bunker2025static,lee2025cablessail,lee2025passivity}, provide the analytical and experimental proof-of-concept results to justify CABLESSail achieving TRL~3.

CABLESSail's path towards TRLs 4 through 6 will require substantially more experimental testing with full-size prototypes to meet the TRL requirements of component and subsystem prototype demonstration in a relevant environment. These tests will ideally be performed with an engineering design unit for a potential technology demonstration flight test. To simulate a relevant environment, gravity offloading may be incorporated into the ground tests. Software-in-the-loop testing of CABLESSail's control allocation, feedback control, and state estimation techniques will also play a role in navigating these middle TRLs.

The recent ACS3 solar sail mission demonstrated that large, undesirable boom deformations can appear upon deployment and during the course of the flight mission~\citep{WilkieKeynote}. Although CABLESSail was originally conceived as a momentum management actuator, it may have an even more significant benefit as a device that can provide occasional coarse corrections to undesired boom deformations. This actuation could be performed in an open-loop fashion, without the need for explicit boom tip deformation estimation. For example, it is possible that in a scenario where upon deployment the booms experience larger-than-expected deformations, CABLESSail can be used in an attempt to adjust or trim out these deformations. After some time, camera images can be used to assess whether further corrections are needed. 

A broader application for the CABLESSail technology presented in this work may include actuation for a drag device that provides fine-tune control to a spacecraft's ballistic coefficient. This could be helpful in the design of low-cost propulsion for small satellites in low-Earth orbit, where traditional propulsion options are limited. Such a device could be combined with recent advances in hardware~\citep{Murbach2010,omar2019hardware} and control technology~\citep{omar2017spacecraft,hayes2023model,hayes2024atmospheric} to enable the precise targeted reentry of drag-modulated spacecraft.

\section{Conclusions}
\label{sec:Conclusions}

This paper presented critical simulation and experimental results towards increasing CABLESSail to TRL~3. Specifically, the deployable small-scale prototype fabricated and tested as part of this work provides evidence that CABLESSail is capable of achieving meaningful deformations of deployable lenticular booms without inducing buckling. Moreover, the control allocation methodology proposed in this work solves the challenge of determining the appropriate boom deformations needed to generate a particular momentum management torque. The robustness of the control allocation method to sail membrane shape uncertainty and the assessment of its range of feasible torques verified that CABLESSail is capable of providing practical momentum management capabilities to solar sails.

\section*{Acknowledgments}
This work was supported by an Early Career Faculty grant from NASA’s Space Technology Research Grants Program under award No. 80NSSC23K0075. The authors also acknowledge assistance from Austin Bodin, Niko Sexton, Chris Thacker, Michael Dallah, and Ryan Levendusky in the fabrication and testing of the CABLESSail prototype.

\bibliographystyle{elsarticle-harv}
\biboptions{authoryear}
\bibliography{Bib}

\begin{thebibliography}{48}
\expandafter\ifx\csname natexlab\endcsname\relax\def\natexlab#1{#1}\fi
\providecommand{\url}[1]{\texttt{#1}}
\providecommand{\href}[2]{#2}
\providecommand{\path}[1]{#1}
\providecommand{\DOIprefix}{doi:}
\providecommand{\ArXivprefix}{arXiv:}
\providecommand{\URLprefix}{URL: }
\providecommand{\Pubmedprefix}{pmid:}
\providecommand{\doi}[1]{\href{http://dx.doi.org/#1}{\path{#1}}}
\providecommand{\Pubmed}[1]{\href{pmid:#1}{\path{#1}}}
\providecommand{\bibinfo}[2]{#2}
\ifx\xfnm\relax \def\xfnm[#1]{\unskip,\space#1}\fi
\bibitem[{Amodio et~al.(2025)Amodio, Visser and
  Heiligers}]{amodio2025dynamical}
\bibinfo{author}{Amodio, A.}, \bibinfo{author}{Visser, P.},
  \bibinfo{author}{Heiligers, J.}, \bibinfo{year}{2025}.
\newblock \bibinfo{title}{Dynamical modelling of {NASA's ACS3} solar sail
  mission}.
\newblock \bibinfo{journal}{Aerospace Science and Technology}
  \bibinfo{volume}{161}, \bibinfo{pages}{110146}.
\newblock \DOIprefix\doi{10.1016/j.ast.2025.110146}.
\bibitem[{Ancona and Kezerashvili(2025)}]{ancona2025recent}
\bibinfo{author}{Ancona, E.}, \bibinfo{author}{Kezerashvili, R.},
  \bibinfo{year}{2025}.
\newblock \bibinfo{title}{Recent advances in space sailing missions and
  technology: Review of the 6th {International Symposium on Space Sailing (ISSS
  2023)}}.
\newblock \bibinfo{journal}{Aeronautics and Aerospace Open Access Journal}
  \bibinfo{volume}{9}, \bibinfo{pages}{62--73}.
\newblock \DOIprefix\doi{10.15406/aaoaj.2025.09.00221}.
\bibitem[{Banik and Murphey(2010)}]{JeremyTRAC}
\bibinfo{author}{Banik, J.}, \bibinfo{author}{Murphey, T.},
  \bibinfo{year}{2010}.
\newblock \bibinfo{title}{Performance validation of the triangular rollable and
  collapsible mast}, in: \bibinfo{booktitle}{24th Annual AIAA/USU Conference on
  Small Satellites}, \bibinfo{address}{Logan, UT}.
\bibitem[{Berthet et~al.(2024)Berthet, Schalkwyk, Sengupta, Fujino, Hein,
  Tenorio, Cardoso~dos Santos, Worden, Mauskopf, Miyazaki, Funaki, Tsuji, Fil
  and Suzuki}]{Berthet2024-mi}
\bibinfo{author}{Berthet, M.}, \bibinfo{author}{Schalkwyk, J.and~Çelik, O.},
  \bibinfo{author}{Sengupta, D.}, \bibinfo{author}{Fujino, K.},
  \bibinfo{author}{Hein, A.}, \bibinfo{author}{Tenorio, L.},
  \bibinfo{author}{Cardoso~dos Santos, J.}, \bibinfo{author}{Worden, S.},
  \bibinfo{author}{Mauskopf, P.}, \bibinfo{author}{Miyazaki, Y.},
  \bibinfo{author}{Funaki, I.}, \bibinfo{author}{Tsuji, S.},
  \bibinfo{author}{Fil, P.}, \bibinfo{author}{Suzuki, K.},
  \bibinfo{year}{2024}.
\newblock \bibinfo{title}{Space sails for achieving major space exploration
  goals: Historical review and future outlook}.
\newblock \bibinfo{journal}{Progress in Aerospace Sciences}
  \bibinfo{volume}{150}, \bibinfo{pages}{101047}.
\newblock \DOIprefix\doi{10.1016/j.paerosci.2024.101047}.
\bibitem[{Bodin et~al.(2025)Bodin, States, Lee, Raab and
  Caverly}]{bodin2025design}
\bibinfo{author}{Bodin, A.}, \bibinfo{author}{States, M.},
  \bibinfo{author}{Lee, S.}, \bibinfo{author}{Raab, N.},
  \bibinfo{author}{Caverly, R.}, \bibinfo{year}{2025}.
\newblock \bibinfo{title}{Design, estimation, and control of a cable-driven
  solar sail boom testbed prototype}, in: \bibinfo{booktitle}{AIAA SciTech
  Forum}, \bibinfo{address}{Orlando, FL}.
\newblock \DOIprefix\doi{10.2514/6.2025-2835}. \bibinfo{note}{{AIAA}
  2025-2835}.
\bibitem[{Boni et~al.(2023)Boni, Bassetto, Niccolai, Mengali, Quarta, Circi,
  Pellegrini and Cavallini}]{Boni2023-mb}
\bibinfo{author}{Boni, L.}, \bibinfo{author}{Bassetto, M.},
  \bibinfo{author}{Niccolai, L.}, \bibinfo{author}{Mengali, G.},
  \bibinfo{author}{Quarta, A.}, \bibinfo{author}{Circi, C.},
  \bibinfo{author}{Pellegrini, R.}, \bibinfo{author}{Cavallini, E.},
  \bibinfo{year}{2023}.
\newblock \bibinfo{title}{Structural response of {Helianthus} solar sail during
  attitude maneuvers}.
\newblock \bibinfo{journal}{Aerospace Science and Technology}
  \bibinfo{volume}{133}, \bibinfo{pages}{108152}.
\newblock \DOIprefix\doi{10.1016/j.ast.2023.108152}.
\bibitem[{Brownell et~al.(2023)Brownell, Sinclair and
  Singla}]{brownell2023time}
\bibinfo{author}{Brownell, M.}, \bibinfo{author}{Sinclair, A.},
  \bibinfo{author}{Singla, P.}, \bibinfo{year}{2023}.
\newblock \bibinfo{title}{A time-varying subspace method for shape estimation
  of a flexible spacecraft membrane}, in: \bibinfo{booktitle}{AIAA SciTech
  Forum}, \bibinfo{address}{National Harbor, MD}.
\newblock \DOIprefix\doi{10.2514/6.2023-2068}. \bibinfo{note}{{AIAA}
  2023-2068}.
\bibitem[{Bunker and Caverly(2024)}]{bunker2024modular}
\bibinfo{author}{Bunker, K.}, \bibinfo{author}{Caverly, R.},
  \bibinfo{year}{2024}.
\newblock \bibinfo{title}{Modular dynamic modeling and simulation of a
  cable-actuated flexible solar sail}, in: \bibinfo{booktitle}{AIAA SciTech
  Forum}, \bibinfo{address}{Orlando, FL}.
\newblock \DOIprefix\doi{10.2514/6.2024-2436}. \bibinfo{note}{{AIAA}
  2024-2436}.
\bibitem[{Bunker and Caverly(2025)}]{bunker2025static}
\bibinfo{author}{Bunker, K.}, \bibinfo{author}{Caverly, R.},
  \bibinfo{year}{2025}.
\newblock \bibinfo{title}{Static and dynamic torque generation analysis of a
  cable-actuated solar sail}.
\newblock \bibinfo{journal}{arXiv preprint arXiv:2501.17336} .
\bibitem[{Caverly et~al.(2023)Caverly, Bunker, Raab, Nguyen, Saner, Chen,
  Douvier, Lyman, Sorby, Sorge, Teshale and Toriseva}]{caverly2023solar}
\bibinfo{author}{Caverly, R.}, \bibinfo{author}{Bunker, K.},
  \bibinfo{author}{Raab, N.}, \bibinfo{author}{Nguyen, V.},
  \bibinfo{author}{Saner, G.}, \bibinfo{author}{Chen, Z.},
  \bibinfo{author}{Douvier, T.}, \bibinfo{author}{Lyman, R.},
  \bibinfo{author}{Sorby, O.}, \bibinfo{author}{Sorge, B.},
  \bibinfo{author}{Teshale, E.}, \bibinfo{author}{Toriseva, B.},
  \bibinfo{year}{2023}.
\newblock \bibinfo{title}{Solar sail attitude control using shape modulation:
  The {Cable-Actuated Bio-inspired Lightweight Elastic Solar Sail (CABLESSail)}
  concept}, in: \bibinfo{booktitle}{6th International Symposium on Space
  Sailing}, \bibinfo{address}{New York, NY}.
\bibitem[{Chen et~al.(2023)Chen, Liu, Cai and You}]{Chen2023-nw}
\bibinfo{author}{Chen, T.Z.}, \bibinfo{author}{Liu, X.}, \bibinfo{author}{Cai,
  G.P.}, \bibinfo{author}{You, C.L.}, \bibinfo{year}{2023}.
\newblock \bibinfo{title}{Attitude and vibration control of a solar sail}.
\newblock \bibinfo{journal}{Advances in Space Research} \bibinfo{volume}{71},
  \bibinfo{pages}{4557--4567}.
\newblock \DOIprefix\doi{10.1016/j.asr.2023.01.039}.
\bibitem[{Fu et~al.(2016)Fu, Sperber and Eke}]{fu2016solar}
\bibinfo{author}{Fu, B.}, \bibinfo{author}{Sperber, E.}, \bibinfo{author}{Eke,
  F.}, \bibinfo{year}{2016}.
\newblock \bibinfo{title}{Solar sail technology---a state of the art review}.
\newblock \bibinfo{journal}{Progress in Aerospace Sciences}
  \bibinfo{volume}{86}, \bibinfo{pages}{1--19}.
\newblock \DOIprefix\doi{10.1016/j.paerosci.2016.07.001}.
\bibitem[{Gauvain and Tyler(2023)}]{gauvain2023solar}
\bibinfo{author}{Gauvain, B.}, \bibinfo{author}{Tyler, D.},
  \bibinfo{year}{2023}.
\newblock \bibinfo{title}{A solar sail shape modeling approach for attitude
  control design and analysis}, in: \bibinfo{booktitle}{6th International
  Symposium on Space Sailing}, \bibinfo{address}{New York, NY}.
\bibitem[{Gong and Macdonald(2019)}]{gong2019review}
\bibinfo{author}{Gong, S.}, \bibinfo{author}{Macdonald, M.},
  \bibinfo{year}{2019}.
\newblock \bibinfo{title}{Review on solar sail technology}.
\newblock \bibinfo{journal}{Astrodynamics} \bibinfo{volume}{3},
  \bibinfo{pages}{93--125}.
\newblock \DOIprefix\doi{10.1007/s42064-019-0038-x}.
\bibitem[{Hayes and Caverly(2023)}]{hayes2023model}
\bibinfo{author}{Hayes, A.}, \bibinfo{author}{Caverly, R.},
  \bibinfo{year}{2023}.
\newblock \bibinfo{title}{Model predictive tracking of spacecraft deorbit
  trajectories using drag modulation}.
\newblock \bibinfo{journal}{Acta Astronautica} \bibinfo{volume}{202},
  \bibinfo{pages}{670--685}.
\newblock \DOIprefix\doi{10.1016/j.actaastro.2022.10.057}.
\bibitem[{Hayes and Caverly(2025)}]{hayes2024atmospheric}
\bibinfo{author}{Hayes, A.}, \bibinfo{author}{Caverly, R.},
  \bibinfo{year}{2025}.
\newblock \bibinfo{title}{Atmospheric density-compensating model predictive
  control for targeted reentry of drag-modulated spacecraft}.
\newblock \bibinfo{journal}{Journal of Guidance, Control, and Dynamics}
  \bibinfo{volume}{48}, \bibinfo{pages}{2541--2556}.
\newblock \DOIprefix\doi{10.2514/1.G008665}.
\bibitem[{Heaton(2023)}]{heaton2023reflectivity}
\bibinfo{author}{Heaton, A.}, \bibinfo{year}{2023}.
\newblock \bibinfo{title}{Reflectivity control device {(RCD)} roll momentum
  management for {Solar Cruiser} and beyond}, in: \bibinfo{booktitle}{6th
  International Symposium on Space Sailing}, \bibinfo{address}{New York, NY}.
\bibitem[{Heaton and Artusio-Glimpse(2015)}]{heaton2015update}
\bibinfo{author}{Heaton, A.}, \bibinfo{author}{Artusio-Glimpse, A.},
  \bibinfo{year}{2015}.
\newblock \bibinfo{title}{An update to the {NASA} reference solar sail thrust
  model}, in: \bibinfo{booktitle}{AIAA SPACE Conference and Exposition},
  \bibinfo{address}{Pasadena, CA}.
\newblock \DOIprefix\doi{10.2514/6.2015-4506}. \bibinfo{note}{{AIAA}
  2015-4506}.
\bibitem[{Heaton et~al.(2023)Heaton, Ramazani and Tyler}]{heaton2023RCD}
\bibinfo{author}{Heaton, A.}, \bibinfo{author}{Ramazani, S.},
  \bibinfo{author}{Tyler, D.}, \bibinfo{year}{2023}.
\newblock \bibinfo{title}{Reflectivity control device {(RCD)} roll momentum
  management for {Solar Cruiser} and beyond}, in: \bibinfo{booktitle}{6th
  International Symposium on Solar Sailing}, \bibinfo{address}{New York, NY}.
\bibitem[{Hibbert and Jordaan(2021)}]{Hibbert2021-xg}
\bibinfo{author}{Hibbert, L.T.}, \bibinfo{author}{Jordaan, H.W.},
  \bibinfo{year}{2021}.
\newblock \bibinfo{title}{Considerations in the design and deployment of
  flexible booms for a solar sail}.
\newblock \bibinfo{journal}{Advances in Space Research} \bibinfo{volume}{67},
  \bibinfo{pages}{2716--2726}.
\newblock \DOIprefix\doi{10.1016/j.asr.2020.01.019}.
\bibitem[{Huang et~al.(2021)Huang, Zeng, Circi, Vulpetti and
  Qiao}]{Huang2021-ba}
\bibinfo{author}{Huang, X.}, \bibinfo{author}{Zeng, X.},
  \bibinfo{author}{Circi, C.}, \bibinfo{author}{Vulpetti, G.},
  \bibinfo{author}{Qiao, D.}, \bibinfo{year}{2021}.
\newblock \bibinfo{title}{Analysis of the solar sail deformation based on the
  point cloud method}.
\newblock \bibinfo{journal}{Advances in Space Research} \bibinfo{volume}{67},
  \bibinfo{pages}{2613--2627}.
\newblock \DOIprefix\doi{10.1016/j.asr.2020.05.008}.
\bibitem[{Inness et~al.(2023)Inness, Tyler, Diedrich, Ramazani and
  Orphee}]{inness2023momentum}
\bibinfo{author}{Inness, J.}, \bibinfo{author}{Tyler, D.},
  \bibinfo{author}{Diedrich, B.}, \bibinfo{author}{Ramazani, S.},
  \bibinfo{author}{Orphee, J.}, \bibinfo{year}{2023}.
\newblock \bibinfo{title}{Momentum management strategies for {Solar Cruiser}
  and beyond}, in: \bibinfo{booktitle}{6th International Symposium on Space
  Sailing}, \bibinfo{address}{New York, NY}.
\bibitem[{Johnson et~al.(2025)Johnson, Akhavan-Tafti, Sood, Szabo and
  Thomas}]{Johnson2025-lt}
\bibinfo{author}{Johnson, L.}, \bibinfo{author}{Akhavan-Tafti, M.},
  \bibinfo{author}{Sood, R.}, \bibinfo{author}{Szabo, A.},
  \bibinfo{author}{Thomas, H.D.}, \bibinfo{year}{2025}.
\newblock \bibinfo{title}{Space weather investigation frontier ({SWIFT})
  mission concept: Continuous, distributed observations of heliospheric
  structures from the vantage points of sun-earth {L1} and sub-{L1}}.
\newblock \bibinfo{journal}{Acta Astronautica} \bibinfo{volume}{236},
  \bibinfo{pages}{684--691}.
\newblock \DOIprefix\doi{10.1016/j.actaastro.2025.07.038}.
\bibitem[{Johnson and Curran(2020)}]{JohnsonLes2020SCTM}
\bibinfo{author}{Johnson, L.}, \bibinfo{author}{Curran, F.},
  \bibinfo{year}{2020}.
\newblock \bibinfo{title}{{Solar Cruiser} Technology Maturation Plans}.
\newblock \bibinfo{type}{Technical Report} \bibinfo{number}{20205003681}. NASA
  Marshall Space Flight Center.
\bibitem[{Johnson et~al.(2011)Johnson, Whorton, Heaton, Pinson, Laue and
  Adams}]{johnson2011nanosail}
\bibinfo{author}{Johnson, L.}, \bibinfo{author}{Whorton, M.},
  \bibinfo{author}{Heaton, A.}, \bibinfo{author}{Pinson, R.},
  \bibinfo{author}{Laue, G.}, \bibinfo{author}{Adams, C.},
  \bibinfo{year}{2011}.
\newblock \bibinfo{title}{{NanoSail-D}: A solar sail demonstration mission}.
\newblock \bibinfo{journal}{Acta Astronautica} \bibinfo{volume}{68},
  \bibinfo{pages}{571--575}.
\newblock \DOIprefix\doi{10.1016/j.actaastro.2010.02.008}.
\bibitem[{Lee et~al.(2025)Lee, Bunker and Caverly}]{lee2025cablessail}
\bibinfo{author}{Lee, S.}, \bibinfo{author}{Bunker, K.R.},
  \bibinfo{author}{Caverly, R.J.}, \bibinfo{year}{2025}.
\newblock \bibinfo{title}{{CABLESSail}: Solar sail momentum management using
  cable-actuated shape control}.
\newblock \bibinfo{journal}{7th International Symposium on Space Sailing} .
\bibitem[{Lee and Caverly(2024)}]{lee2024robust}
\bibinfo{author}{Lee, S.}, \bibinfo{author}{Caverly, R.}, \bibinfo{year}{2024}.
\newblock \bibinfo{title}{Robust cable-actuated shape control of a flexible
  solar sail boom for the {CABLESSail} concept}, in: \bibinfo{booktitle}{AAS
  Guidance, Navigation and Control Conference}, \bibinfo{address}{Breckenridge,
  CO}.
\newblock \bibinfo{note}{AAS 24-071}.
\bibitem[{Lee and Caverly(2026)}]{lee2025passivity}
\bibinfo{author}{Lee, S.}, \bibinfo{author}{Caverly, R.}, \bibinfo{year}{2026}.
\newblock \bibinfo{title}{Passivity-based robust shape control of a
  cable-driven solar sail boom for the {CABLESSail} concept}.
\newblock \bibinfo{journal}{Acta Astronautica} \bibinfo{volume}{238, Part B},
  \bibinfo{pages}{602--611}.
\newblock \DOIprefix\doi{10.1016/j.actaastro.2025.10.034}.
\bibitem[{Lockett et~al.(2020)Lockett, Castillo-Rogez, Johnson, Matus,
  Lightholder, Marinan and Few}]{lockett2020near}
\bibinfo{author}{Lockett, T.}, \bibinfo{author}{Castillo-Rogez, J.},
  \bibinfo{author}{Johnson, L.}, \bibinfo{author}{Matus, J.},
  \bibinfo{author}{Lightholder, J.}, \bibinfo{author}{Marinan, A.},
  \bibinfo{author}{Few, A.}, \bibinfo{year}{2020}.
\newblock \bibinfo{title}{{Near-Earth Asteroid Scout} flight mission}.
\newblock \bibinfo{journal}{IEEE Aerospace and Electronic Systems Magazine}
  \bibinfo{volume}{35}, \bibinfo{pages}{20--29}.
\newblock \DOIprefix\doi{10.1109/MAES.2019.2958729}.
\bibitem[{Murbach et~al.(2010)Murbach, Boronowsky, Benton, White and
  Fritzler}]{Murbach2010}
\bibinfo{author}{Murbach, M.}, \bibinfo{author}{Boronowsky, K.},
  \bibinfo{author}{Benton, J.}, \bibinfo{author}{White, B.},
  \bibinfo{author}{Fritzler, E.}, \bibinfo{year}{2010}.
\newblock \bibinfo{title}{Options for returning payloads from the {ISS} after
  the termination of {STS} flights}, in: \bibinfo{booktitle}{40th International
  Conference on Environmental Systems}, \bibinfo{address}{Barcelona, Spain}. p.
  \bibinfo{pages}{6223}.
\newblock \DOIprefix\doi{10.2514/6.2010-6223}.
\bibitem[{Nguyen et~al.(2023)Nguyen, Medina, McConnel and
  Lake}]{nguyen2023solar}
\bibinfo{author}{Nguyen, L.}, \bibinfo{author}{Medina, K.},
  \bibinfo{author}{McConnel, Z.}, \bibinfo{author}{Lake, M.S.},
  \bibinfo{year}{2023}.
\newblock \bibinfo{title}{{Solar Cruiser TRAC} boom development}, in:
  \bibinfo{booktitle}{AIAA SciTech Forum}, p. \bibinfo{pages}{1507}.
\bibitem[{Omar and Bevilacqua(2019)}]{omar2019hardware}
\bibinfo{author}{Omar, S.}, \bibinfo{author}{Bevilacqua, R.},
  \bibinfo{year}{2019}.
\newblock \bibinfo{title}{Hardware and {GNC} solutions for controlled
  spacecraft re-entry using aerodynamic drag}.
\newblock \bibinfo{journal}{Acta Astronautica} \bibinfo{volume}{159},
  \bibinfo{pages}{49--64}.
\newblock \DOIprefix\doi{10.1016/j.actaastro.2019.03.051}.
\bibitem[{Omar et~al.(2017)Omar, Bevilacqua, Guglielmo, Fineberg, Treptow,
  Clark and Johnson}]{omar2017spacecraft}
\bibinfo{author}{Omar, S.}, \bibinfo{author}{Bevilacqua, R.},
  \bibinfo{author}{Guglielmo, D.}, \bibinfo{author}{Fineberg, L.},
  \bibinfo{author}{Treptow, J.}, \bibinfo{author}{Clark, S.},
  \bibinfo{author}{Johnson, Y.}, \bibinfo{year}{2017}.
\newblock \bibinfo{title}{Spacecraft deorbit point targeting using aerodynamic
  drag}.
\newblock \bibinfo{journal}{Journal of Guidance, Control, and Dynamics}
  \bibinfo{volume}{40}, \bibinfo{pages}{2646--2652}.
\newblock \DOIprefix\doi{10.2514/1.G002612}.
\bibitem[{Pezent et~al.(2021a)Pezent, Sood and Heaton}]{pezent2021contingency}
\bibinfo{author}{Pezent, J.}, \bibinfo{author}{Sood, R.},
  \bibinfo{author}{Heaton, A.}, \bibinfo{year}{2021}a.
\newblock \bibinfo{title}{Contingency target assessment, trajectory design, and
  analysis for {NASA’s NEA Scout} solar sail mission}.
\newblock \bibinfo{journal}{Advances in Space Research} \bibinfo{volume}{67},
  \bibinfo{pages}{2890--2898}.
\newblock \DOIprefix\doi{10.1016/j.asr.2020.02.004}.
\bibitem[{Pezent et~al.(2021b)Pezent, Sood, Heaton, Miller and
  Johnson}]{pezent2021preliminary}
\bibinfo{author}{Pezent, J.}, \bibinfo{author}{Sood, R.},
  \bibinfo{author}{Heaton, A.}, \bibinfo{author}{Miller, K.},
  \bibinfo{author}{Johnson, L.}, \bibinfo{year}{2021}b.
\newblock \bibinfo{title}{Preliminary trajectory design for {NASA's Solar
  Cruiser}: A technology demonstration mission}.
\newblock \bibinfo{journal}{Acta Astronautica} \bibinfo{volume}{183},
  \bibinfo{pages}{134--140}.
\newblock \DOIprefix\doi{10.1016/j.actaastro.2021.03.006}.
\bibitem[{Pimienta-Penalver et~al.(2019)Pimienta-Penalver, Tsai, Juang and
  Crassidis}]{pimienta2019heliogyro}
\bibinfo{author}{Pimienta-Penalver, A.}, \bibinfo{author}{Tsai, L.W.},
  \bibinfo{author}{Juang, J.N.}, \bibinfo{author}{Crassidis, J.},
  \bibinfo{year}{2019}.
\newblock \bibinfo{title}{Heliogyro solar sail structural dynamics and
  stability}.
\newblock \bibinfo{journal}{Journal of Guidance, Control, and Dynamics}
  \bibinfo{volume}{42}, \bibinfo{pages}{1645--1657}.
\newblock \DOIprefix\doi{10.2514/1.G003758}.
\bibitem[{Shen and Caverly(2025a)}]{shenISSS2025}
\bibinfo{author}{Shen, P.Y.}, \bibinfo{author}{Caverly, R.},
  \bibinfo{year}{2025}a.
\newblock \bibinfo{title}{{Solar Cruiser} momentum management using model
  predictive control}, in: \bibinfo{booktitle}{7th International Symposium on
  Space Sailing}, \bibinfo{address}{Delft, The Netherlands}.
\bibitem[{Shen and Caverly(2025b)}]{shen2025solar}
\bibinfo{author}{Shen, P.Y.}, \bibinfo{author}{Caverly, R.},
  \bibinfo{year}{2025}b.
\newblock \bibinfo{title}{Solar sail momentum management with mass translation
  and reflectivity devices using predictive control}.
\newblock \bibinfo{journal}{arXiv preprint arXiv:2503.12643} .
\bibitem[{Spencer et~al.(2021)Spencer, Betts, Bellardo, Diaz, Plante and
  Mansell}]{spencer2021lightsail}
\bibinfo{author}{Spencer, D.}, \bibinfo{author}{Betts, B.},
  \bibinfo{author}{Bellardo, J.}, \bibinfo{author}{Diaz, A.},
  \bibinfo{author}{Plante, B.}, \bibinfo{author}{Mansell, J.},
  \bibinfo{year}{2021}.
\newblock \bibinfo{title}{The {LightSail 2} solar sailing technology
  demonstration}.
\newblock \bibinfo{journal}{Advances in Space Research} \bibinfo{volume}{67},
  \bibinfo{pages}{2878--2889}.
\newblock \DOIprefix\doi{10.1016/j.asr.2020.06.029}.
\bibitem[{Spencer et~al.(2019)Spencer, Johnson and Long}]{spencer2019solar}
\bibinfo{author}{Spencer, D.}, \bibinfo{author}{Johnson, L.},
  \bibinfo{author}{Long, A.C.}, \bibinfo{year}{2019}.
\newblock \bibinfo{title}{Solar sailing technology challenges}.
\newblock \bibinfo{journal}{Aerospace Science and Technology}
  \bibinfo{volume}{93}, \bibinfo{pages}{105276}.
\newblock \DOIprefix\doi{10.1016/j.ast.2019.07.009}.
\bibitem[{Thomas et~al.(2020)Thomas, Baysinger, Sutherlin, Bean, Clements,
  Kobayashi, Garcia, Fabisinski and Capizzo}]{thomas2020solar}
\bibinfo{author}{Thomas, D.}, \bibinfo{author}{Baysinger, M.},
  \bibinfo{author}{Sutherlin, S.}, \bibinfo{author}{Bean, Q.},
  \bibinfo{author}{Clements, K.}, \bibinfo{author}{Kobayashi, K.},
  \bibinfo{author}{Garcia, J.}, \bibinfo{author}{Fabisinski, L.},
  \bibinfo{author}{Capizzo, P.}, \bibinfo{year}{2020}.
\newblock \bibinfo{title}{{Solar Polar Imager} concept}, in:
  \bibinfo{booktitle}{ASCEND}. \bibinfo{address}{Virtual Event}.
\newblock \DOIprefix\doi{10.2514/6.2020-4060}. \bibinfo{note}{{AIAA}
  2020-4060}.
\bibitem[{Tsuda et~al.(2013)Tsuda, Mori, Funase, Sawada, Yamamoto, Saiki, Endo,
  Yonekura, Hoshino and Kawaguchi}]{tsuda2013achievement}
\bibinfo{author}{Tsuda, Y.}, \bibinfo{author}{Mori, O.},
  \bibinfo{author}{Funase, R.}, \bibinfo{author}{Sawada, H.},
  \bibinfo{author}{Yamamoto, T.}, \bibinfo{author}{Saiki, T.},
  \bibinfo{author}{Endo, T.}, \bibinfo{author}{Yonekura, K.},
  \bibinfo{author}{Hoshino, H.}, \bibinfo{author}{Kawaguchi, J.},
  \bibinfo{year}{2013}.
\newblock \bibinfo{title}{Achievement of {IKAROS—Japanese} deep space solar
  sail demonstration mission}.
\newblock \bibinfo{journal}{Acta Astronautica} \bibinfo{volume}{82},
  \bibinfo{pages}{183--188}.
\newblock \DOIprefix\doi{10.1016/j.actaastro.2012.03.032}.
\bibitem[{Tyler et~al.(2023)Tyler, Diedrich, Gauvain, Inness, Heaton and
  Orphee}]{Tyler2024}
\bibinfo{author}{Tyler, D.}, \bibinfo{author}{Diedrich, B.},
  \bibinfo{author}{Gauvain, B.}, \bibinfo{author}{Inness, J.},
  \bibinfo{author}{Heaton, A.}, \bibinfo{author}{Orphee, J.},
  \bibinfo{year}{2023}.
\newblock \bibinfo{title}{Attitude control approach for {Solar Cruiser}, a
  large, deep space solar sail mission}, in: \bibinfo{booktitle}{AAS Guidance,
  Navigation and Control Conference}, \bibinfo{address}{Breckenridge, CO}.
\bibitem[{Wang et~al.(2025)Wang, Cheng, He and Yuan}]{Wang2025-ql}
\bibinfo{author}{Wang, J.}, \bibinfo{author}{Cheng, Z.}, \bibinfo{author}{He,
  G.}, \bibinfo{author}{Yuan, H.}, \bibinfo{year}{2025}.
\newblock \bibinfo{title}{Uncertainty characterization of solar sail thrust
  with a multiscale modeling method}.
\newblock \bibinfo{journal}{Advances in Space Research} \bibinfo{volume}{75},
  \bibinfo{pages}{5640--5655}.
\newblock \DOIprefix\doi{10.1016/j.asr.2025.01.026}.
\bibitem[{Wilkie et~al.(2025)Wilkie, Fernandez, Stohlman, Warren, Schneider,
  Dean, Turczynski, Denkins, Kang, Aquilina, Shih, Li, Perez, Rhodes,
  Saravanan, Tomer and Hickman}]{WilkieKeynote}
\bibinfo{author}{Wilkie, K.}, \bibinfo{author}{Fernandez, J.},
  \bibinfo{author}{Stohlman, O.}, \bibinfo{author}{Warren, J.},
  \bibinfo{author}{Schneider, N.}, \bibinfo{author}{Dean, G.},
  \bibinfo{author}{Turczynski, C.}, \bibinfo{author}{Denkins, T.},
  \bibinfo{author}{Kang, J.H.}, \bibinfo{author}{Aquilina, R.},
  \bibinfo{author}{Shih, P.}, \bibinfo{author}{Li, D.}, \bibinfo{author}{Perez,
  M.}, \bibinfo{author}{Rhodes, A.}, \bibinfo{author}{Saravanan, P.},
  \bibinfo{author}{Tomer, S.}, \bibinfo{author}{Hickman, T.},
  \bibinfo{year}{2025}.
\newblock \bibinfo{title}{Adventures in solar sailing: Lessons learned from the
  {Advanced Solar Sail System (ACS3)} mission - volume {I}}, in:
  \bibinfo{booktitle}{7th International Symposium on Space Sailing},
  \bibinfo{address}{Delft, The Netherlands}.
\bibitem[{Wilkie et~al.(2021)Wilkie, Fernandez, Stohlman, Schneider, Dean,
  Kang, Warren, Cook, Brown, Denkins, Horner, Tapio, Straubel, Richter and
  Heiligers}]{wilkie2021overview}
\bibinfo{author}{Wilkie, W.}, \bibinfo{author}{Fernandez, J.},
  \bibinfo{author}{Stohlman, O.}, \bibinfo{author}{Schneider, N.},
  \bibinfo{author}{Dean, G.}, \bibinfo{author}{Kang, J.},
  \bibinfo{author}{Warren, J.}, \bibinfo{author}{Cook, S.},
  \bibinfo{author}{Brown, P.}, \bibinfo{author}{Denkins, T.},
  \bibinfo{author}{Horner, S.}, \bibinfo{author}{Tapio, E.},
  \bibinfo{author}{Straubel, M.}, \bibinfo{author}{Richter, M.},
  \bibinfo{author}{Heiligers, J.}, \bibinfo{year}{2021}.
\newblock \bibinfo{title}{Overview of the {NASA Advanced Composite Solar Sail
  System (ACS3)} technology demonstration project}, in:
  \bibinfo{booktitle}{AIAA SciTech Forum}, \bibinfo{address}{Virtual Event}.
\newblock \DOIprefix\doi{10.2514/6.2021-1260}. \bibinfo{note}{{AIAA}
  2021-1260}.
\bibitem[{Zhang et~al.(2021a)Zhang, Gong and Baoyin}]{zhang2021three}
\bibinfo{author}{Zhang, F.}, \bibinfo{author}{Gong, S.},
  \bibinfo{author}{Baoyin, H.}, \bibinfo{year}{2021}a.
\newblock \bibinfo{title}{Three-axes attitude control of solar sail based on
  shape variation of booms}.
\newblock \bibinfo{journal}{Aerospace} \bibinfo{volume}{8},
  \bibinfo{pages}{198}.
\newblock \DOIprefix\doi{10.3390/aerospace8080198}.
\bibitem[{Zhang et~al.(2021b)Zhang, Shengping, Haoran and
  Baoyin}]{zhang2021solar}
\bibinfo{author}{Zhang, F.}, \bibinfo{author}{Shengping, G.},
  \bibinfo{author}{Haoran, G.}, \bibinfo{author}{Baoyin, H.},
  \bibinfo{year}{2021}b.
\newblock \bibinfo{title}{Solar sail attitude control using shape variation of
  booms}.
\newblock \bibinfo{journal}{Chinese Journal of Aeronautics}
  \bibinfo{volume}{35}, \bibinfo{pages}{326--336}.
\newblock \DOIprefix\doi{10.1016/j.cja.2021.10.036}.

\end{thebibliography}

\end{document}